\documentclass[reprint,amsmath,amssymb,aps,nofootinbib,longbibliography]{revtex4-2}
\usepackage{graphicx}
\usepackage{amsmath}
\usepackage{dcolumn}
\usepackage{bm}

\usepackage{mathtools}
\usepackage{blkarray}
\usepackage{footnote}
\usepackage{hyperref}
\hypersetup{
    colorlinks,
    citecolor=blue,
    filecolor=blue,
    linkcolor=blue,
    urlcolor=blue,
    pagebackref=true,
    pdfdisplaydoctitle=true}
\usepackage[toc,page]{appendix}
\usepackage{adjustbox}
\usepackage{float}
\usepackage{color}

\usepackage{scalerel}
\usepackage{tikz}
\usetikzlibrary{svg.path}

\definecolor{orcidlogocol}{HTML}{A6CE39}
\tikzset{
  orcidlogo/.pic={
    \fill[orcidlogocol] svg{M256,128c0,70.7-57.3,128-128,128C57.3,256,0,198.7,0,128C0,57.3,57.3,0,128,0C198.7,0,256,57.3,256,128z};
    \fill[white] svg{M86.3,186.2H70.9V79.1h15.4v48.4V186.2z}
                 svg{M108.9,79.1h41.6c39.6,0,57,28.3,57,53.6c0,27.5-21.5,53.6-56.8,53.6h-41.8V79.1z M124.3,172.4h24.5c34.9,0,42.9-26.5,42.9-39.7c0-21.5-13.7-39.7-43.7-39.7h-23.7V172.4z}
                 svg{M88.7,56.8c0,5.5-4.5,10.1-10.1,10.1c-5.6,0-10.1-4.6-10.1-10.1c0-5.6,4.5-10.1,10.1-10.1C84.2,46.7,88.7,51.3,88.7,56.8z};
  }
}

\newcommand\orcidicon[1]{\href{https://orcid.org/#1}{\mbox{\scalerel*{
\begin{tikzpicture}[yscale=-1,transform shape]
\pic{orcidlogo};
\end{tikzpicture}
}{|}}}}

\definecolor{myred}{rgb}{1,0,0}
\definecolor{myblue}{rgb}{0,0,1}
\colorlet{lightred}{myred!60!white}
\colorlet{lightblue}{myblue!60!white}
\colorlet{lightgreen}{green!60!white}
\colorlet{darkred}{myred!60!black}

\definecolor{mypurple}{rgb}{1,0,1}
\colorlet{lavender}{mypurple!50!white}
\colorlet{lightgray}{gray!30!white}

\newcommand{\todounseen}[1]{}

\newcommand{\tosortunseen}[1]{}

\newcommand{\mycaption}[1]{\textcolor{black}{#1}}

\newcommand{\jbm}[1]{\textcolor{black}{#1}}
\newcommand{\jbmb}[1]{\textcolor{black}{#1}}
\newcommand{\jbmc}[1]{\textcolor{black}{#1}}
\newcommand{\jbmd}[1]{\textcolor{black}{#1}}
\newcommand{\jbme}[1]{\textcolor{black}{#1}}
\newcommand{\jbmf}[1]{\textcolor{black}{#1}}

\usepackage{mathptmx}
\usepackage{mathrsfs}
\DeclareMathAlphabet{\mathcal}{OMS}{cmsy}{m}{n}

\usepackage[english]{babel}

\begin{document}

\title{
Universal Chemical Formula Dependence of \textit{Ab Initio} Low-Energy Effective Hamiltonian in Single-Layer Carrier Doped Cuprate Superconductors --- Study by Hierarchical Dependence Extraction Algorithm
}
\author{
Jean-Baptiste Mor\'ee$^{1}$ \orcidicon{0000-0002-0710-9880} 
and Ryotaro Arita$^{1,2}$ \orcidicon{0000-0001-5725-072X}
}
\affiliation{
$^{1}$ RIKEN Center for Emergent Matter Science, Wako, Saitama 351-0198, Japan \looseness=-1 \\
$^{2}$ Research Center for Advanced Science and Technology, University of Tokyo, Komaba, Meguro-ku, Tokyo 153-8904, Japan
}

\begin{abstract}
We explore the possibility to control the superconducting (SC) transition temperature at optimal hole doping $T_{c}^{\rm opt}$ in cuprates by tuning the chemical formula (CF).
$T_{c}^{\rm opt}$ can be theoretically predicted from the parameters of the \textit{ab initio} low-energy effective Hamiltonian (LEH) with one antibonding (AB) Cu$3d_{x^2-y^2}$/O$2p_{\sigma}$ orbital per Cu atom in the CuO$_2$ plane,
notably the nearest neighbor hopping amplitude $|t_1|$ and the ratio $u=U/|t_1|$, where $U$ is the onsite effective Coulomb repulsion. However, the CF dependence of $|t_1|$ and $u$ is a highly nontrivial question.
In this paper, we propose the universal dependence of $|t_1|$ and $u$ on the CF and structural features in hole doped cuprates with a single CuO$_2$ layer sandwiched between block layers. To do so, we perform extensive \textit{ab initio} calculations of $|t_1|$ and $u$ and analyze the results by employing a machine learning method called Hierarchical Dependence Extraction (HDE). The main results are the following: (a) $|t_1|$ has a main-order dependence on the radii $R_{\rm X}$ and $R_{\rm A}$ of the apical anion X and cation A in the block layer. ($|t_1|$ increases when $R_{\rm X}$ or $R_{\rm A}$ decreases.) (b) $u$ has a main-order dependence on the negative ionic charge $Z_{\rm X}$ of X and the hole doping $\delta$ of the AB orbital. ($u$ decreases when $|Z_{\rm X}|$ increases or $\delta$ increases.) We elucidate and discuss the microscopic mechanism of (a,b). We demonstrate the predictive power of the HDE by showing the consistency between (a,b) and results from previous works. The present results provide a basis for optimizing SC properties in cuprates and possibly akin materials. 
Also, the HDE method offers a general platform to identify dependencies between physical quantities.
\end{abstract}

\maketitle

\renewcommand{\figurename}{FIG.}
\renewcommand{\tablename}{TABLE}

\section{Introduction}
\label{sec:intro}

\jbmc{One of the grand challenges in condensed matter physics is the design of superconductors (SCs) with high transition temperature $T_{c}$.
The diverse distribution of $T_{c}^{\rm opt}$ (the experimental $T_{c}$ at optimal hole doping) in carrier doped SC cuprates 
provides useful insights into such design.
}
\jbm{
In carrier doped SC cuprates, we have}
$T_{c}^{\rm opt} \simeq 10-13\jbmd{8}$ K at ambient pressure 
\cite{Torrance1988,Gao1994,Yamamoto2015,Putilin1993,Singh1994,Grant1987,Attfield1998,AlMamouri1994,Slater1995,Tatsuki1996,Hiroi1994,Kim2006,Zenitani1996} 
and up to $T_{c}^{\rm opt} \simeq \jbm{166}$ K under pressure in HgBa$_2$Ca$_2$Cu$_3$O$_8$ (Hg1223)~\cite{Gao1994,Yamamoto2015}.
This diverse distribution is already present in 
single-layer cuprates,
in which $T_{c}^{\rm opt} \simeq 10-94$ K at ambient pressure and up to $T_{c}^{\rm opt}  \simeq 110$ K in HgBa$_2$CuO$_4$ (Hg1201) under pressure~\cite{Gao1994}.
Thus, single-layer carrier doped cuprates are a platform of choice to investigate %
the microscopic mechanism and origin of the materials dependence of $T_{c}^{\rm opt}$.

The diverse distribution of $T_{c}^{\rm opt}$ can be described by the materials dependent %
AB LEH parameters, especially $|t_1|$ and \jbmc{$u=U/|t_1|$}~\cite{Schmid2023}. %
Indeed, the scaling
\begin{equation}
T_{c}^{\rm opt} \simeq 0.16 |t_1|F_{\rm SC}
\label{eq:tcopt}
\end{equation}
was proposed~\cite{Schmid2023}, in which \jbmc{the dimensionless SC order parameter} $F_{\rm SC}$ mainly depends on $u$.
The $u$ dependence of $F_{\rm SC}$ 
\jbmf{is reminded in Appendix~\ref{app:ufsc}}
and is summarized below.
$F_{\rm SC}$ is zero for $u \lesssim 6.5$, 
and increases sharply with increasing $u \simeq 6.5-8.0$ (weak-coupling regime), 
reaching a maximum at $u_{\rm opt} \simeq 8.0-8.5$ (optimal regime);
then, $F_{\rm SC}$ decreases with increasing $u \gtrsim 9.0$ (strong-coupling regime).
For a given material, $|t_1|$ and $u$ can be calculated %
by using the multiscale \textit{ab initio} scheme for correlated electrons (MACE) \cite{Imada2010,Hirayama2013,Hirayama2018,Hirayama2019},
\jbmd{which allowed to establish Eq.~\eqref{eq:tcopt}.}

Thus, a key point for materials design of higher-$T_{c}^{\rm opt}$ cuprates is to elucidate the universal \jbm{chemical formula (CF)} dependence of $|t_1|$ and $u$.
In previous works on cuprates such as Hg1201, Bi$_2$Sr$_2$CuO$_6$ (Bi2201), Bi$_2$Sr$_2$CaCu$_2$O$_8$ (Bi2212) and CaCuO$_2$~\cite{Moree2022} as well as Hg1223~\cite{Moree2023Hg1223}, the non\jbmd{trivial} dependence of $|t_1|$ and $u$ on the interatomic distances and the CF has been partly clarified.
However, the more general CF dependence of $|t_1|$ and $u$ is required to obtain a thorough understanding of the CF dependence of $T_{c}^{\rm opt}$.
\\

The goal of this paper is twofold.
First, we propose \jbmd{a machine learning} procedure that is tailor-made to extract the nonlinear dependencies of a given quantity $y$ on other quantities $x_i$
from the main-order to the higher-order.
This procedure is denoted as \jbmb{Hierarchical Dependence Extraction (HDE)}.
Second, we propose the universal CF dependence of $|t_1|$ and $u$ in single-layer cuprates,
by performing explicit \textit{ab initio} calculations of the AB LEH for a training set that is representative of single-layer cuprates (including copper oxides, oxychlorides and oxyfluorides), and applying the HDE to analyze the results and construct expressions of $|t_1|$ and $u$.
We generalize the existing MACE procedure to obtain the crystal parameters as a function of the chemical variables (the radii and charges of the cations and anions in the block layer), and \textit{in fine} the AB LEH as a function of the chemical variables.
The combination of the generalized MACE (gMACE) procedure with the analysis of the results by the HDE is denoted as gMACE+HDE.
We demonstrate the predictive power of the HDE by showing the consistency between the universal CF dependencies of $|t_1|$ and $u$ obtained \jbmd{by employing} the gMACE+HDE and previous results on Hg1201, Bi2201, Bi2212, CaCuO$_2$ and Hg1223~\cite{Moree2022,Moree2023Hg1223}. %
\\

This paper is organized as follows.
Section~\ref{sec:overview} gives an overview of the main-order dependence (MOD) of $|t_1|$ and $u$ on the chemical variables.
Section~\ref{sec:meth} describes \jbm{the HDE} \jbmd{and} the gMACE methodolog\jbmd{ies} employed in this paper, and how the HDE is applied to analyze the results of the gMACE calculation in the gMACE+HDE.
Section~\ref{sec:results} details the results on the MOD of $|t_1|$ and $u$ on the chemical variables,
and proposes the microscopic mechanism underlying to this MOD.
Section~\ref{sec:disc} discusses the results from the perspective of Eq.~\eqref{eq:tcopt}, and proposes guidelines to optimize the value of $T_{c}^{\rm opt}$ in future design of single-layer cuprates \jbmd{for which the gMACE calculation is performed}.
Section~\ref{sec:concl} is the conclusion.
\jbmf{Appendix~\ref{app:ufsc} reminds the $u$ dependence of $F_{\rm SC}$ from Ref.~\cite{Schmid2023}.}
Appendix~\ref{app:sc} details the choice of the training set of single-layer cuprates.
Appendix~\ref{app:meth-sr} gives %
\jbmd{details} on the HDE. 
Appendix~\ref{app:meth-mace} gives %
\jbmd{details on} the gMACE procedure and the values of the intermediary quantities \jbmd{obtained in} the gMACE \jbmd{calculation}.
Appendix~\ref{app:HDEinterp} gives the analysis of the competition between variables in the MOD.
\jbmc{Appendix~\ref{app:xfcl} gives details on the hole doping dependence of the screening for each compound in the training set, and possible implications on the SC properties.}
\jbmd{Appendix~\ref{app:doseo} gives complements on the density of states near the Fermi level in hole-doped oxychlorides.}

\section{Overview: Main-order dependence of AB LEH parameters on chemical formula}
\label{sec:overview}

\begin{figure}[!htb]
\includegraphics[scale=0.13]{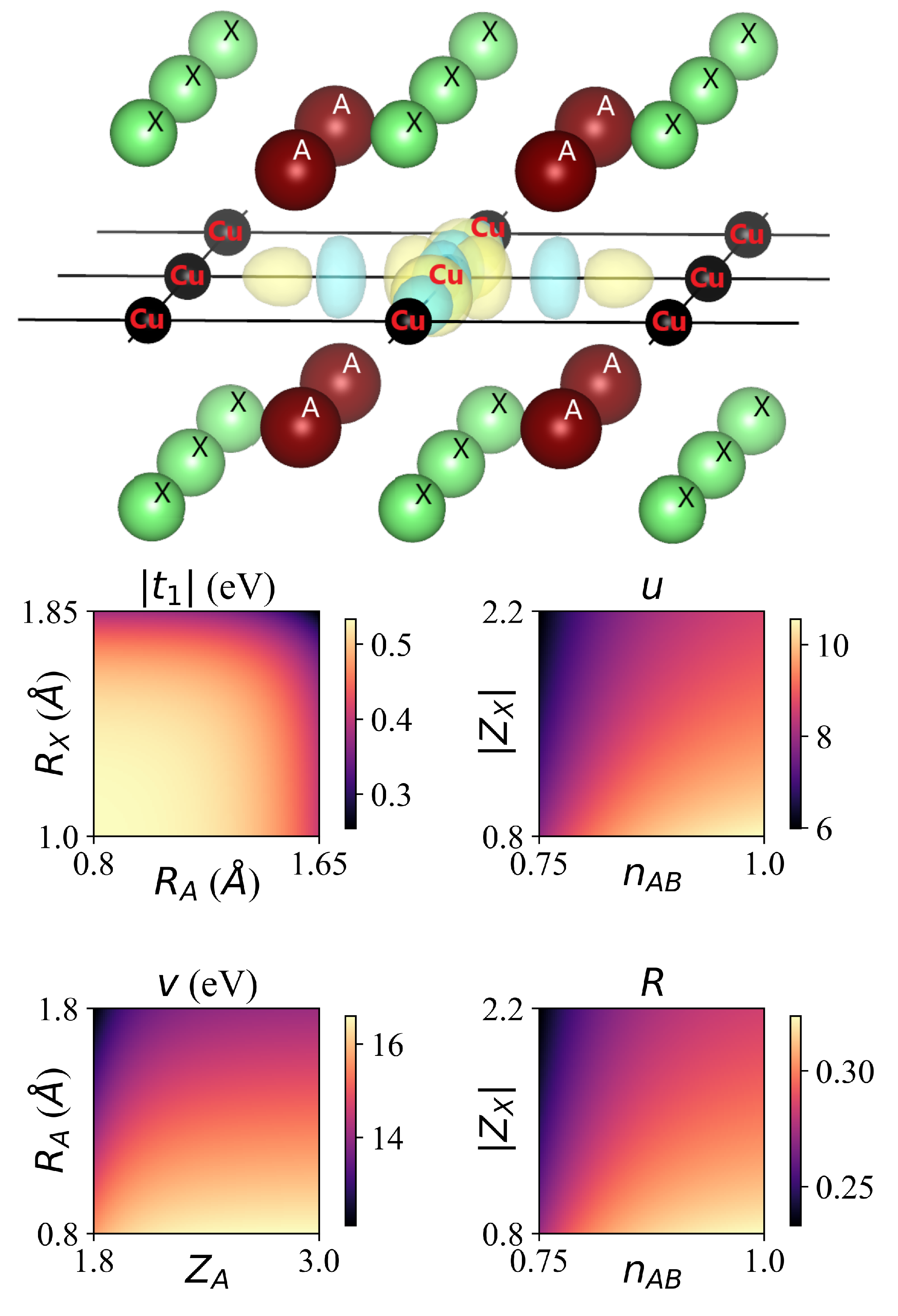}
\caption{\mycaption{
Upper panel: 
Simplified representation of the CuO$_2$ plane and the surrounding crystalline environment.
We show the square lattice formed by the Cu atoms in the CuO$_2$ plane (the in-plane O atoms are not shown),
and the isosurface of the AB orbital centered on one of the Cu atoms \jbmd{(yellow is positive, blue is negative)}.
We also show the A cations and apical X anions near the CuO$_2$ plane.
Lower panel: 
Representation of the MODs of the AB LEH parameters on the CF obtained from gMACE+HDE.
We show the MOD of $|t_1|$ [Eq.~\eqref{eq:intro-t1}] and $u=U/|t_1|$ [Eq.~\eqref{eq:intro-u}] together with the onsite bare Coulomb interaction $v$ [Eq.~\eqref{eq:intro-v}] and the screening ratio $R=U/v$ [Eq.~\eqref{eq:intro-R}].
$R_{\rm A}$ and $Z_{\rm A}$ ($R_{\rm X}$ and $Z_{\rm X}$) are the crystal ionic radius %
and ionic charge of the A cation (apical X anion).
$n_{\rm AB}$ is the average number of electrons per AB orbital.
}}
\label{fig:summary}
\end{figure}

Here, we give an overview of the MOD of AB LEH parameters on the CF
that is obtained by applying the gMACE+HDE.
(Details on the results are given later in Sec.~\ref{sec:results},
and prescriptions to optimize $T_{c}^{\rm opt}$ based on these results are proposed in Sec.~\ref{sec:disc}.)
The MODs of $|t_1|$ and $u$ are summarized below in (I,II) and illustrated in Fig.~\ref{fig:summary}.
\\

(I) $|t_1|$ mainly depends on the \jbmd{crystal} ionic radii $R_{\rm X}$ and $R_{\rm A}$ of the apical anion \jbmd{X} and cation \jbmd{A} in the block layer that separates two CuO$_2$ layers.
\jbmd{(See Fig.~\ref{fig:summary} for an illustration of X and A.)}
The MOD up to the second order (MOD2) is
\footnote{In the MOD2s given in Eqs.~\eqref{eq:intro-t1},~\eqref{eq:intro-u},~\eqref{eq:intro-v} and~\eqref{eq:intro-R},
the unit of $|t_1|$ and $v$ is eV, whereas $u$ and $R$ are dimensionless.
The ranges of values obtained in our \textit{ab initio} calculations are 
$|t_1| = 0.40-0.57$ eV,
$u = 7.2-10.6$,
$v = 13.3-16.4$ eV,
and $R = 0.23-0.34$.
These ranges of values are reproduced by the MOD2s,
as seen in Fig.~\ref{fig:summary}.}
\begin{equation}
|t_1|_{{\rm MOD}2} = 0.534 - 0.000327 \Big[ R_{\rm X}^{9.99} + 7.99 R_{\rm A}^{7.76} \Big].
\label{eq:intro-t1}
\end{equation}
Qualitatively, $|t_1|$ increases when $R_{\rm X}$ or $R_{\rm A}$ decreases (see Fig.~\ref{fig:summary}).
The microscopic mechanism is summarized as follows:
Reducing $R_{\rm X}$ and/or $R_{\rm A}$ reduces the chemical pressure that pushes atoms apart from each other inside the crystal.
This reduces the cell parameter $a$ and thus the distance between Cu atoms in the CuO$_2$ plane,
which increases the overlap between AB orbitals located on neighboring Cu sites, and thus $|t_1|$.
This result is consistent with that in~\cite{Moree2023Hg1223}, in which $|t_1|$ increases when applying physical uniaxial pressure $P_{a}$ along $a$ direction. (The application of $P_a$ reduces $a$.)
\\

(II) $u$ mainly depends on the negative ionic charge $Z_{\rm X}$ of the apical anion
and the average number of electrons $n_{\rm AB}$ in the AB orbital.
(We have $n_{\rm AB} = 1 - \delta$, where $\delta$ is the hole doping.)
The MOD2 is
\begin{equation}
u_{{\rm MOD}2} = 200.72\jbmb{5} - 190.29\jbmb{4} \Big[|Z_{\rm X}|^{0.01} +  0.00155 n_{\rm AB}^{-7.99}\Big].
\label{eq:intro-u}
\end{equation}
Qualitatively, $u$ increases when $|Z_{\rm X}|$ decreases or $n_{\rm AB}$ increases (see Fig.~\ref{fig:summary}).
To understand the origin of the MOD of $u$ \jbmd{in Eq.~\eqref{eq:intro-u}}, we decompose $u=vR/|t_1|$,
where $v$ is the onsite bare Coulomb interaction and $R=U/v$ is the screening ratio,
and we examine the MOD\jbmd{s} of $v$ and $R$ in (III) and (IV) below.
(As shown below, the MOD of $u$ is mainly determined by the MOD of $R$.)
\\

(III) $v$ mainly depends on $R_{\rm A}$ and the positive ionic charge $Z_{\rm A}$ of the cation.
The MOD2 is
\begin{equation}
v_{{\rm MOD}2} = 18.874 -  2.787 R_{\rm A}^{0.92} \Big[ 1 + 87.9\jbmb{0} Z_{\rm A}^{-9.12} \Big].
\label{eq:intro-v}
\end{equation}
Qualitatively, $v$ increases when $R_{\rm A}$ decreases or $Z_{\rm A}$ increases (see Fig.~\ref{fig:summary}).
The microscopic mechanism is summarized as follows: 
Reducing $R_{\rm A}$ or increasing $Z_{\rm A}$ modifies the crystal electric field
(namely, the Madelung potential created by cations and anions in the crystal) 
felt by the Cu$3d_{x^2-y^2}$ and O$2p_{\sigma}$ electrons.
This stabilizes the in-plane O$2p_{\sigma}$ orbitals with respect to the Cu$3d_{x^2-y^2}$ orbitals, 
which increases the Cu$3d_{x^2-y^2}$/O$2p_{\sigma}$ charge transfer energy $\Delta E_{xp}$. 
(Details are given later in Sec.~\ref{sec:results}.)
This reduces the Cu$3d_{x^2-y^2}$/O$2p_{\sigma}$ hybridization,
which increases the Cu$3d_{x^2-y^2}$ atomic character of the AB orbital.
This increases the localization of the AB orbital and thus $v$.
This result is consistent with~\cite{Moree2023Hg1223}.
\\

(IV) The CF dependence of $R$ is more complex than that of $|t_1|$ and $v$, but we identify a \jbmb{rough} MOD2 of $R$ on $Z_{\rm X}$ and $n_{\rm AB}$,
which is
\begin{equation}
R_{{\rm MOD}2} = \jbmb{4.224} - \jbmb{3.906} \Big[|Z_{\rm X}|^{0.01} + 0.00078 n_{\rm AB}^{-9.99}\Big].
\label{eq:intro-R}
\end{equation}
Qualitatively, $R$ increases when (i) $|Z_{\rm X}|$ decreases or (ii) $n_{\rm AB}$ increases (see Fig.~\ref{fig:summary}).
The microscopic mechanism of (i) and (ii) is briefly summarized below. 
(Details are given later in Sec.~\ref{sec:results}.)

(i) Decreasing $|Z_{\rm X}|$ reduces the negative charge of the apical anion.
This reduces the negative Madelung potential (MP$^{-}$) 
created by the apical anion and felt by the electrons in the nearby CuO$_2$ plane.
This reduces the energy of the electrons in the CuO$_2$ plane, and also reduces the Fermi energy.
As a consequence, the empty states become higher in energy relative to the Fermi level.
This reduces the screening from empty states, and thus, increases $R$.

(ii) The decrease in $R$ with decreasing $n_{\rm AB}$ (increasing $\delta$) is consistent with~\cite{Moree2022},
in which the increase in $\delta$ causes the rapid decrease in $R$ and thus $u$.
\jbmd{(}In~\cite{Moree2022}, calculations were made at fixed $|Z_{\rm X}|=2$ and varying $n_{\rm AB}=1.0$, $0.9$ and $0.8$, which corresponds to $\delta=0.0$, $0.1$ and $0.2$.\jbmd{)}
The rapid decrease in $u$ eventually suppresses $F_{\rm SC}$~\cite{Schmid2023} so that the system ends up in the metallic state,
in agreement with the experimental ground state in the overdoped region. \todounseen{~\cite{?}}
\\

Remarkably, the MOD2 of $R$ [Eq.~\eqref{eq:intro-R}] is very similar to the MOD2 of $u$ [Eq.~\eqref{eq:intro-u}].
Thus, the MOD2 of $u=vR/|t_1|$ is dominated by the MOD2 of $R$.
Consistently, in the \textit{ab initio} result,
the diverse distribution of $u \simeq 7-10.5$
originates mainly (albeit not exclusively) from the diverse distribution of $R$.
Indeed, the relative variation between the minimum and maximum \textit{ab initio} values
is $27\%$ for $v \simeq 13.5-16.5$ eV, $30\%$ for $|t_1| \simeq 0.40-0.55$ eV, $48\%$ for $R \simeq 0.22-0.34$,
and $50\%$ for $u \simeq 7-10.5$.
(The variation in $u$ is reproduced by that in $R$.)
Still, the relative variation in $|t_1|$ and $v$ is nonnegligible compared to that in $R$,
so that the CF dependencies of $|t_1|$ and $v$ also contribute to the CF dependence of $u$
beyond the MOD2 %
\jbmd{in Eq.~\eqref{eq:intro-u}.}
In Sec.~\ref{sec:results}, we will decompose $u=vR/|t_1|$, and we will discuss in detail the CF dependence of $|t_1|$, $v$ and $R$.

\section{Methodology}
\label{sec:meth}

\subsection{Framework of HDE}
\label{sec:meth-algo}

Here, we summarize the HDE methodology
to construct a descriptor for a given physical quantity $y$ as a function of other given quantities in the variable space $\mathcal{V}=\{ x_i, i=1..N_{\mathcal{V}} \}$\jbm{, where $i$ is the variable index.}
(We make complete abstraction of the physical meaning of these variables.)
We restrict the presentation to the essence of the procedure; %
complementary discussions and computational details are given in Appendix~\ref{app:meth-sr}\jbm{, and the motivation of the HDE procedure is discussed in detail in Appendix~\ref{app:meth-sr-motiv}.}
\\

\paragraph*{General problem ---}
Starting from the variable space $\mathcal{V}$, we define the candidate descriptor space
\begin{equation}
\mathcal{C}_{\mathcal{V}}^{} = \{ x^{}_{}({\bf p}), {\bf p}_{} \}.
\label{eq:d}
\end{equation}
The elements $x({\bf p})$ of $\mathcal{C}_{\mathcal{V}}$ are called "candidate descriptors",
and are functions of \jbm{the variables $x_i$ in $\mathcal{V}$}; their analytic expression depends on variational parameters which are encoded in the vector ${\bf p}_{}$.
[The expression of $x({\bf p})$ and definition of variational parameters are given later.]
The general problem consists in finding ${\bf p}^{\rm opt}$ such that $x^{\rm opt}=x({\bf p}^{\rm opt})$ is the best candidate descriptor for $y$ among the elements of $\mathcal{C}_{\mathcal{V}}$, i.e.
\begin{equation}
f[y,x^{}({\bf p}^{\rm opt})] = {\rm max}_{{\bf p}} f[y,x({\bf p})],
\label{eq:ftilde}
\end{equation}
where $f_{}$ is a fitness function that describes how well $y$ is described by $x^{}({\bf p}^{\rm opt})$. 
\\

\paragraph*{Fitness function ---}
We choose the fitness function 
\begin{equation}
f[y,x({\bf p})] = |\rho[y,x({\bf p})]|,
\label{eq:deffit}
\end{equation}
where $\rho$ is the Pearson correlation coefficient.
(The definition of $\rho$ is reminded in Appendix~\ref{app:meth-sr-fit}.)
Further discussions on the choice of $f$ are given in Appendix~\ref{app:meth-sr-fit}, and key points are summarized below.
Eq.~\eqref{eq:deffit} encodes the affine dependence of $y$ on $x({\bf p})$, which is the relevant information.
The values of $f[y,x({\bf p})]$ are between 0 and 1.
The candidate descriptor $x({\bf p})$ is deemed perfect 
if $f[y,x({\bf p})] = 1$:
In this case, there exist $k_0$, $k_1$ such that we have rigorously
$y = k_0 + k_1 x^{}({\bf p})$.
The closer $f[y,x({\bf p})]$ is to $1$, the more accurate the affine dependence of $y$ on $x^{}({\bf p})$ is.
Note that %
$f$ has an invariance property
\begin{equation}
f[y,k_0 + k_1x({\bf p})] = f[y,x({\bf p})]
\label{eq:finv}
\end{equation}
(for $k_1 \neq 0$), which is prominently used in the HDE procedure as discussed later.
The value of $f$ is insensitive to the values of $k_0$ and $k_1$ in Eq.~\eqref{eq:finv},
and has an intuitive interpretation irrespective of the values of $k_0$ and $k_1$.
(See Appendix~\ref{app:meth-sr-fit} for details.)
\\

\paragraph*{Expression of the candidate descriptor ---}
The descriptor is constructed iteratively by adding factors that contain dependencies of $y$ on $x_i$ from the \jbmd{lowest} order to the highest order.
The \jbmd{expressions of the} candidate descriptors \jbmd{that are considered in this paper} at generation $g=1$ and $g \geq 2$ %
\jbmd{are,}
respectively,
\begin{align}
x_{(1)} & = x_{i_1}^{\alpha_1}, \label{eq:editer1} \\
x_{(g)} & = x^{\rm opt}_{(g-1)} \star_{(\zeta_{\jbm{g}},\beta_{\jbm{g}},\alpha_{\jbm{g}})} x_{i_g}, \label{eq:editer}
\end{align}
where $x^{\rm opt}_{(g-1)}$ is the best candidate descriptor at generation $g-1$, 
$i_g$ is one of the indices between $1$ and $N_{\mathcal{V}}$, %
and
the %
parametric operator $\star_{(\zeta,\beta,\alpha)}$ (dubbed hereafter as the wildcard operator) is defined as
\begin{equation}
x \star_{(\zeta,\beta,\alpha)} x' = x \Bigg[ 1 + \zeta \frac{x^{' \alpha}} {x^{\beta}} \Bigg].
\label{eq:op}
\end{equation}
The variational parameter vector at $g$ is ${\bf p}_g = (i_g,\alpha_g,\zeta_g,\beta_g)$ [${\bf p}_1 = (i_1,\alpha_1)$ \jbmd{at} $g=1$].
For $g=1..N$, we obtain ${\bf p}_{(N)}=({\bf p}_1,{\bf p}_2,...,{\bf p}_N)$ that corresponds to ${\bf p}$ in Eq.~\eqref{eq:ftilde}.
\jbmd{At $g$, the} values of variational parameters that maximize the fitness function are denoted as  ${\bf p}^{\rm opt}_g = (i_g^{\rm opt},\alpha_g^{\rm opt},\zeta_g^{\rm opt},\beta_g^{\rm opt})$.

The wildcard operator in Eq.~\eqref{eq:op} is versatile and can represent any algebraic operation depending on the values of $(\zeta,\beta,\alpha)$, \jbmb{as discussed in detail in Appendix~\ref{app:meth-sr-gen}.}
In the practical procedure, 
($\zeta_{\jbm{g}},\beta_{\jbm{g}},\alpha_{\jbm{g}}$) are optimized together with $i_{\jbm{g}}$
to maximize $f[y,x_{(g)}]$, so that the character of the wildcard operator is automatically adjusted to describe $y$ as accurately as possible.
Also, note that the wildcard operator can mimick the identity operator for specific values of $(\zeta,\beta,\alpha)$ (\jbmb{see Appendix~\ref{app:meth-sr-gen}})\jbmd{, so that $x^{\rm opt}_{(g-1)}$ is included in the candidate descriptor space at $g$. This} guarantees that
\begin{equation}
f[y,x^{\rm opt}_{(g+1)}] \geq f[y,x^{\rm opt}_{(g)}].
\label{eq:fincreases}
\end{equation}
Finally, note that even though $y$ has an affine dependence on $x({\bf p})$ when $f[y,x\jbmd{({\bf p}})] = 1$, 
$x({\bf p})$ has a nonlinear dependence on $x_i$ in the general case.
Thus, nonlinearity in the dependence of $y$ on $x_i$ may be described by the above formalism.
\\

\paragraph*{Procedure to construct the descriptor for $y$ as a function of $x_i$ ---}
To obtain $x^{\rm opt}_{(N)}$,
we employ the following procedure, which is denoted as $\mathcal{P}{[}y,\mathcal{V}{]}$.
First, at $g=1$, we determine $\jbmd{{\bf p}^{\rm opt}_1 = (i^{\rm opt}_1,\alpha^{\rm opt}_1)}$ in Eq.~\eqref{eq:editer1} that maximize\jbmd{s} 
\jbmb{$f_{}[y,x_{(1)}]$}, and we obtain $x^{\rm opt}_{(1)}$.
Then, we increment $g$, \jbmd{we} determine $\jbmd{{\bf p}^{\rm opt}_g = (i^{\rm opt}_g,\alpha^{\rm opt}_g,\zeta^{\rm opt}_g,\beta^{\rm opt}_g)}$ in Eq.~\eqref{eq:editer} that maximize\jbmd{s} 
\jbmb{$f_{}[y,x_{(g)}]$}, and we obtain $x^{\rm opt}_{(g)}$.
We iterate up to $g=N$.
(The computational details of the optimization are given in Appendix~\ref{app:meth-sr-cd}.)
We obtain
\begin{equation}
x^{\rm opt}_{(N)} = (x_{i^{\rm opt}_1})^{\alpha^{\rm opt}_1} \prod_{g=2}^{N} \Bigg[ 1 + \zeta^{\rm opt}_g \frac{(x_{i^{\rm opt}_g})^{\alpha^{\rm opt}_g}}{{\Big[x^{\rm opt}_{(g-1)}\Big]}^{\beta^{\rm opt}_g}} \Bigg].
\label{eq:xN}
\end{equation}
From this descriptor, we deduce the estimated expression of $y$ as a function of $x_i$, as
\begin{equation}
y_{N} = k_0 + k_1 x^{\rm opt}_{(N)},
\label{eq:ff}
\end{equation}
in which the coefficients $k_0$ and $k_1$ are calculated by an affine regression of $y$ on $x^{\rm opt}_{(N)}$.
\\

\paragraph*{Completeness of the dependence of $y$ on $x_i$ ---}
The procedure $\mathcal{P}{[}y,\mathcal{V}{]}$ allows to probe the completeness of the dependence of $y$ on $x_i$.
When we perform $\mathcal{P}{[}y,\mathcal{V}{]}$,
we assume the following conjecture:
$\mathcal{D}_{}{[}y,\mathcal{V}{]}$ The dependence of $y$ is entirely contained in $\mathcal{V}=\{ x_i \}$.
[This implies that there exists ${\bf p}$ such that $f[y,x({\bf p})] = 1$.]
After performing $\mathcal{P}{[}y,\mathcal{V}{]}$, the validity of $\mathcal{D}_{}{[}y,\mathcal{V}{]}$ can be checked by examining the value of
\begin{equation}
f_{\infty}[y,\mathcal{V}] = {\rm lim}_{g \rightarrow \infty} f_{(g)}[y,\mathcal{V}],
\label{eq:finfty}
\end{equation}
in which
\begin{equation}
f_{(g)}[y,\mathcal{V}] = f[y,x^{\rm opt}_{(g)}].
\end{equation}
If $f_{\infty}[y,\mathcal{V}]$ is close to 1, %
\jbmb{it} is the proof that $\mathcal{D}_{}{[}y,\mathcal{V}{]}$ is correct.
If not, there are two possibilities:
either $\mathcal{D}_{}{[}y,\mathcal{V}{]}$ is incorrect,
or $\mathcal{D}_{}{[}y,\mathcal{V}{]}$ is correct but the procedure $\mathcal{P}{[}y,\mathcal{V}{]}$ is insufficient to capture the whole dependence of $y$ on \jbmd{the variables $x_i$ in $\mathcal{V}$.} %
In the scope of this paper, we assume $\mathcal{D}_{}{[}y,\mathcal{V}{]}$ is incorrect.

In case $\mathcal{D}_{}{[}y,\mathcal{V}{]}$ is incorrect, it is possible to improve the descriptor by replacing $\mathcal{V}$ with a superset of $\mathcal{V}$, as mentioned in the next paragraph.
In case $\mathcal{D}_{}{[}y,\mathcal{V}{]}$ is correct, 
we obtain $x^{\rm opt}_{(N)}$ on which $y$ has an affine dependence,
and we 
\jbmd{obtain}
the expression $y_{N}$ of $y$ as a function of $x_i$ \jbmd{in Eq.~\eqref{eq:ff}}.
[The correctness of $\mathcal{D}_{}{[}y,\mathcal{V}{]}$ is the guarantee that the affine interpolation \jbmd{in Eq.~\eqref{eq:ff}} is accurate.]
\tosortunseen{Note that, in Eq.~\eqref{eq:ff}, the invariance property~\eqref{eq:finv} of the fitness function is not used,
because we need to obtain the values of $y_0$ and $y_1 \neq 0$ in Eq.~\eqref{eq:ff}.
These can be determined without computational effort by the affine regression.
}
\\

\paragraph*{Main-order dependence of $y$ on $x_i$ ---} %
The expression of $x^{\rm opt}_{(N)}$ [Eq.~\eqref{eq:xN}]
has a hierarchical structure which
reveals the hierarchy in the dependencies of $y$ on $x_i$:
The lower (higher) values of $g$ correspond to the lower-order (higher-order) dependencies of $y$ on $x_i$,
and the variable $x_{i_g}$ contains the $g^{\rm th}$-order dependence of $y$.
Note that, when incrementing $g$, we allow only up to \jbm{one variable} to be introduced in Eq.~\eqref{eq:xN}:
This allows to select the variable which corresponds to the $g^{\rm th}$-order dependence.
Also, note that several variables may be in close competition in the $g^{\rm th}$ order dependence,
as discussed later.

In this paper, we mainly discuss the MODs that are contained in $g \lesssim 2-3$;
the full list of dependencies for $g$ from $1$ to $N$ is given in Sec.~S1 of Supplemental Material (SM)~\cite{Moree2024Supplemental_PRX}.
For $g \lesssim 2-3$, we define the MOD of $y$ on $x_i$ up to the $g^{\rm th}$ order [MOD$g$], as
$y_{{\rm MOD}g} = y_{g}$,
which is Eq.~\eqref{eq:ff} with $N=g$. %
Because $y_{{\rm MOD}g}$ \jbmd{depends on} no more than $g$ different variables in $\mathcal{V}$, %
it is possible to represent graphically $y_{{\rm MOD}g}$ as a function of $x_i$ for $g \leq 2$ by using e.g. a color map, as done in Fig.~\ref{fig:summary}.

The accuracy of the MOD$g$ is quantified in $f_{(g)}[y,\mathcal{V}]$.
Note that, in the general case, the MOD$g$ of $y$ on $x_i$ is but a rough description of $y$:
Typically, $f_{(g)}[y,\mathcal{V}] \simeq 0.65-0.95$ \jbmd{for $g \lesssim 3$}.
The accurate description requires to take into account the higher-order dependencies of $y$ on $x_i$ beyond $g \lesssim 3$ as well. %
Nonetheless, the MOD$g$ contains the principal mechanism of the dependence of $y$;
also, for $g \leq 2$, we have $f_{(g)}[y,\mathcal{V}] \simeq 1$ in some particular cases, as seen later.
\\

\paragraph*{Competition between the $x_i$ in the dependence of $y$ ---}
At the generation $g$, we determine the variable $x_i$ that corresponds to $x_{i^{\rm opt}_{g}}$, as mentioned before.
We also examine the competition between variables as detailed below. 
We define the maximal fitness of the variable $x_i$ at the generation $g=1$ and $g \geq 2$ as, respectively,
\begin{align}
\tilde{f}^{}_{(1)}[y,x_i] & = {\rm max}_{\alpha_1} f_{(1)}[y,x_{i}^{\alpha_1}], \label{eq:mfit1}\\
\tilde{f}^{}_{(g)}[y,x_i] & = {\rm max}_{(\alpha_g,\beta_g,\zeta_g)} f_{(g)}[y,x^{\rm opt}_{(g-1)} \star_{(\zeta_{\jbm{g}},\beta_{\jbm{g}},\alpha_{\jbm{g}})} x_{i}], \label{eq:mfitg}
\end{align}
which is the maximal value of the fitness function that is obtained at $g$ by enforcing $x_{i^{\rm opt}_g} = x_i$.
Also, we define the score of the variable $x_i$ at the generation $g$ as
\begin{equation}
s_{(g)}[y,x_i] =  \frac{\tilde{f}_{(g)}[y,x_i] - {\rm min}_{i'} \tilde{f}_{(g)}[y,x_{i'}]}{{\rm max}_{i'} \tilde{f}_{(g)}[y,x_{i'}]  - {\rm min}_{i'} \tilde{f}_{(g)}[y,x_{i'}]}.
\label{eq:score}
\end{equation}
We have $s_{(g)}[y,x_i] = 1$ if  $x_i$ corresponds to $x^{\rm opt}_{i_g}$ \jbmd{that is} found in the optimization,
and $s_{(g)}[y,x_i] = 0$ if $x_i$ has the lowest $\tilde{f}^{}_{(g)}[y,x_i]$ among the variables in $\mathcal{V}$.
Complementary discussions are given in Appendix~\ref{app:HDEinterp}.
In the following, the analysis of the competition between variables is performed in Appendix~\ref{app:HDEinterp} and briefly mentioned in the main text.
\\

\paragraph*{Practical HDE procedure ---}
In practice, given $y$ and $\mathcal{V}=\{ x_i \}$,
the HDE is denoted as ${\rm HDE}_{}{[}y,\mathcal{V}{]}$ and is employed as follows.
We perform $\mathcal{P}{[}y,\mathcal{V}{]}$.
The output quantities are $f_{\infty}[y,\mathcal{V}]$ and \jbmd{${\bf p}^{\rm opt}_{g}$}) for $g$ from $1$ to $N$.
Then, we check the validity of $\mathcal{D}_{}{[}y,\mathcal{V}{]}$ by examining the value of $f_{\infty}[y,\mathcal{V}]$.
If $\mathcal{D}_{}{[}y,\mathcal{V}{]}$ is incorrect, we \jbmd{may} attempt to replace $\mathcal{V}$ by a superset of $\mathcal{V}$ and relaunch the procedure.

If $\mathcal{D}_{}{[}y,\mathcal{V}{]}$ is correct,
we perform the below restricted procedure, denoted as r${\rm HDE}_{}{[}y,\mathcal{V}{]}$.
We attempt to simplify the dependence of $y$ by eliminating the variables on which $y$ has a higher-order dependence,
in decreasing order.
Namely, we take the optimized variable indices $i^{\rm \jbmd{opt}}_1,i^{\rm \jbmd{opt}}_2,...,i^{\rm \jbmd{opt}}_N$ obtained at $g=1,2,...,N$ in the ${\rm HDE}_{}{[}y,\mathcal{V}{]}$.
By using these notations, ${\rm HDE}_{}{[}y,\mathcal{V}{]}$ is equivalent to ${\rm HDE}_{}{[}y,\{ x_{i^{\rm \jbmd{opt}}_1}, ..., x_{i^{\rm \jbmd{opt}}_N}\}{]}$.
Then, we perform ${\rm HDE}_{}{[}y,\{ x_{i^{\rm \jbmd{opt}}_1}, ..., x_{i^{\rm \jbmd{opt}}_{N-j}}\}{]}$ by starting from $j = 1$ and incrementing $j$.
(Each time we increment $j$, we remove the variable that corresponds to the highest-order dependence.)
We check whether $\mathcal{D}_{j} = \mathcal{D}_{}{[}y,\{ x_{i^{\rm \jbmd{opt}}_1}, ..., x_{i^{\rm \jbmd{opt}}_{N-j}}\}{]}$ is correct.
If $\mathcal{D}_{j}$ is correct but $\mathcal{D}_{j+1}$ is incorrect,
we conclude that $\{ x_{i^{\rm \jbmd{opt}}_1}, ..., x_{i^{\rm \jbmd{opt}}_{N-j}}\}$ is the minimal subset of $\mathcal{V}$ that describes $y$ entirely.

\subsection{Framework of gMACE}
\label{sec:meth-desc}

Next, we summarize the \textit{ab initio} MACE scheme and its generalization to the gMACE in the present paper. [See Fig.~\ref{fig:nnmace} for an illustration.]
We restrict the presentation to the essence of the procedure;
details are given in Appendix~\ref{app:meth-mace}.
\\

\paragraph*{Chemical formula dependence of crystal parameters ---}

Obtaining the CF dependence of the AB LEH requires to extend the standard MACE scheme~\cite{Imada2010,Hirayama2013,Hirayama2015,Hirayama2017,Hirayama2018,Hirayama2019,Moree2022,Hirayama2022silverarxiv,Moree2023Hg1223}.
The latter allows to calculate the AB LEH by starting from a given CF together with the crystal symmetry and \jbmd{crystal parameter (CP)} values.
(The CP values are usually taken from experiment.)
To calculate the AB LEH as a function of the CF,
the missing step is to calculate the CP as a function of the CF.
We add this missing step to MACE 
by introducing the CF variables (the ionic radii and charges of cations and anions in the crystal) 
and calculating the CP as a function of the CF by performing the structural optimization,
instead of relying on the experimental CP values.
[Even though the experimental CP may be more accurate than the optimized CP, 
the experimental CP is not always available, and the structural optimization allows to obtain the systematic CF dependence of the CP.]
We only assume the symmetry of the primitive cell during the structural optimization (see Appendix~\ref{app:meth-mace-cd} for details).
\\

\paragraph*{Calculation of the AB LEH ---}

After obtaining the CP for a given CF by employing the structural optimization, 
the AB LEH is calculated by following the MACE procedure
as employed in~\cite{Moree2023Hg1223},
which is summarized below in the successive steps (i-v).
This procedure \jbm{combines the generalized gradient approximation (GGA)~\cite{Perdew1996} and the constrained random phase approximation (cRPA)~\cite{Aryasetiawan2004,Aryasetiawan2006}, and} is denoted as GGA+cRPA.

(i) \jbmb{Starting from the CF and the CP values, we first} perform a DFT calculation\jbmb{. This allows} to obtain the DFT electronic structure at the \jbm{GGA level.}

(ii) From the DFT electronic structure, we compute the maximally localized Wannier orbitals (MLWOs)~\cite{Marzari1997,Souza2001} that span the medium-energy (M) space. 
The M space consists in the 17 bands with Cu$3d$-like, in-plane O$2p$-like and apical X$2p$-like character near the Fermi level (see Fig.~\ref{fig:nnmace}). 
The band with the highest energy in the M space contains most of the AB character,
and the bands outside the M space form the high-energy (H) space.
The MLWO centered on the atom $l$ and with $m$-like orbital character is denoted as $(l,m)$,
where the atom index $l,l'$ takes the values ${\rm Cu,O,O',X}$
(O and O' denote the two in-plane O atoms in the unit cell, and the atomic positions are given in Appendix~\ref{app:meth-mace-cd}).
The orbital index $m,m'$ takes the values 
$x$, $z$, $xy$ and $yz$ for the Cu$3d_{x^2-y^2}$, Cu$3d_{3z^2-r^2}$, Cu$3d_{xy}$ and Cu$3d_{yz}$ orbitals
\jbmd{(the Cu$3d_{zx}$ and Cu$3d_{yz}$ orbitals are equivalent)},
$p$, $p_{\pi}$ and $p_{z}$ for the in-plane O$2p_{\sigma}$, O$2p_{\pi}$ and O$2p_{z}$ orbitals,
and $p_x$, $p_z$ for the apical X$2p_{x}$ and X$2p_{z}$ orbitals
\jbmd{(the X$2p_{y}$ and X$2p_{x}$ orbitals are equivalent)}.
Note that \jbmd{O$2p_{\sigma}$, O'$2p_{\sigma}$, O$2p_{\pi}$ and O'$2p_{\pi}$ correspond respectively to O$2p_{x}$, O'$2p_{y}$, O$2p_{y}$ and O'$2p_{x}$.}

(iii) From the M space, we compute the AB MLWO
by using
\jbm{a procedure} that is detailed in Appendix~\ref{app:meth-mace}.
The AB MLWO centered on the Cu atom located in the unit cell at ${\bf R}$ is denoted as $w_{{\bf R}}$.
We also obtain the AB band that corresponds to the dispersion of the AB MLWO.
Then, the other 16 bands in the M space are disentangled~\cite{Miyake2009} from the AB band.
(See Fig.~\ref{fig:nnmace} for an illustration of the AB band and disentangled M band dispersions.)

(iv) From the AB MLWO, we compute
\begin{equation}
t_1^{} = \int_{\Omega} dr w^{*}_{{\bf 0}}(r) h^{\rm GGA}(r) w_{{\bf R_1}}(r),
\end{equation}
in which $\Omega$ is the unit cell,
$R_1=[100]$,
and $h^{\rm GGA}$ is the one-particle part at the GGA level.
We also compute the onsite bare Coulomb interaction as
\begin{equation}
v^{} = \int_{\Omega} dr \int_{\Omega} dr' w_{{\bf 0}}^{*}(r) w_{{\bf 0}}^{*}(r') v(r,r') w_{{\bf 0}}^{}(r) w_{{\bf 0}}^{}(r'),
\label{eq:v}
\end{equation}
in which $v(r,r')$ is the bare Coulomb interaction.

(v) From the AB band, the disentangled M bands and the H bands, we compute the cRPA screening ratio $R$ (and also $u=U/|t_1|$) as follows.
We compute the cRPA effective interaction $W_{\rm H}(r,r')$ at zero frequency, whose expression is found in
\jbmd{Eqs.~\eqref{eq:chiou} and~\eqref{eq:wh} in Appendix~\ref{app:meth-mace-cd}.}
We deduce the onsite effective Coulomb interaction $U$ 
by replacing $v(r,r')$ with $W_{\rm H}(r,r')$ in Eq.~\eqref{eq:v}.
We deduce $R^{}=U^{}/v^{}$ and $u=U/|t_1|$.

\subsection{Training set of cuprates and gMACE+HDE procedure}
\label{sec:gMACE+HDE}

Below, we define the training set of cuprates that is considered in this paper.
\jbmd{(}For each CF in the training set, we apply the gMACE procedure described in Sec.~\ref{sec:meth-desc}.\jbmd{)}
Then, we describe how the HDE is applied to analyze the results of the gMACE calculation and extract the CF dependence of the AB LEH parameters.
\\

\begin{figure}[!htp]
\includegraphics[scale=0.14]{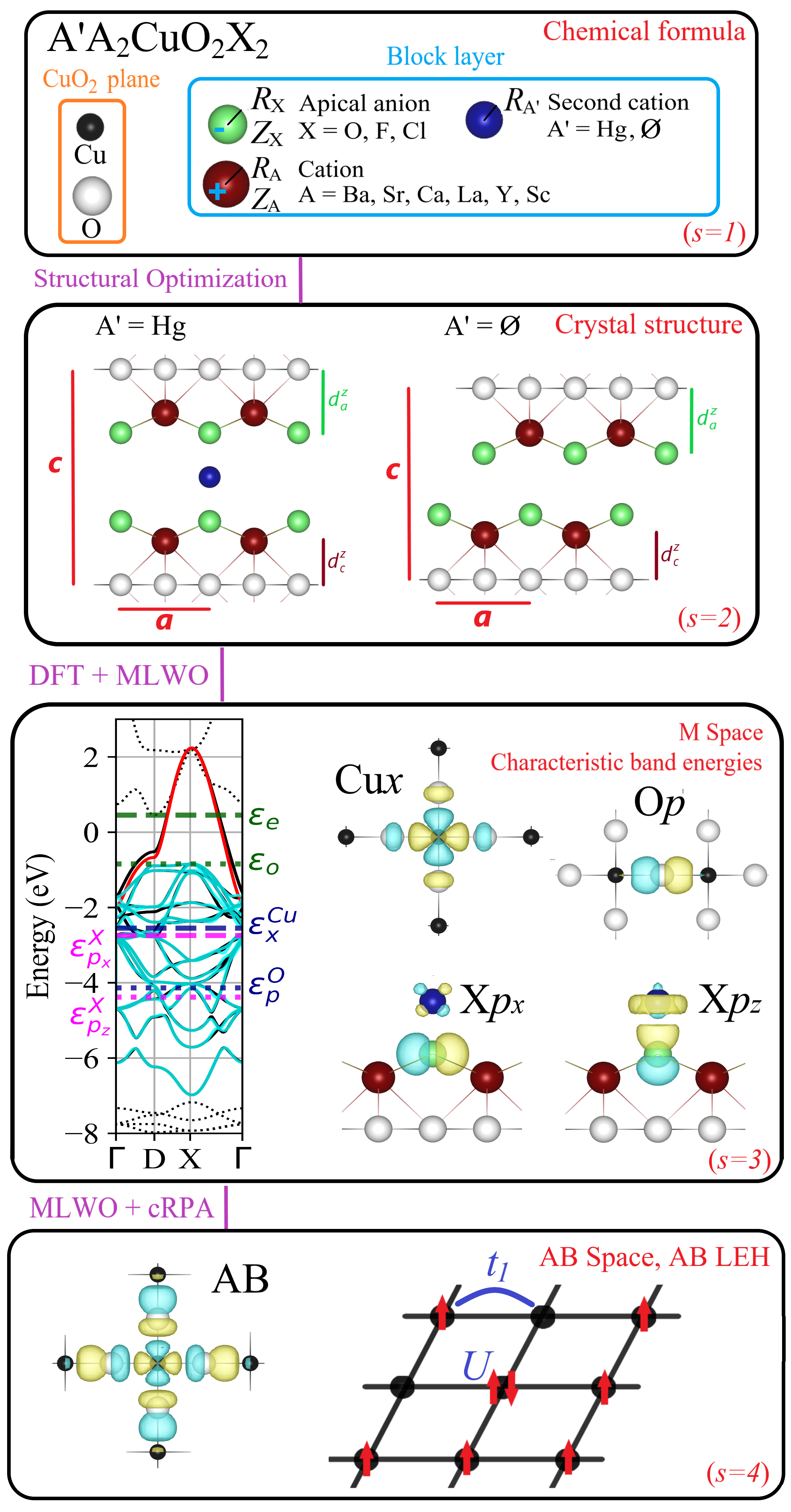}
\caption{\mycaption{
Summary and illustration of the gMACE scheme and variables that are considered in the gMACE+HDE procedure. 
\jbmb{
On $s=2$, the primitive cell vectors and atomic positions are given in Appendix~\ref{app:meth-mace-cd}.
On $s=3$, we show the band structure in the case of HgSr$_2$CuO$_4$ 
[the high-symmetry points are $\Gamma = (0, \ 0, \ 0)$, D$=(1/2, \ 0, \ 0)$ and X$=(1/2, \ 1/2, \ 0)$ in coordinates of the reciprocal lattice].
The dashed (solid) black bands are those outside (inside) the M space.
The red band is the AB band,
and the cyan bands are the 16 other M bands after disentanglement from the AB band.
We also show the principal characteristic energies that are discussed in the main text: 
Energy $\epsilon_{e}$ ($\epsilon_{o}$) of the lowest empty band (highest occupied disentangled M band) outside the AB band, 
and onsite energies $\epsilon^{\rm Cu}_{x}$, $\epsilon^{\rm O}_{p}$, $\epsilon^{\rm X}_{p_x}$ and $\epsilon^{\rm X}_{p_z}$
of the maximally localized Wannier orbitals (MLWOs) whose character is Cu$3d_{x^2-y^2}$ (abbreviated as Cu$x$), O$2p_{\sigma}$ (O$p$), X$2p_{x}$ (X$p_{x}$) and X$2p_{z}$ (X$p_{z}$). 
(The Cu$x$/O$p$ charge transfer energy is $\Delta E_{xp}=\epsilon^{\rm Cu}_{x}-\epsilon^{\rm O}_{p}$).
We also show the isosurfaces of the Cu$x$, O$p$, X$p_x$ and X$p_z$ MLWOs \jbmd{(yellow is positive, blue is negative)}.
On $s=4$, we show the isosurface of the AB MLWO centered on the Cu atom, and a schematic illustration of the square lattice formed by the Cu atoms in the CuO$_2$ plane.
}
}}
\label{fig:nnmace}
\end{figure}

\paragraph*{Training set ---}

The training set of cuprates is defined below and illustrated in Fig.~\ref{fig:chart-comp}.
(Detailed discussions on the choice of the training set are given in Appendix~\ref{app:sc}.)
The training set includes $N_{\rm tr}=36$ CFs, including both experimentally confirmed and hypothetical SC compounds.
The general CF is A'A$_{2}$CuO$_2$X$_2$
and the block layer consists in A'A$_{2}$X$_2$.
For the undoped compound,
we have A' = Hg, $\varnothing$,
A = Ba, Sr, Ca, La, Y, Sc,
and X = O, F, Cl.
For the doped compound, we use the same procedure as in~\cite{Moree2022,Moree2023Hg1223}:
We use the \jbm{virtual crystal approximation}~\cite{nordheim1931electron} to substitute part of the A or A' chemical element
by the chemical element whose atomic number is that of A or A' minus one.
We consider hole doping $\delta=0.0$, $0.1$ and $0.2$ (that is, up to $20\%$ hole doping).
This range includes the experimental range in which the SC state is observed.
We have $Z_{\rm A'}=2-\delta$ if A' = Hg$_{1-\delta}$Au$_{\delta}$ and $Z_{\rm A'}=0$ if A' = $\varnothing$,
$Z_{\rm A} = 2-\delta/2$ if A = Ba, Sr, Ca and $Z_{\rm A} = 3-\delta/2$ if A = La, Y, Sc,
$Z_{\rm X} = -2$ if X = O and $Z_{\rm X}=-1$ if X = F, Cl.
The ionic charges are related to $\delta$ and $n_{\rm AB}$ as follows:
\begin{equation}
Z_{\rm A'} + 2 Z_{\rm A} + 2 Z_{\rm X} = 2 - \delta = 1 + n_{\rm AB}.
\end{equation}
\jbmd{Note that all CFs in the training set do not need to correspond to experimentally confirmed SC cuprates.
For a given CF, %
the gMACE result can be used as part of the data that is analyzed by the HDE procedure
to infer the systematic CF dependence of $|t_1|$ and $u$,
by making complete abstraction of whether or not the corresponding crystal structure can be stabilized in experiment.}
\\

\begin{figure}[!htb]
\includegraphics[scale=0.51]{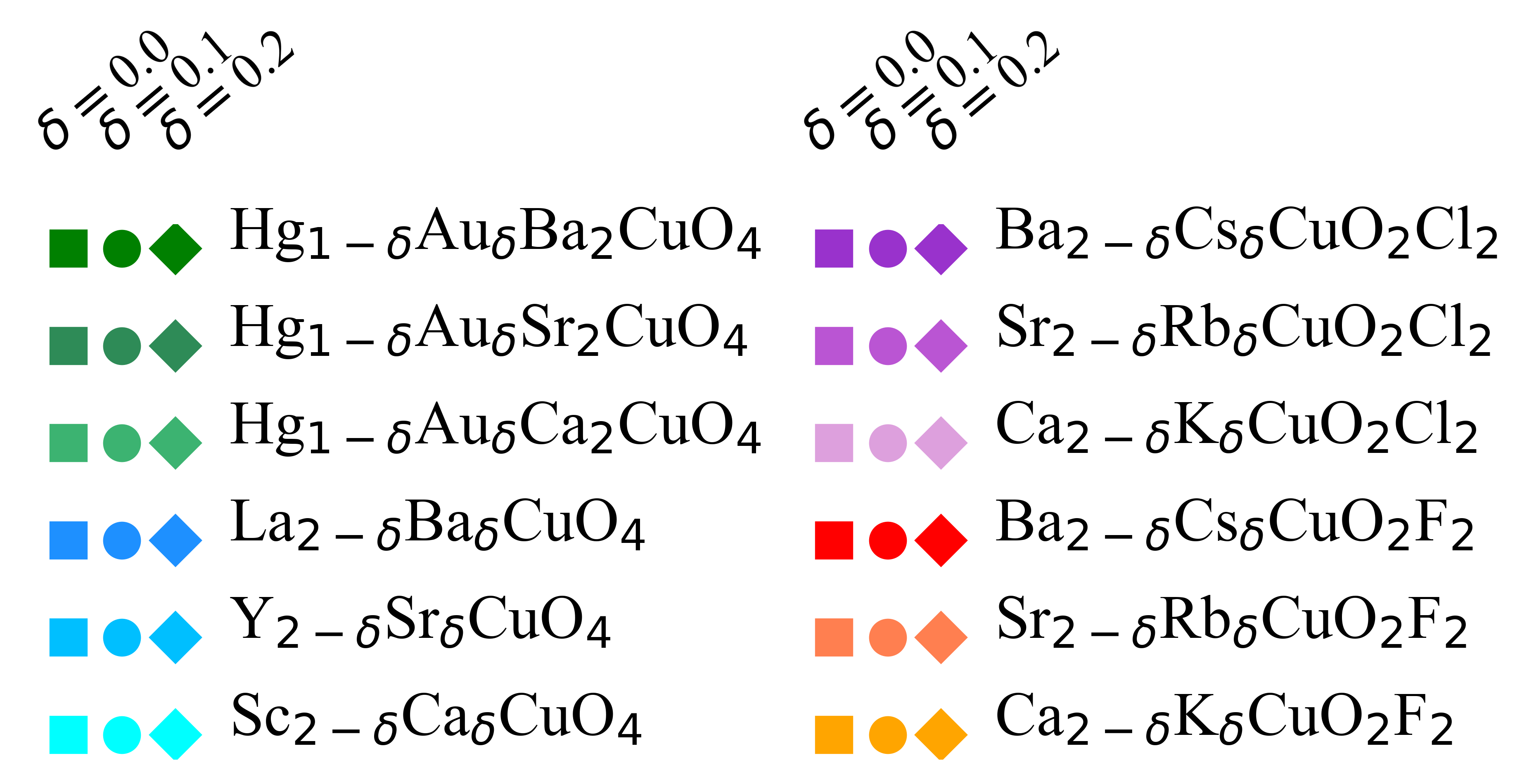}
\caption{\mycaption{
List of the CFs in the training set \jbmd{of cuprates that is considered in this paper}.
We consider $N_t=36$ CFs in total.
\jbmb{The general CF is A'A$_{2}$CuO$_2$X$_2$.}
The color points that correspond to each CF are used in the subsequent figures.
}}
\label{fig:chart-comp}
\end{figure}

\paragraph*{\jbmd{gMACE+HDE procedure} ---}

We apply the HDE to express the AB LEH parameters as a function of the CF variables.
In addition, to gain further insights on the underlying microscopic mechanism, %
we apply the HDE to express the intermediate quantities within the gMACE as functions of each other.
Namely, we consider $N_{s}=4$ levels of variables whose definition is guided by the hierarchical structure of gMACE as illustrated in Fig.~\ref{fig:nnmace}.
At each step $s$, we define the variable space
$\mathcal{V}_{s} = \{ x_{i}^{s}, i=1..N_{\mathcal{V}_{s}} \}$,
and we use the HDE to express $x_{i}^{s}$ as a function of the variables in $\mathcal{V}_{s'}$ (with $s' < s$).
For each $s$, the variables in $\mathcal{V}_{s}$ are chosen as follows.
(\jbmb{These} variables are illustrated in Fig.~\ref{fig:nnmace}.)

\begin{description}%

\item[Chemical formula ($s=1$)]
We consider $\mathcal{V}_{1} = \{ R_{\rm A}, R_{\rm X}, Z_{\rm A}, |Z_{\rm X}|, R_{\rm A'}, n_{\rm AB} \}$.
$R_{\rm A'}$ is the \jbmd{crystal} ionic radius of the second cation A' = Hg$_{1-x}$Au$_x$ in the block layer
($R_{\rm A'}$ is set to $0$ if A' = $\varnothing$).
Values of the variables in $\mathcal{V}_{1}$ that are considered in this paper are given in Appendix~\ref{app:int}.
$R_{\rm A}$, $R_{\rm X}$ and $R_{\rm A'}$ are expressed in \AA \ throughout this paper.

\item[Crystal parameters ($s=2$)]
We consider $\mathcal{V}_{2}=\{a,c,d^{z}_{A},d^{z}_{X}, c_{\perp} \}$,
where $a$\jbmd{, $c$ and $c_{\perp}$} are the cell parameters, and $d^{z}_{A}$ ($d^{z}_{X}$) is the distance between 
the CuO$_2$ plane and the A cation (apical X anion).
The coordinates of primitive vectors and atoms in the unit cell
are given and discussed in Appendix~\ref{app:meth-mace-cd}.
All variables in $\mathcal{V}_{2}$ are expressed in \AA \ throughout this paper.

\item[DFT band structure ($s=3$)]
We consider
\jbmd{
 $\mathcal{V}_{3} = \{ 
 |\jbmd{\epsilon}^{l}_{m}|, 
 |t^{\rm l,l'}_{m,m'}|,
 \Delta E_{xp}$, 
 $W_{\rm M}$, 
 $|\epsilon_{o}|$, $\epsilon_{e} \}$.}
These variables consist in characteristic energies within the M space;
they are defined below,
and their choice 
is further discussed in Appendix~\ref{app:v3}.
First, we include the \jbmd{absolute value of the} onsite energies $\jbmd{\epsilon}^{l}_{m}$ of all MLWOs $(l,m)$ in the M space.
\jbmd{(Note that $\jbmd{\epsilon}^{l}_{m} < 0$ because the onsite energy is below the Fermi level.)}
Second, we include the nonzero hopping amplitudes
$|t^{\rm l,l'}_{m,m'}|$ between 
the MLWOs $(l,m)$ and $(l',m')$ in the unit cell,
where $(l,m) = ({\rm Cu},x)$, $({\rm O},p)$.
We do not take into account hoppings in the unit cell that do not involve $({\rm Cu},x)$ or $({\rm O},p)$ MLWOs,
neither hoppings beyond the unit cell.
We use the abbreviation
$|t_{xp}|=|t^{\rm Cu,O}_{x,p}|$.
Third, we include other characteristic energies:
$\Delta E_{xp} = \jbmd{\epsilon}^{\rm Cu}_{x} - \jbmd{\epsilon}^{\rm O}_{p}$ is the charge transfer energy between the $({\rm Cu},x)$ and $({\rm O},p)$ MLWOs,
$W_{\rm M}$ is the bandwidth of the M space,
\jbmd{and} $\epsilon_{o}$ ($\epsilon_{e}$) is the energy of the highest occupied band in the M space (lowest empty band in the H space) outside the AB band.
\jbmd{(Note that $\epsilon_{o} < 0$ and $\epsilon_{e} > 0$.)}
All variables in $\mathcal{V}_{3}$ are expressed in eV throughout this paper.

\item[AB LEH parameters ($s=4$)]
We consider $\mathcal{V}_{4}=\{|t_1|,\jbmd{u,}v,R\}$ 
as discussed in Sec.~\ref{sec:overview}.
$|t_1|$ and $v$ are expressed in eV throughout this paper.
\end{description}

\section{Results}
\label{sec:results}

\begin{figure}[!htb]
\includegraphics[scale=0.12]{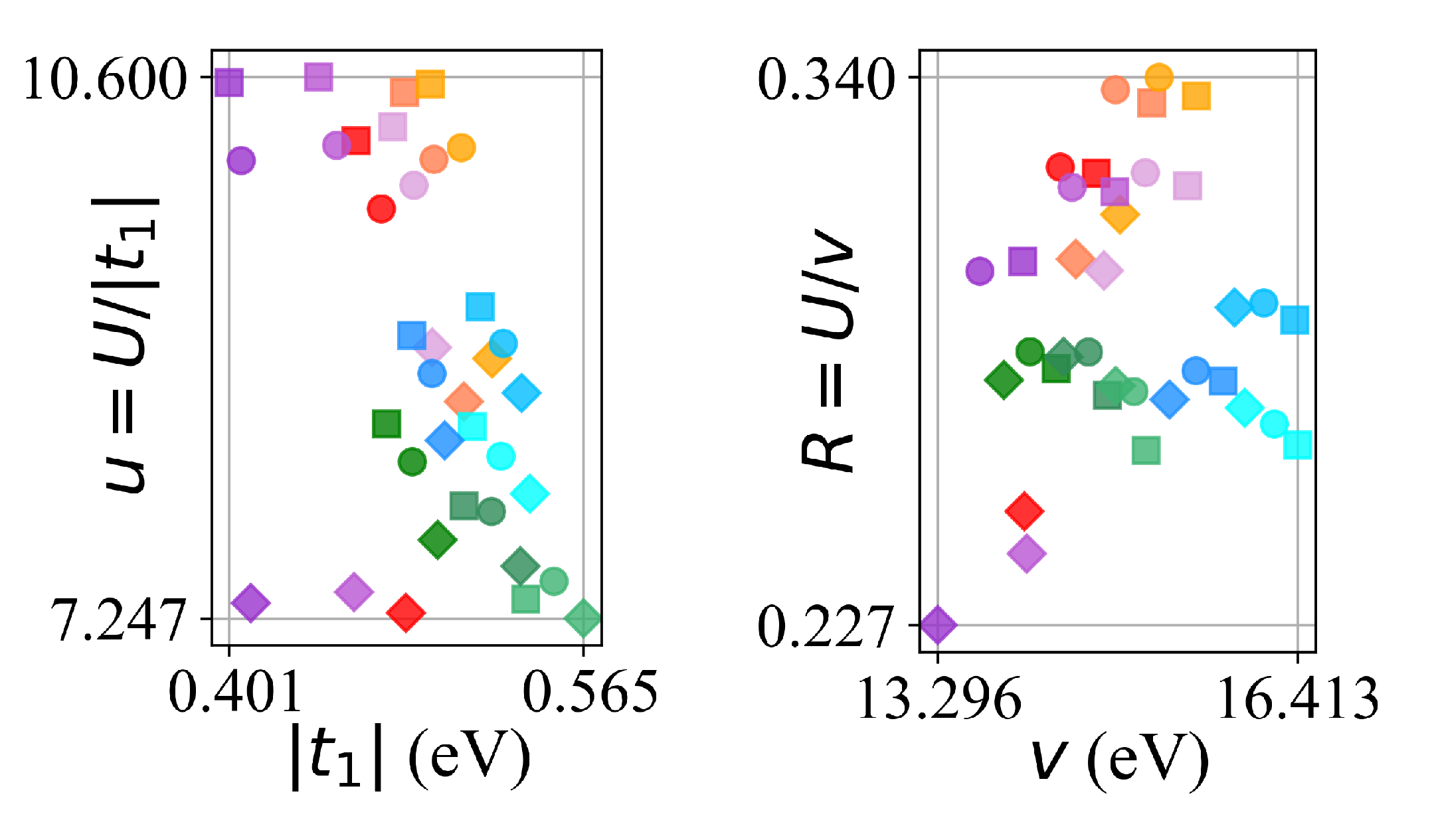}
\caption{\mycaption{
\textit{Ab initio} values of $|t_1|$, $u=U/|t_1|$, $v$ and $R=U/v$ 
obtained by \jbmd{employing} the gMACE \jbmd{for} all CFs in the training set.
For each color point, the corresponding CF is shown in Fig.~\ref{fig:chart-comp}.
}}
\label{fig:chart-t1u}
\end{figure}

\begin{figure}[!htb]
\includegraphics[scale=0.28]{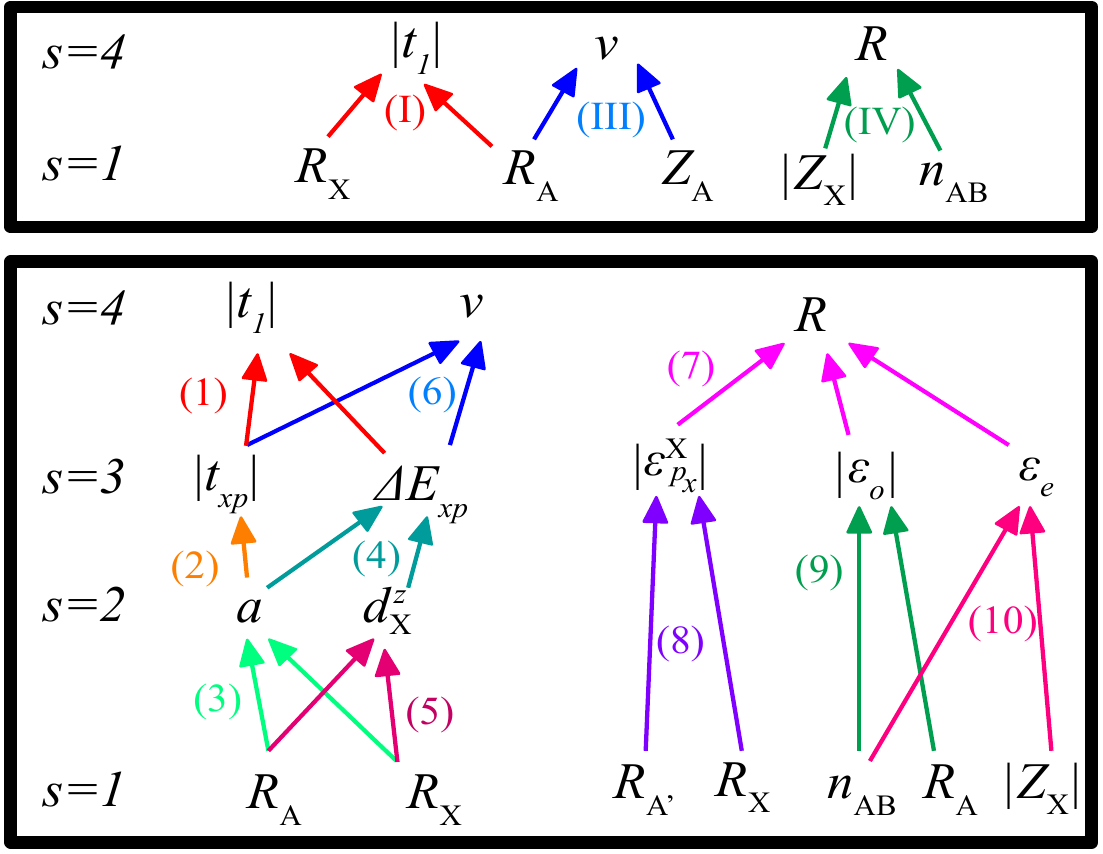}
\caption{\mycaption{
Summary of the MODs between quantities obtained within the gMACE+HDE.
Upper panel: MOD2s of $|t_1|$, $v$ and $R$ on the CF variables in $\mathcal{V}_{1}$. 
[The items (I), (III) and (IV) have been summarized in Sec.~\ref{sec:overview}, and are discussed in detail in Sec.~\ref{sec:results}.]
Lower panel: MODs between intermediate quantities within the gMACE.
[The items (1-10) are discussed in detail in Sec.~\ref{sec:results}.]
}}
\label{fig:treeplot}
\end{figure}

The \textit{ab initio} values of $|t_1|$ and $u$ together with those of $v$ and $R$ are summarized in Fig.~\ref{fig:chart-t1u}.
In this section, we detail the dependencies of $|t_1|$, $v$ and $R$ on $\mathcal{V}_{1}$.
These dependencies correspond to the items (I), (III) and (IV) discussed in Sec.~\ref{sec:overview} and shown in Fig.~\ref{fig:treeplot}.
We also detail the microscopic mechanism of (I), (III) and (IV)
by detailing the dependencies between intermediate quantities within the gMACE 
[the items (1-10) in Fig.~\ref{fig:treeplot}].

\subsection{Chemical formula dependence of $|t_1|$}
\label{sec:results-t1}

\begin{figure}[!htb]
\includegraphics[scale=0.12]{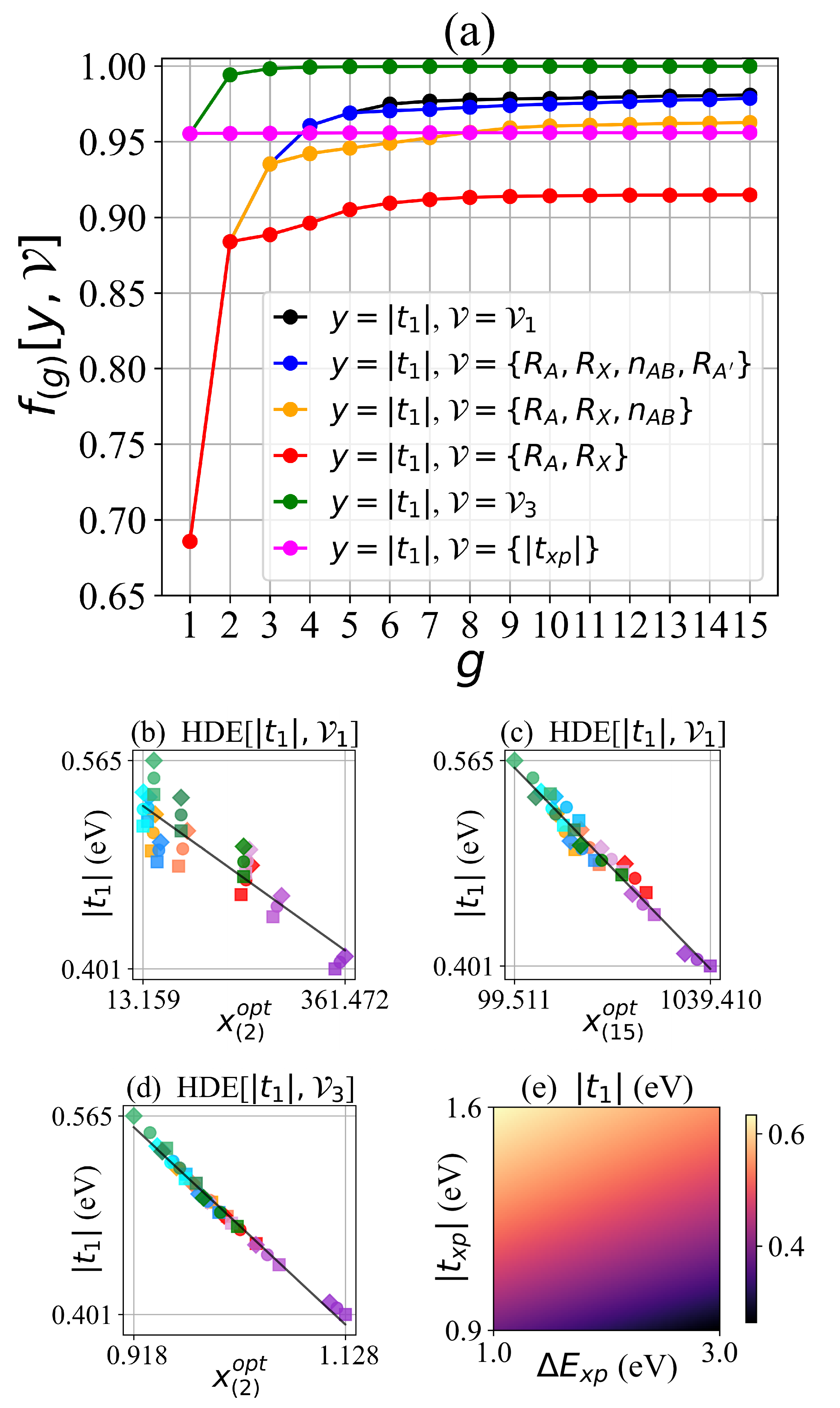}
\caption{
\mycaption{
Details on the dependence of $|t_1|$ on $\mathcal{V}_{1}$ and $\mathcal{V}_{3}$.
Panel (a): Values of $f_{(g)}[|t_1|,\mathcal{V}_{1}]$
in the ${\rm HDE}{[}|t_1|,\mathcal{V}_{1}{]}$,
the ${\rm rHDE}{[}|t_1|,\mathcal{V}_{1}{]}$,
the ${\rm HDE}{[}|t_1|,\mathcal{V}_{3}{]}$,
and the ${\rm rHDE}{[}|t_1|,\mathcal{V}_{3}{]}$.
Panels (b-c): Dependence of $|t_1|$ on $x^{\rm opt}_{(2)}$ \jbmd{[corresponding to Eq.~\eqref{eq:intro-t1}]} and $x^{\rm opt}_{(15)}$
in the ${\rm HDE}{[}|t_1|,\mathcal{V}_{1}{]}$.
Panel (d): Dependence of $|t_1|$ on $x^{\rm opt}_{(2)}$ in Eq.~\eqref{eq:t1s3}.
Panel (e): Representation of the MOD2 of $|t_1|$ on $\{ |t_{xp}|, \Delta E_{xp} \}$ in Eq.~\eqref{eq:t1s3}.
\jbmb{In the panels (b,c,d), the CF that corresponds to each color point is shown in Fig.~\ref{fig:chart-comp},
and the solid black line shows the linear interpolation.}
}}
\label{fig:results-t1}
\end{figure}

\begin{figure*}[!htb]

\includegraphics[scale=0.135]{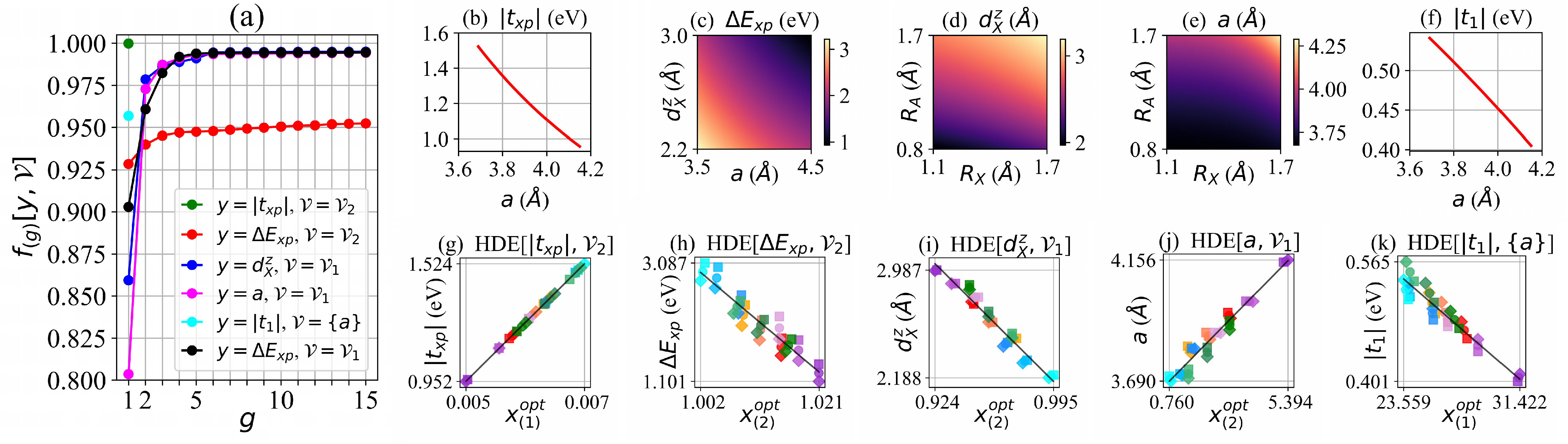}
\caption{
\mycaption{%
Details on the dependence of 
$|t_{xp}|$ on $\mathcal{V}_{2}$, 
$\Delta E_{xp}$ on $\mathcal{V}_{2}$,
$a$ on $\mathcal{V}_{1}$, 
\jbmd{$d^{z}_{\rm X}$ on $\mathcal{V}_{1}$, and also $|t_1|$ on $\mathcal{V}_{1}$ and $\Delta E_{xp}$ on $\mathcal{V}_{1}$ for completeness}.
Panel (a): Values of $f_{(g)}[y,\mathcal{V}]$ in
the ${\rm HDE}{[}|t_{xp}|$,$\mathcal{V}_{2}$], 
the ${\rm HDE}{[}\Delta E_{xp}$,$\mathcal{V}_{2}$], 
the ${\rm HDE}{[}a$,$\mathcal{V}_{1}$], 
the ${\rm HDE}{[}d^{z}_{\rm X}$,$\mathcal{V}_{1}$],
\jbmd{the ${\rm HDE}{[}|t_1|$,$\{ a \}$],
and the
${\rm HDE}{[}\Delta E_{xp}$,$\mathcal{V}_{1}$].} 
Panels (b-f): Representation of the MOD1 of 
$|t_{xp}|$ on $\mathcal{V}_{2}$ [Eq.~\eqref{eq:main_txp_a}],
the MOD2 of $\Delta E_{xp}$ on $\mathcal{V}_{2}$ [Eq.~\eqref{eq:main_dexp_vs2_0}], 
the MOD2 of $d^{z}_{\rm X}$ on $\mathcal{V}_{1}$ [Eq.~\eqref{eq:main_dza_vs1_0}],
the MOD2 of $a$ on $\mathcal{V}_{1}$ [Eq.~\eqref{eq:main_a_vs1}],
and the MOD1 of $|t_1|$ on $\{a \}$ [Eq.~\eqref{eq:t1a}].
Panels (g-k): Dependence of
$|t_{xp}|$ \jbmd{and $|t_1|$ on their respective} $x^{\rm opt}_{(1)}$,
\jbmd{and} $\Delta E_{xp}$, $d^{z}_{\rm X}$ and $a$ on their respective $x^{\rm opt}_{(2)}$.
[The panels (g) to (k) correspond to the panels (b) to (f), respectively.]
\jbmb{\jbmd{T}he CF that corresponds to each color point is shown in Fig.~\ref{fig:chart-comp},
and the solid black line shows the linear interpolation.}
}}
\label{fig:results-t1-interm}
\end{figure*}

Here, we first detail (I). Then, we detail the items (1-\jbmb{5}) in Fig.~\ref{fig:treeplot}.
\\

\paragraph*{$({\rm I})$ Dependence of $|t_1|$ on $\mathcal{V}_{1}$ ---} 
First, $|t_1|$ depends entirely on the variables in $\mathcal{V}_{1}$.
The ${\rm HDE}[|t_1|,\mathcal{V}_{1}]$ yields $f_{\infty}[|t_1|,\mathcal{V}_{1}] = 0.981$ [see Fig.~\ref{fig:results-t1}(a)],
so that $\mathcal{D}_{}{[}|t_1|$,$\mathcal{V}_{1}$] is correct.
Consistently, at $g=15$, the dependence of $|t_1|$ on $x^{\rm opt}_{(15)}$ is almost linear [see Fig.~\ref{fig:results-t1}(c)].

Second, the dependence of $|t_1|$ on $\mathcal{V}_{1}$ can be restricted to the subset $\{ R_{\rm A}, R_{\rm X}, n_{\rm AB}, R_{\rm A'}\}$ of $\mathcal{V}_{1}$.
In the ${\rm HDE}{[}|t_1|$,$\mathcal{V}_{1}$], 
the variables $R_{\rm X}$, $R_{\rm A}$, $R_{\rm A'}$ and $n_{\rm AB}$ correspond to $x_{i^{\rm opt}_{1}}$, $x_{i^{\rm opt}_{2}}$, $x_{i^{\rm opt}_{3}}$ and $x_{i^{\rm opt}_{4}}$, respectively.
The r${\rm HDE}{[}|t_1|$,$\mathcal{V}_{1}$] yields
$f_{\infty}[|t_1|,\{ R_{\rm A}, R_{\rm X},  R_{\rm A'}, n_{\rm AB}\}] \jbmd{=} 0.979$
and $f_{\infty}[|t_1|,\{ R_{\rm A}, R_{\rm X}, R_{\rm A'}\}] \jbmd{=} 0.963$ [see Fig.~\ref{fig:results-t1}(a)].
Thus, $\mathcal{D}_{}{[}|t_1|,\{ R_{\rm A}, R_{\rm X}, R_{\rm A'}, n_{\rm AB}\}$] is correct, 
but $\mathcal{D}_{}{[}|t_1|,\{ R_{\rm A}, R_{\rm X}, R_{\rm A'}\}$] is not.
Thus, $\{ R_{\rm A}, R_{\rm X}, R_{\rm A'}, n_{\rm AB}\}$ is the minimal subset of $\mathcal{V}_{1}$ that describes $|t_1|$.

Third, we discuss details of the MOD2 of $|t_1|$ on $\mathcal{V}_{1}$,
which was given in %
Eq.~\eqref{eq:intro-t1}. 
The MOD2 is not sufficient to describe $|t_1|$ entirely, but the main-order dependence of $|t_1|$ is captured.
We have $f_{(2)}[|t_1|,\mathcal{V}_{1}] \jbmd{=} 0.884$ [see Fig.~\ref{fig:results-t1}(a)],
and the dependence of $|t_1|$ on $x^{\rm opt}_{(2)}$ is shown in Fig.~\ref{fig:results-t1}(b).
Note that the dependence of $|t_1|$ on $R_{\rm A}$ is equally important to that on $R_{\rm X}$ in Eq.~\eqref{eq:intro-t1},
even though $R_{\rm X}$ corresponds to $x_{i^{\rm opt}_{1}}$
(see the score analysis in Appendix~\ref{app:HDEinterp}).
\\

\paragraph*{$(1)$ Dependence of $|t_1|$ on $\mathcal{V}_{3}$ ---} 
First, $|t_1|$ is entirely determined by $|t_{xp}|$ and $\Delta E_{xp}$ irrespective of other variables in $\mathcal{V}_{3}$.
The ${\rm HDE}{[}|t_1|$,$\mathcal{V}_{3}$] yields 
$f_{\infty}$[$|t_1|,\mathcal{V}_{3}$]$=1.000$ [see Fig.~\ref{fig:results-t1}(a)].
The variables $|t_{xp}|$ and $\Delta E_{xp}$ correspond to $x_{i^{\rm opt}_{1}}$ and $x_{i^{\rm opt}_{2}}$, respectively.
The r${\rm HDE}{[}|t_1|$,$\mathcal{V}_{3}$] yields
$f_{\infty}$[$|t_1|$,$\{ |t_{xp}|, \Delta E_{xp} \}$]$=0.999$
and $f_{\infty}$[$|t_1|$,$\{ |t_{xp}| \}$]$=0.956$,
so that $\mathcal{D}_{}{[}|t_1|$,$\{ |t_{xp}|, \Delta E_{xp} \}$] is correct
but $\mathcal{D}_{}{[}|t_1|$,$\{ |t_{xp}|\}$] is not.
Thus, $\{ |t_{xp}|, \Delta E_{xp} \}$ is the subset of $\mathcal{V}_{3}$ that describes $|t_1|$.

Second, the MOD2 of $|t_1|$ on $\{ |t_{xp}|, \Delta E_{xp} \}$ is sufficient to describe accurately $|t_1|$.
Indeed, we have $f_{(2)}$[$|t_1|$,$\{ |t_{xp}|, \Delta E_{xp} \}$] = 0.994,
and the dependence of $|t_1|$ on $x^{\rm opt}_{(2)}$ is almost linear [see Fig.~\ref{fig:results-t1}(\jbmd{d})].
The MOD2 is
\begin{equation}
|t_1|_{\rm MOD2} = 1.269 -0.777 \Big[ |t_{xp}|^{-0.58} [1 + 0.0739 \Delta E_{xp}^{0.98}] \Big]
\label{eq:t1s3}
\end{equation}
and is represented in Fig.~\ref{fig:results-t1}(\jbmd{d,e}).
Note that $|t_{xp}|$ dominates over $\Delta E_{xp}$ in the MOD2 %
(see the score analysis in Appendix~\ref{app:HDEinterp}),
which is also visible in Fig.~\ref{fig:results-t1}(\jbmd{e}):
The color map has a horizontal-like pattern.

Qualitatively, $|t_1|_{\rm MOD2}$ increases with increasing $|t_{xp}|$ and decreasing $\Delta E_{xp}$
in Eq.~\eqref{eq:t1s3},
and this is consistent with previous works.
In the case of Hg1223, we have
$|t_1| \propto |t_{xp}|^2 /  \Delta E_{xp}$
in Fig.~12 of~\cite{Moree2023Hg1223}.
(The latter result was obtained by modifying $a$ artificially without modifying other CPs.)
The result in Eq.~\eqref{eq:t1s3} is more general,
because Eq.~\eqref{eq:t1s3} is established for $N_{\rm tr}=36$ CFs rather than one CF, 
and it accounts for the CF dependence of all CPs. %

The MOD2 in Eq.~\eqref{eq:t1s3} can be interpreted as follows.
$|t_1|$ represents the hopping amplitude between neighboring AB orbitals,
and thus, $|t_1|$ mainly depends on the overlap between neighboring AB orbitals.
Since the AB orbital is formed by the Cu$3d_{x^2-y^2}$ and in-plane O$2p_{\sigma}$ orbitals,
the overlap between the neighboring AB orbitals
is determined by the overlap between the neighboring Cu$3d_{x^2-y^2}$ and in-plane O$2p_{\sigma}$ orbitals,
which is mainly encoded in $|t_{xp}|$.
Thus, it is natural that $|t_1|$ mainly depends on $|t_{xp}|$.
In addition, decreasing $\Delta E_{xp}$ reduces the localization of the AB orbital (as discussed later in Sec.~\ref{sec:results-v}):
The delocalization of the AB orbital within the CuO$_2$ plane
contributes to increase the overlap between neighboring AB orbitals and thus $|t_1|$.
\\
	
\paragraph*{$(2)$ Dependence of $|t_{xp}|$ on $\mathcal{V}_{2}$ ---} 	
$|t_{xp}|$ is entirely determined by the cell parameter $a$ irrespective of other variables in $\mathcal{V}_{2}$,
and the MOD1 describes $|t_{xp}|$ perfectly.
Indeed, the ${\rm HDE}{[}|t_{xp}|$,$\mathcal{V}_{2}$] yields 
$f_{(1)}$[$|t_{xp}|$, $\mathcal{V}_{2}$] = 1.000 [see Fig.~\ref{fig:results-t1-interm}(a)],
and $x_{i^{\rm opt}_{1}}$ corresponds to $a$.
The MOD1 is
\begin{equation}
|t_{xp}|_{\rm MOD1} = -0.063\jbmb{2} + 212.23\jbmb{4} a^{-3.75}
\label{eq:main_txp_a}
\end{equation}
and is represented in Fig.~\ref{fig:results-t1-interm}(\jbmd{b,g}).
The score analysis shows that the dominance of $a$ in the dependence of $|t_{xp}|$ is unambiguous (see Appendix~\ref{app:HDEinterp}).

The $a$ dependence of $|t_{xp}|$ is consistent with results on Hg1223~\cite{Moree2023Hg1223}.
In particular, the exponent $-3.75$ in Eq.~\eqref{eq:main_txp_a} is very close to that obtained for Hg1223, in which $|t_{xp}| \propto a^{-3.86}$ (see \cite{Moree2023Hg1223}, Fig. 12).
This suggests $|t_{xp}|$ scales as $a^{-3.75}$ universally and irrespective of the crystalline environment outside the CuO$_2$ plane.
\jbmd{This is intuitive because the Cu$3d_{x^2-y^2}$ and O$2p_{\sigma}$ orbitals extend mainly in the CuO$_2$ plane as illustrated in Fig.~\ref{fig:nnmace}.}

In addition, $|t_1|$ increases when $a$ decreases
according to Eqs.~\eqref{eq:main_txp_a} and~\eqref{eq:t1s3}.
This is consistent with~\cite{Moree2023Hg1223}, in which the pressure-induced decrease in $a$ is the main cause of the pressure-induced increase in $|t_1|$. 
For completeness, we perform the ${\rm HDE}[|t_1|,a]$:
We obtain $f_{\infty}$[$|t_1|$,$\{a\}$] = $f_{(1)}$[$|t_1|$,$\{a\}$] =  $0.957$,
so that $a$ describes $|t_1|$ reasonably, but not perfectly.
This is because $|t_1|$ has not only a dominant dependence on $|t_{xp}|$ but also a small dependence on $\Delta E_{xp}$, 
and $\Delta E_{xp}$ is not described entirely by $a$ as seen later in (4).
The MOD1 of $|t_1|$ on $a$ is
\begin{equation}
|t_1|_{\rm MOD1} = 0.956 - 0.01\jbmb{76} a^{2.42},
\label{eq:t1a}
\end{equation} 
and is represented in Fig.~\ref{fig:results-t1-interm}(\jbmd{f,k}).
Qualitatively, $|t_1|_{\rm MOD1}$ increases when $a$ decreases, which is consistent with~\cite{Moree2023Hg1223}
and also with Eqs.~\eqref{eq:main_txp_a} and~\eqref{eq:t1s3}.
For completeness, note that there is a quantitative difference between Eq.~\eqref{eq:t1a} and~\cite{Moree2023Hg1223}:
In the latter, we have $|t_{1}| \propto a^{-2.88}$ (see \cite{Moree2023Hg1223}, Fig. 12).
\jbmd{The difference may be explained as follows.
$|t_1|$ depends on both $|t_{xp}|$ and $\Delta E_{xp}$ [Eq.~\eqref{eq:t1s3}], 
and $\Delta E_{xp}$ depends on the crystalline environment outside the CuO$_2$ plane contrary to $|t_{xp}|$.
[For instance, as seen later in (5), the MOD2 of $\Delta E_{xp}$ depends not only on $a$ but also on $d^{z}_{\rm X}$.]
The result in~\cite{Moree2023Hg1223} captures the $a$ dependence of $|t_1|$ by fixing the other CP values; 
the present result is more general because it accounts for the materials dependence of other CP values \textit{via} the structural optimization.
}
\\

\paragraph*{$(3)$ Dependence of $a$ on $\mathcal{V}_{1}$ ---} 	
$a$ is determined entirely by $R_{\rm A}$ and $R_{\rm X}$ in $\mathcal{V}_{1}$.
The ${\rm HDE}{[}a$,$\mathcal{V}_{1}$] yields 
$f_{\infty}$[$a$,$\mathcal{V}_{1}$] = 0.994,
and the r${\rm HDE}{[}a$,$\mathcal{V}_{1}$] yields 
$f_{\infty}$[$a$,$\{ R_{\rm A}, R_{\rm X} \}$] = 0.985.
On the MOD2, we have $f_{(2)}$[$a$,$\{ R_{\rm A}, R_{\rm X} \}$] = 0.973,
and
\begin{equation}
a_{\rm MOD2} = 3.613 + 0.100 \Big[ R_{\rm A}^{2.71} [1 + 0.00711 R_{\rm X}^{9.14}] \Big],
\label{eq:main_a_vs1}
\end{equation}
\jbmd{which} is represented in Fig.~\ref{fig:results-t1-interm}(\jbmd{e,j}).
The score analysis confirms that $R_{\rm A}$ and $R_{\rm X}$ correspond respectively to $x_{i^{\rm opt}_{1}}$ and $x_{i^{\rm opt}_{2}}$ 
(see Appendix~\ref{app:HDEinterp}).

\begin{figure}
\includegraphics[scale=0.13]{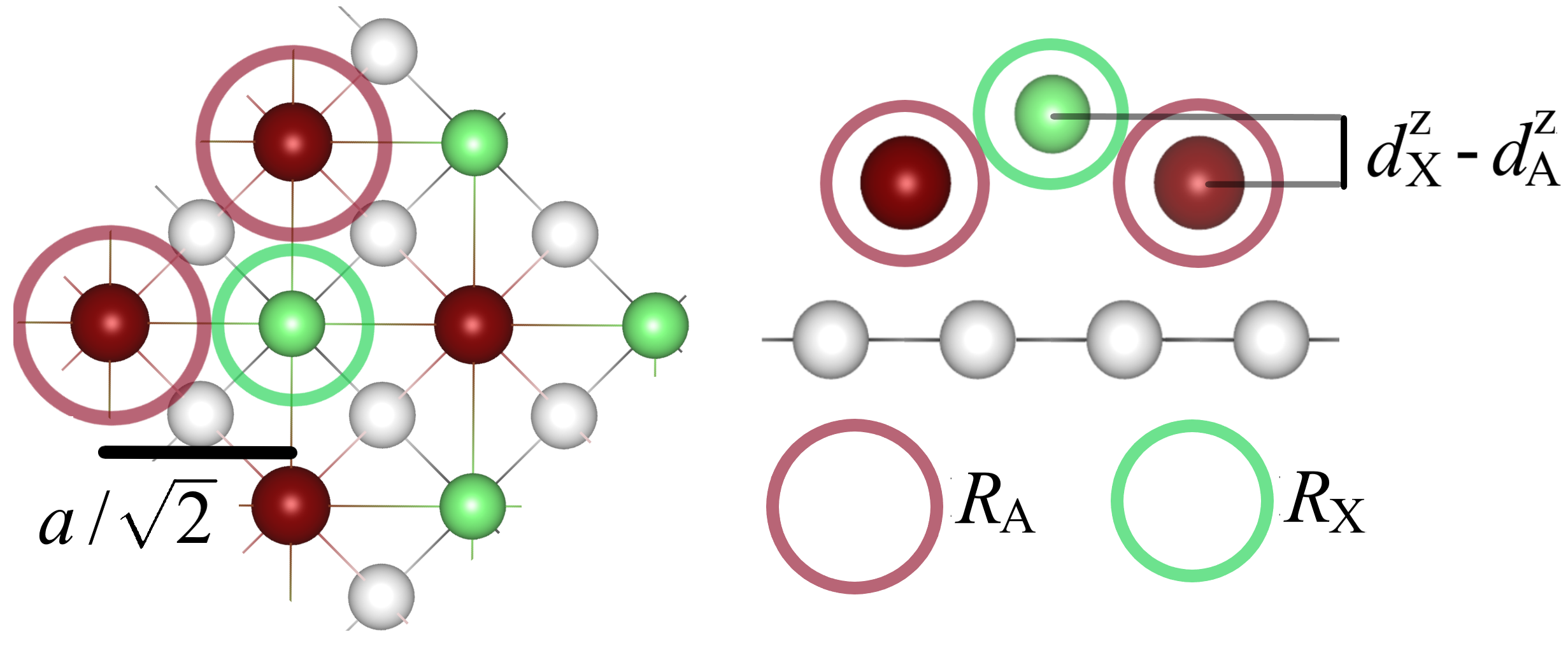}
\caption{\mycaption{Illustration of the hard sphere picture for the crystal of the single-layer cuprate.
\jbmd{In this picture, c}ations and apical anions are assumed to be rigid spheres that touch each other\jbmd{, and} we have $R_{\rm A} + R_{\rm X} = \sqrt{(a/\sqrt{2})^2 + (d^{z}_{X}-d^{z}_{A})^2}$.
}}
\label{fig:hardsphere}
\end{figure}
	
The MOD2 in Eq.~\eqref{eq:main_a_vs1} is interpreted as follows.
Qualitatively, $a_{\rm MOD2}$ increases when $R_{\rm A}$ or $R_{\rm X}$ increases,
which is consistent with the hard sphere picture illustrated in Fig.~\ref{fig:hardsphere}:
The interatomic distances and thus the cell parameter $a$ increase when the ionic radii are larger.
Thus, the hard-sphere picture reproduces the qualitative dependence of $a$ on the ionic radii.
Furthermore, the MOD2 in Eq.~\eqref{eq:main_a_vs1} is qualitatively consistent with experiment,
\jbme{in which $a$ increases with increasing $R_{\rm A}$.
For instance, for X = Cl, we have 
$a = 3.87$ \AA \ for A = Ca~\cite{Grande1977,Argyriou1995}, 
$a=3.97-3.98$ \AA \ for A = Sr~\cite{Grande1975,Miller1990}, 
and $a = 4.10$ \AA \ for A = Ba~\cite{Tatsuki1996}.
(These values are in correct agreement with the values $a = 3.88$ \AA \ for A = Ca, $a = 4.00$ \AA \ for A = Sr and $a = 4.15$ \AA \ for A = Ba obtained in this paper by employing the structural optimization.)
And, we have $R_{\rm Ca} = 1.14$ \AA, $R_{\rm Sr} = 1.32$ \AA \ and $R_{\rm Ba} = 1.49$ \AA~\cite{Shannon1976}.
(The values of $R_{\rm A}$ are given in Appendix~\ref{app:int}, and the $R_{\rm A}$ dependence of experimental $a$ is also emphasized in~\cite{Tatsuki1996}.)
Also, for X = O, we have  %
$a \simeq 3.78$ \AA \ in La$_2$CuO$_4$~\cite{Jorgensen1987} 
and $a \simeq 3.88$ \AA \ in HgBa$_2$CuO$_4$~\cite{Putilin1993}.
(These values are in correct agreement with the values $a \simeq 3.82$ in La$_2$CuO$_4$ and $a \simeq 3.92$ \AA \ in HgBa$_2$CuO$_4$ obtained in this paper by employing the structural optimization.)
And, we have $R_{\rm La} = 1.17$ \AA \ and $R_{\rm Ba} = 1.49$ \AA~\cite{Shannon1976}.
}
\\

\paragraph*{$(4)$ Dependence of $\Delta E_{xp}$ on $\mathcal{V}_{2}$ ---} 	
$\Delta E_{xp}$ is not determined entirely by $\mathcal{V}_{2}$, but the MOD2 reveals the main-order mechanism that controls the value of $\Delta E_{xp}$.
The	${\rm HDE}{[}\Delta E_{xp}$,$\mathcal{V}_{2}$] yields 
$f_{\infty}$[$\Delta E_{xp}$,$\mathcal{V}_{2}$] = 0.952 [see Fig.~\ref{fig:results-t1-interm}(a)],
so that $\mathcal{D}_{}{[}\Delta E_{xp}$,$\mathcal{V}_{2}$] is not completely correct.
As for the MOD2,
we have $f_{(2)}$[$\Delta E_{xp}$,$\mathcal{V}_{2}$] = 0.940;
the MOD2 is
\begin{equation}
\Delta E_{xp{\rm MOD2}} = 88.343 - 85.265 \Big[ (d^{z}_{X})^{0.04} - 0.4\jbmb{79} a^{-2.12} \Big]
\label{eq:main_dexp_vs2_0}
\end{equation}
and is represented in Fig.~\ref{fig:results-t1-interm}(\jbmd{c,h}).
Even though $d^{z}_{X}$ corresponds to $x_{i^{\rm opt}_{1}}$,
the dependence of $\Delta E_{xp}$ on $d^{z}_{X}$ is equally important to that on $a$
(see the score analysis in Appendix~\ref{app:HDEinterp}).
Consistently, in Fig.~\ref{fig:results-t1-interm}(c), the color map has a diagonal-like pattern.

\begin{figure}[!htb]
\includegraphics[scale=0.24]{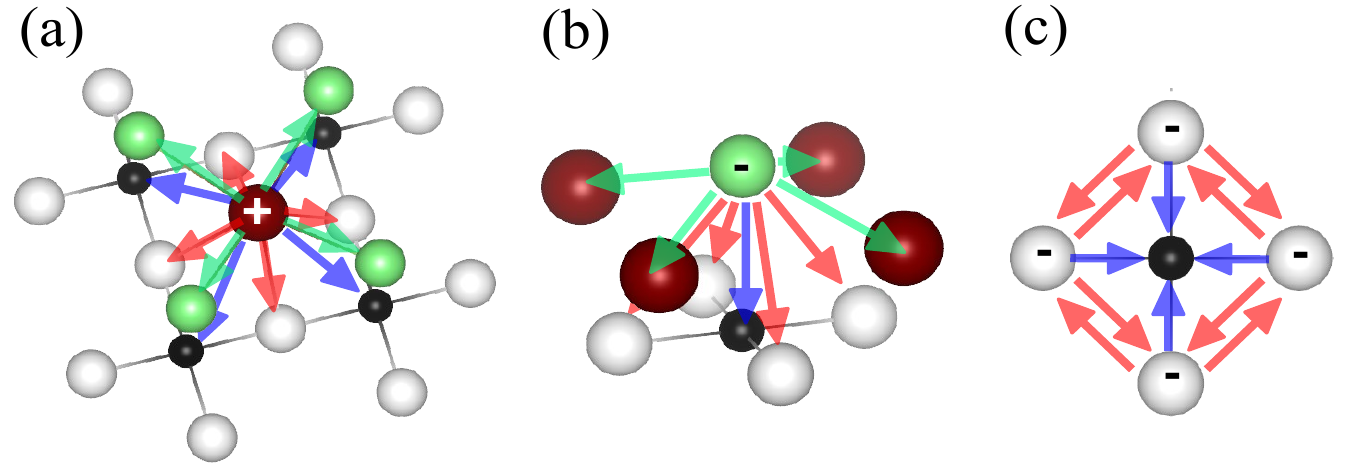}
\caption{\mycaption{Schematic illustration of the positive Madelung potential (MP$^{+}$)
and negative Madelung potential (MP$^{-}$) created by the cations and anions within the crystal.
Panel (a): MP$^{+}$ created by the cation A
and felt by the nearest neighbor apical X (green arrows), in-plane O (red arrows), and Cu (blue arrows).
Panel (b): MP$^{-}$ created by the apical X anion
and felt by the nearest neighbor Cu (blue arrow), in-plane O (red arrows), and cations (\jbmd{green} arrows).
Panel (c): MP$^{-}$ created by the in-plane O anion and felt by the nearest neighbor Cu (blue arrows) and other in-plane O (red arrows).
In the panel (a),
the MP$^{+}$ created by A and felt at a distance $d$ of A scales as $V_{\rm A}(d) = Z_{\rm A}/d$.
The distance $d_{\rm A,X} = \sqrt{a^2/2 + (d^{z}_{X} - d^{z}_{A})^2}$ between A and X
is smaller than the distance $d_{\rm A,O} = \sqrt{a^2/4 + (d^{z}_{A})^2}$] between A and in-plane O,
which is smaller than the distance $d_{\rm A,Cu} = \sqrt{a^2/2 + (d^{z}_{A})^2}$ between A and Cu.
Thus, we have $V_{\rm A}(d_{\rm A,X}) > V_{\rm A}(d_{\rm A,O}) > V_{\rm A}(d_{\rm A,Cu})$.
In the panel (b),
we have $|V_{\rm X}(d_{\rm X,A})| > |V_{\rm X}(d_{\rm X,Cu})| > |V_{\rm X}(d_{\rm X,O})|$.
In the panel (c),
we have $|V_{\rm O}(d_{\rm O,Cu})| > |V_{\rm O}(d_{\rm O,O'})|$.
}}
\label{fig:cation_madelung}
\end{figure}

Qualitatively, $\Delta E_{xp}$ increases when $d^{z}_{X}$ decreases or $a$ decreases in Eq.~\eqref{eq:main_dexp_vs2_0}. 
\jbmb{The} increase in $\Delta E_{xp}$ with decreasing $d^{z}_{X}$ can be understood as follows.
When $d^{z}_{X}$ decreases, the distance between the apical X and the CuO$_2$ plane decreases,
so that the MP$^{-}$ created by the apical X anion and felt by the Cu and in-plane O is stronger.
This increases the energy of both Cu$3d_{x^2-y^2}$ and O$2p_{\sigma}$ electrons.
The MP$^{-}$ felt by Cu is the strongest, because the Cu is closer to the apical X compared to the in-plane O.
[See Fig.~\ref{fig:cation_madelung}(b) for an illustration.]
Thus, the energy of Cu$3d_{x^2-y^2}$ electrons increases more than that of in-plane O$2p_{\sigma}$ electrons.
As a consequence, $\Delta E_{xp}$ increases.

The increase in $\Delta E_{xp}$ with decreasing $a$ is consistent with~\cite{Moree2023Hg1223},
and the mechanism is reminded here.
When $a$ decreases, the distance between the in-plane O and Cu is reduced,
so that the MP$^{-}$ created by the in-plane O anions and felt by the Cu is stronger.
[See Fig.~\ref{fig:cation_madelung}(c) for an illustration.]
This increases the energy of Cu$3d_{x^2-y^2}$ electrons with respect to that of in-plane O$2p_{\sigma}$ electrons\jbmd{, which increases $\Delta E_{xp}$.}
\\

\paragraph*{(5) Dependence of $d^{z}_{X}$ on $\mathcal{V}_{1}$ ---} 	
$d^{z}_{X}$ is determined entirely by $\mathcal{V}_{1}$.
The ${\rm HDE}{[}d^{z}_{X}$,$\mathcal{V}_{1}$] yields $f_{\infty}$[$d^{z}_{X}$,$\mathcal{V}_{1}$] = 0.995.
As for the MOD2 of $d^{z}_{X}$ on $\mathcal{V}_{1}$,
we have $f_{(2)}$[$d^{z}_{X}$,$\mathcal{V}_{1}$] = 0.979,
and %
\begin{equation}
d^{z}_{\rm XMOD2} = 14.35\jbmb{3} - 12.25\jbmb{4} \Big[ R_{\rm A}^{-0.10} [1 - 0.00897 R_{\rm X}^{2.80} ] \Big]
\label{eq:main_dza_vs1_0}
\end{equation} 
\jbmd{is} represented in Fig.~\ref{fig:results-t1-interm}(\jbmd{d,i}).
The score analysis confirms that $R_{\rm A}$ and $R_{\rm X}$ correspond respectively to $x_{i^{\rm opt}_{1}}$ and $x_{i^{\rm opt}_{2}}$ 
(see Appendix~\ref{app:HDEinterp}).

Qualitatively, $d^{z}_{\rm XMOD2}$ increases when $R_{\rm A}$ or $R_{\rm X}$ increases in Eq.~\eqref{eq:main_dza_vs1_0}.
This can also be understood by considering the hard sphere picture
(see the right panel in Fig.~\ref{fig:hardsphere}).
In the \textit{ab initio} result, we always have $d^{z}_{\rm A} < d^{z}_{\rm X}$ 
(the values \jbmd{of $d^{z}_{\rm A}$ and $d^{z}_{\rm X}$} are given in \jbmd{Appendix~\ref{app:int}}), 
so that the apical X is farther from the CuO$_2$ plane compared to the A cation.
In the hard sphere picture, increasing the ionic radius $R_{\rm A}$ of the A cation pushes the apical X even farther from the CuO$_2$ plane, 
which increases $d^{z}_{\rm X}$. 
The same mechanism occurs when $R_{\rm X}$ increases.
\\

\paragraph*{Dependence of $\Delta E_{xp}$ on $\mathcal{V}_{1}$ ---} 	
On (4,5), for completeness, we discuss the dependence of $\Delta E_{xp}$ on $\mathcal{V}_{1}$.
$\Delta E_{xp}$ is determined entirely by the subset $\{R_{\rm A},R_{\rm X},n_{\rm AB}\}$ of $\mathcal{V}_{1}$.
The ${\rm HDE}{[}\Delta E_{xp}$,$\mathcal{V}_{1}$] yields
$f_{\infty}$[$\Delta E_{xp}$,$\mathcal{V}_{1}$] = 0.995,
and the r${\rm HDE}{[}\Delta E_{xp}$,$\mathcal{V}_{1}$] yields
$f_{\infty}$[$\Delta E_{xp}$,$\{ R_{\rm A},R_{\rm X},n_{\rm AB}\}$] = 0.988
and $f_{\infty}$[$\Delta E_{xp}$,$\{R_{\rm A},R_{\rm X}\}$] = 0.967
[see Fig.~\ref{fig:results-t1-interm}(a)].
Thus, $\mathcal{D}_{}{[}\Delta E_{xp}$,$\{ R_{\rm A},R_{\rm X},n_{\rm AB}\}$] is correct, but $\mathcal{D}_{}{[}\Delta E_{xp}$,$\{ R_{\rm A},R_{\rm X}\}$] is not.
We have $f_{(2)}$[$\Delta E_{xp}$,$\mathcal{V}_{1}$] = 0.961
and $f_{(3)}$[$\Delta E_{xp}$,$\mathcal{V}_{1}$] = 0.982,
and the MOD3 is
\begin{align}
\Delta E_{xp{\rm MOD3}} = \ & 9.71\jbmb{0} - 7.5\jbmb{65} \Big[ R_{\rm A}^{0.33} [1 + 0.000411R_{\rm X}^{9.33}] \nonumber \\
& - 0.0999 n_{\rm AB}^{1.77} \Big].
\label{eq:main_dexp_vs1}
\end{align} 
The score analysis confirms that $R_{\rm A}$ and $R_{\rm X}$ correspond respectively to $x_{i^{\rm opt}_{1}}$ and $x_{i^{\rm opt}_{2}}$,
and $n_{\rm AB}$ corresponds to $x_{i^{\rm opt}_{3}}$ but is slightly in competition with $Z_{\rm A}$
(see Appendix~\ref{app:HDEinterp}).

Qualitatively, $\Delta E_{xp{\rm MOD3}}$ increases when $R_{\rm A}$ decreases, $R_{\rm X}$ decreases or $n_{\rm AB}$ increases in Eq.~\eqref{eq:main_dexp_vs1}.
\jbmd{This is consistent with Eqs.~\eqref{eq:main_dexp_vs2_0},~\eqref{eq:main_dza_vs1_0} and~\eqref{eq:main_a_vs1}.}
Also, $\Delta E_{xp}$ decreases when $n_{\rm AB}$ decreases \jbmd{(the hole doping $\delta$ increases)}.
This is consistent with~\cite{Moree2022} \jbmd{and explained as follows.}
\jbmd{
When $\delta$ increases, the hole doping of O sites increases.
(The holes localize on O sites to form the Zhang-Rice singlet.)
This reduces the negative charge of in-plane O anions.
This reduces the MP$^{-}$ created by the in-plane O and felt by the nearby Cu.
[See Fig.~\ref{fig:cation_madelung}(c) for an illustration.]
This reduces the energy of Cu$3d$ electrons, and also reduces the Fermi energy.
(Indeed, the AB band at the Fermi level has Cu$3d_{x^2-y^2}$ character.)
On the other hand, the O$2p$ electrons are less affected.
Thus, the energy difference $\Delta E_{xp} = \epsilon^{\rm Cu}_{x} - \epsilon^{\rm O}_{p}$ increases.
}

\subsection{Chemical formula dependence of $v$}
\label{sec:results-v}

Here, we first detail (III). Then, we detail the item (6) in Fig.~\ref{fig:treeplot}.
[The items (2-5) have already been discussed in the previous section].
\\

\begin{figure}[!htb]
\includegraphics[scale=0.115]{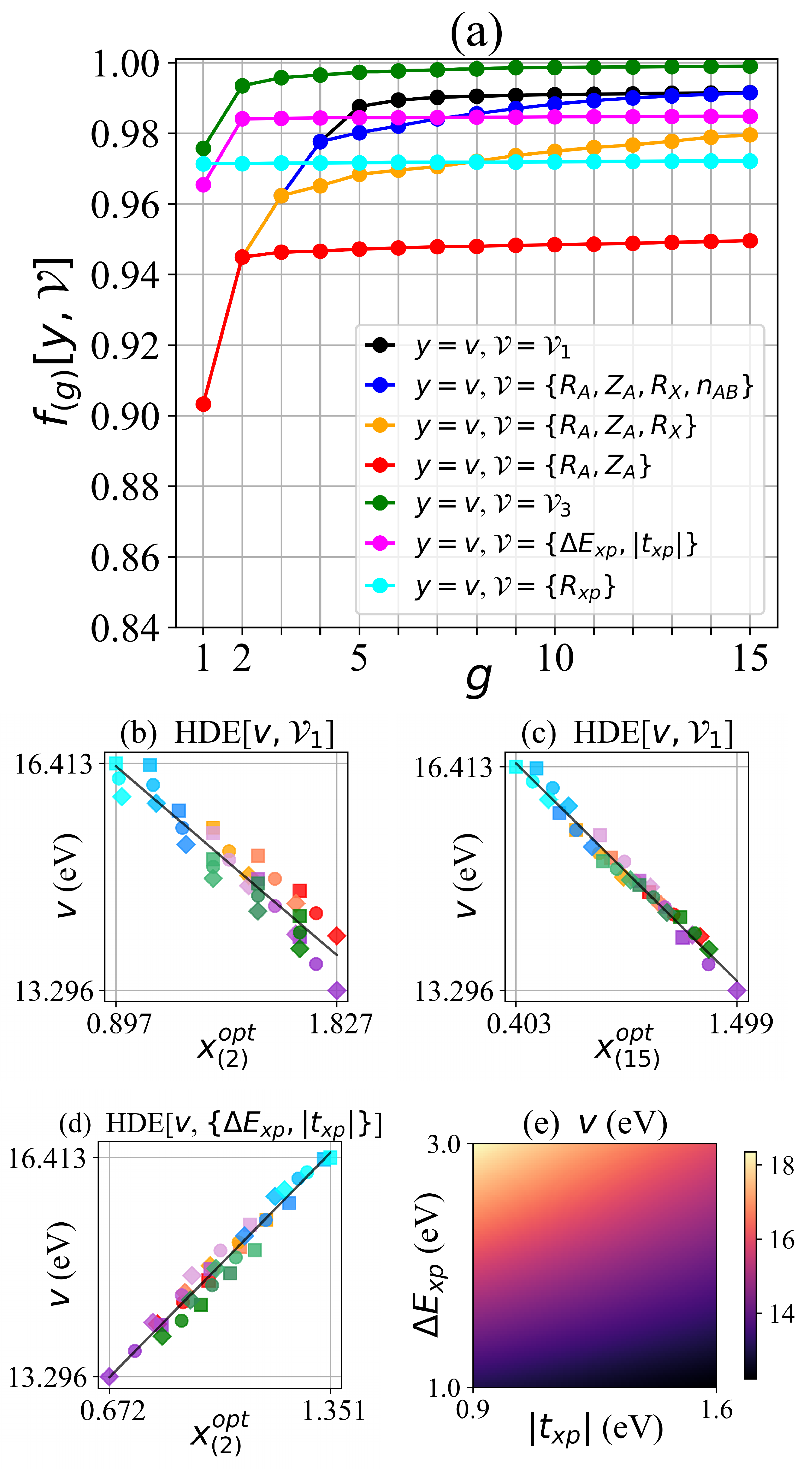}

\caption{\mycaption{
Dependence of $v$ on $\mathcal{V}_{1}$ and $\mathcal{V}_{3}$.
Panel (a): Values of $f_{(g)}[v,\mathcal{V}]$ %
in the ${\rm HDE}{[}v$,$\mathcal{V}_{1}$],
 the r${\rm HDE}{[}v$,$\mathcal{V}_{1}$],
 the ${\rm HDE}{[}v$,$\mathcal{V}_{3}$],
 \jbmd{the ${\rm HDE}{[}v$,$\{ \Delta E_{xp}, |t_{xp}| \}$] and the ${\rm HDE}{[}v$,$\{ R_{xp}=\Delta E_{xp}/|t_{xp}| \}$]}.
Panels (b-c): Dependence of $|t_1|$ on $x^{\rm opt}_{(2)}$ \jbmd{[corresponding to Eq.~\eqref{eq:intro-v}]} and $x^{\rm opt}_{(15)}$ in the ${\rm HDE}{[}v$,$\mathcal{V}_{1}$].
Panel (d): Dependence of $|t_1|$ on $x^{\rm opt}_{(2)}$ in the ${\rm HDE}{[}v$,\jbmd{$\{ \Delta E_{xp}, |t_{xp}| \}$}].
The panel (d) corresponds to the MOD2 in Eq.~\eqref{eq:main_v_vs3}.
Panel (e): Representation of the MOD2 of $v$ on $\Delta E_{xp}$ and $|t_{xp}|$.
\jbmb{In the panels (b-d), the CF that corresponds to each color point is shown in Fig.~\ref{fig:chart-comp},
and the solid black line shows the linear interpolation.}
}}
\label{fig:results-v}
\end{figure}

\paragraph*{$({\rm II})$ Dependence of $v$ on $\mathcal{V}_{1}$ ---} 	
First, $v$ depends entirely on the variables in $\mathcal{V}_{1}$.
The ${\rm HDE}{[}v$,$\mathcal{V}_{1}$] yields $f_{\infty}$[$v$,$\mathcal{V}_{1}$] = 0.992
[see Fig.~\ref{fig:results-v}(a)],
so that $\mathcal{D}_{}{[}v$,$\mathcal{V}_{1}$] is correct.
Second, the dependence of $v$ can be restricted to the subset $\{R_{\rm A},Z_{\rm A},R_{\rm X}\}$ of $\mathcal{V}_{1}$.
The variables $R_{\rm A}$, $Z_{\rm A}$, $R_{\rm X}$, $n_{\rm AB}$ correspond to $x_{i^{\rm opt}_{g}}$ at $g=1,2,3,4$, respectively,
and the ${\rm rHDE}$[v,$\mathcal{V}_{1}$] yields
$f_{\infty}$[$v$,$\{ R_{\rm A}, Z_{\rm A}, R_{\rm X}, n_{\rm AB} \}$] = 0.9\jbmb{91},
$f_{\infty}$[$v$,$\{ R_{\rm A}, Z_{\rm A}, R_{\rm X} \}$] = 0.9\jbmb{80},
and $f_{\infty}$[$v$,$\{ R_{\rm A}, Z_{\rm A} \}$] = 0.95\jbmb{0},
so that $\mathcal{D}_{}{[}v$,$\{ R_{\rm A}, Z_{\rm A}, R_{\rm X} \}$] is correct
but $\mathcal{D}_{}{[}v$,$\{ R_{\rm A}, Z_{\rm A} \}$] is not.
\jbmd{The MOD2 of $v$ on $\mathcal{V}_{1}$ is given in Eq.~\eqref{eq:intro-v}.}
The score analysis shows that $R_{\rm A}$ and $Z_{\rm A}$ correspond to $x_{i^{\rm opt}_{1}}$ and $x_{i^{\rm opt}_{2}}$ unambiguously 
(see Appendix~\ref{app:HDEinterp}).
\\

\paragraph*{$(6)$ Dependence of $v$ on $\mathcal{V}_{3}$ ---} 	
$v$ is entirely determined by $|t_{xp}|$ and $\Delta E_{xp}$ in $\mathcal{V}_{3}$,
but the clarification of this dependence is a bit more subtle and we discuss it in detail here.
The ${\rm HDE}{[}v$,$\mathcal{V}_{3}$] yields $f_{\infty}$[$v$,$\mathcal{V}_{3}$] = 0.999
[see Fig.~\ref{fig:results-v}(a)].
The variable $|\jbmd{\epsilon}^{\rm Cu}_{p_z}|$ corresponds to $x_{i^{\rm opt}_1}$, and we have $f_{(1)}$[$v$,$\mathcal{V}_{3}$] = 0.976.
However, the physical MOD1 of $v$ is not on $|\jbmd{\epsilon}^{\rm Cu}_{p_z}|$ but rather on $\Delta E_{xp}$, as discussed below.
The score analysis in Appendix~\ref{app:HDEinterp}
shows that the three variables $|\epsilon^{\rm O}_{p}|$, $|\epsilon^{\rm O}_{p_\pi}|$ and $\Delta E_{xp} = \jbmd{\epsilon}^{\rm Cu}_{x^2} - \jbmd{\epsilon}^{\rm O}_{p}$
are in very close competition with $|\jbmd{\epsilon}^{\rm Cu}_{p_z}|$ at $g=1$.
These four variables have a common point: 
They are all related to the onsite energies of in-plane O$2p$ orbitals.
The information that can be extracted from the above result is the following:
\jbmd{$v$ is primarily controlled by the energy of the in-plane O$2p$ orbitals.}
We pinpoint the physical dependence as that on $\Delta E_{xp}$
by considering the \jbmd{result} in~\cite{Moree2023Hg1223}:
$v$ mainly depends on $R_{xp}=|t_{xp}|/\Delta E_{xp}$,
and $v$ increases when $R_{xp}$ decreases (that is, when $|t_{xp}|$ decreases or $\Delta E_{xp}$ increases).
The interpretation is reminded here:
Decreasing $R_{xp}$ reduces the Cu$3d_{x^2-y^2}$/O$2p_{\sigma}$ hybridization,
which increases the localization of the AB orbital and thus $v$.\footnote{If $|t_{xp}| \rightarrow 0$, the Cu$3d_{x^2-y^2}$ and O$2p_{\sigma}$ orbitals do not overlap and thus do not hybridize. 
If $\Delta E_{xp} \rightarrow +\infty$, the difference between the Cu$3d_{x^2-y^2}$ and O$2p_{\sigma}$ energy levels becomes very large, so that the hybridization becomes negligible.}
To confirm the consistency with~\cite{Moree2023Hg1223},
we perform the ${\rm HDE}_{}[v,\{ \Delta E_{xp}, |t_{xp}| \}]$.
We obtain $f_{(2)}$[$v$,$\{ \Delta E_{xp}, |t_{xp}| \}$] = 0.985,
so that $\mathcal{D}_{}{[}v$,$\{ \Delta E_{xp}, |t_{xp}| \}$] is correct
and the MOD2 of $v$ on $\{ \Delta E_{xp}, |t_{xp}| \}$ is accurate.
The MOD2 is
\begin{equation}
v_{\rm MOD2} = 10.119 + 4.710 \Big[ \Delta E_{xp}^{0.93} [ 1 - 0.399 |t_{xp}^{0.69}|] \Big]
\label{eq:main_v_vs3}
\end{equation}
and is illustrated in Fig.~\ref{fig:results-v}(\jbmd{d}).
We choose to keep Eq.~\eqref{eq:main_v_vs3} as the final result for the dependence of $v$ on $\mathcal{V}_{3}$.
For completeness,
we also perform the
${\rm HDE}{[}v$,$\{ R_{xp} \}$].
We obtain  $f_{\infty}$[$v$,$\{ R_{xp} \}$] = 0.972
and $f_{(1)}$[$v$,$\{ R_{xp} \}$] = 0.971,
and the MOD1 is
\begin{equation}
v_{\rm MOD1} = 6.824 + 5.941 R_{xp}^{-0.67}.
\label{eq:main_v_vs3_rxp}
\end{equation}
\jbmd{According to} both Eq.~\eqref{eq:main_v_vs3} and Eq.~\eqref{eq:main_v_vs3_rxp},
\jbmd{$v$ increases}
when $|t_{xp}|$ decreases or $\Delta E_{xp}$ increases,
which is consistent with~\cite{Moree2023Hg1223}.

\subsection{Chemical formula dependence of $R$}
\label{sec:results-R}

Here, we first detail (IV). Then, we discuss the items (7-10) in Fig.~\ref{fig:treeplot}.
\\

\begin{figure}[!htb]
\includegraphics[scale=0.135]{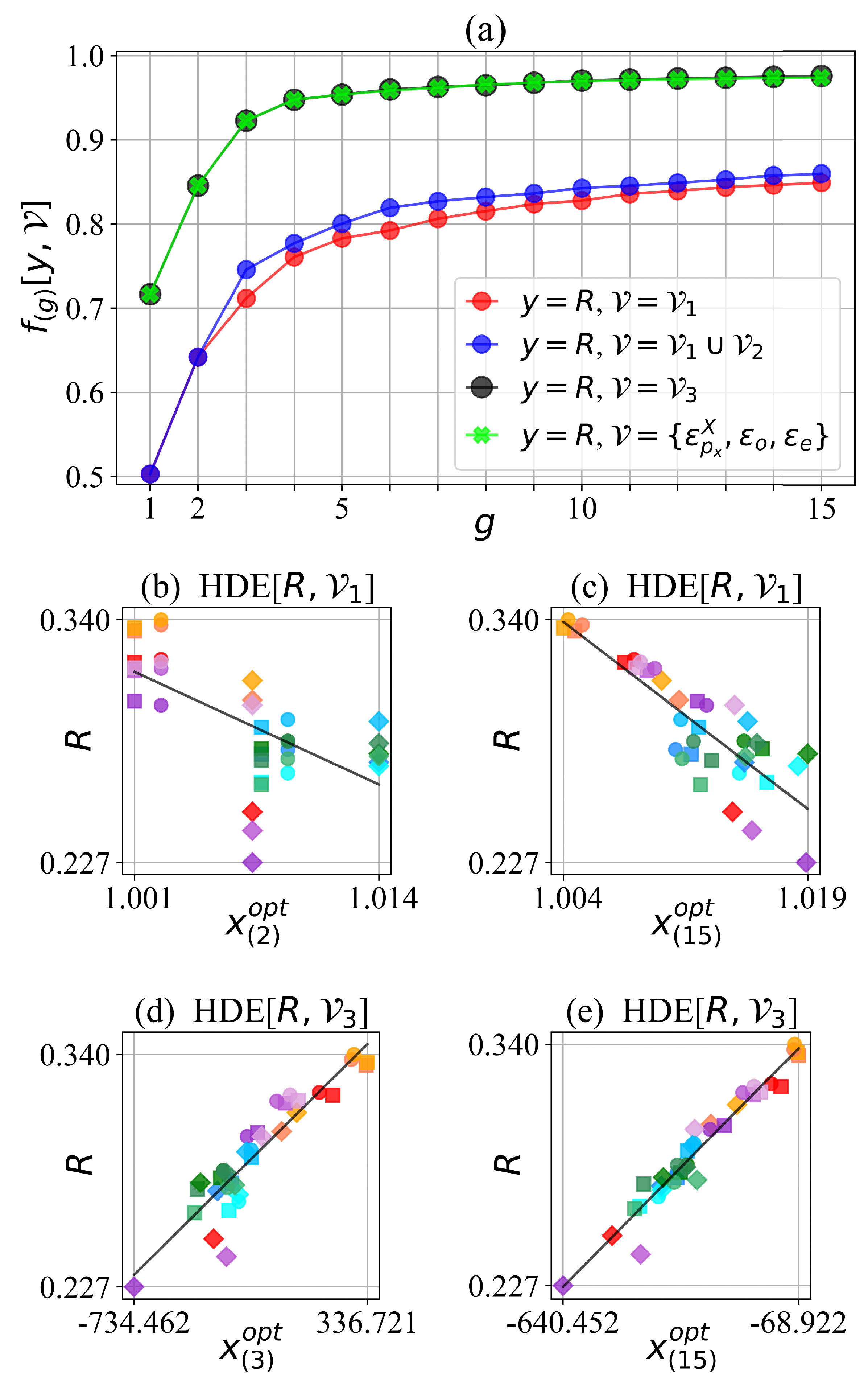}
\caption{\mycaption{
Dependence of $R$ on $\mathcal{V}_{1}$ and $\mathcal{V}_{3}$.
Panel (a): Values of $f_{(g)}[R,\mathcal{V}]$
in the ${\rm HDE}{[}R$,$\mathcal{V}_{1}$], 
\jbmd{the ${\rm HDE}{[}R$,$\mathcal{V}_{1}\cup\mathcal{V}_{2}$],} 
the ${\rm HDE}{[}R$,$\mathcal{V}_{3}$] 
and the r${\rm HDE}{[}R$,$\mathcal{V}_{3}$].
Panels (b,c): Dependence of $R$ on $x^{\rm opt}_{(2)}$ \jbmb{[corresponding to Eq.~\eqref{eq:intro-R}]} and $x^{\rm opt}_{(15)}$ in the ${\rm HDE}{[}R$,$\mathcal{V}_{1}$].
Panels (d,e): Dependence of $R$ on $x^{\rm opt}_{(3)}$ \jbmb{[corresponding to Eq.~\eqref{eq:R-3var}]} and $x^{\rm opt}_{(15)}$ in the ${\rm HDE}{[}R$,$\mathcal{V}_{3}$].
\jbmb{In the panels (b-e), the CF that corresponds to each color point is shown in Fig.~\ref{fig:chart-comp},
and the solid black line shows the linear interpolation.}
}}
\label{fig:results-R}
\end{figure}

\paragraph*{$({\rm III})$ Dependence of $R$ on $\mathcal{V}_{1}$ ---} 	
The dependence of $R$ is more complex than that of $|t_1|$ and $v$:
$R$ is not entirely determined by $\mathcal{V}_{1}$ or even $\mathcal{V}_{1} \cup \mathcal{V}_{2}$.
The ${\rm HDE}{[}R$,$\mathcal{V}_{1}$] yields $f_{\infty}$[$R$,$\mathcal{V}_{1}$] = 0.849, 
so that $\mathcal{D}_{}{[}R$,$\mathcal{V}_{1}$] is incorrect.
The ${\rm HDE}{[}R$,$\mathcal{V}_{1}+\mathcal{V}_{2}$] yields $f_{\infty}$[$R$,$\mathcal{V}_{1}+\mathcal{V}_{2}$] = 0.860, 
so that $\mathcal{D}_{}{[}R$,$\mathcal{V}_{1}\cup\mathcal{V}_{2}$] is still incorrect.

Even though $\mathcal{D}_{}{[}R$,$\mathcal{V}_{1}$] is incorrect,
the MOD2 of $R$ on $\mathcal{V}_{1}$ [Eq.~\eqref{eq:intro-R}] reveals the main-order mechanism of the dependence of $R$.
Namely, $R$ has a very rough MOD2 on $|Z_{\rm X}|$ and $n_{\rm AB}$,
which is consistent with the below discussion.
The score analysis shows that $Z_{\rm X}$ and $n_{\rm AB}$ correspond respectively to $x_{i^{\rm opt}_{1}}$ and $x_{i^{\rm opt}_{2}}$ unambiguously 
(see Appendix~\ref{app:HDEinterp}).

Qualitatively, $R$ increases when (i) $|Z_{\rm X}|$ decreases or (ii) $n_{\rm AB}$ increases in Eq.~\eqref{eq:intro-R} (see also Fig.~\ref{fig:summary}).
The microscopic mechanism of (i) and (ii) is detailed below.

(i) Decreasing $|Z_{\rm X}|$ reduces the negative charge of the apical anion.
This reduces the MP$^{-}$
created by the apical anion and felt by the electrons in the nearby CuO$_2$ plane.
[See Fig.~\ref{fig:cation_madelung}(b) for an illustration.]
This reduces the energy of the electrons in the CuO$_2$ plane, and also reduces the Fermi energy.
(Indeed, the electrons in the CuO$_2$ plane are near the Fermi level, so that the Fermi level is determined by the energy of the electrons in the CuO$_2$ plane.)
On the other hand, the empty states are less affected, and their energy does not change substantially.
However, because the Fermi level is reduced as discussed above, the empty states become higher in energy relative to the Fermi level \jbmd{(so that $\epsilon_{e}$ increases)}.
This reduces the screening from empty states, and thus, increases $R$.

(ii) The decrease in $R$ with decreasing $n_{\rm AB}$ (increasing $\delta$) is discussed in~\cite{Moree2022},
and the microscopic mechanism is reminded here.
When $\delta$ increases, the hole doping of O sites increases.
This reduces the negative charge of in-plane O anions.
This reduces the MP$^{-}$ created by the in-plane O and felt by the nearby Cu.
[See Fig.~\ref{fig:cation_madelung}(c) for an illustration.]
This reduces the energy of Cu$3d$ electrons, and also reduces the Fermi energy.
(Indeed, the AB band at the Fermi level has Cu$3d_{x^2-y^2}$ character.)
On the other hand, the O$2p$ electrons are less affected.
However, because the Fermi level is reduced, the occupied O$2p$ states become closer to the Fermi level \jbmd{(so that $|\epsilon_{o}|$ decreases)}.
This increases the screening from occupied states, and thus, reduces $R$.

\jbmd{For completeness, the hole doping dependence of $R$ is discussed in detail in Appendix~\ref{app:xfcl}, which is summarized here.
Even though $R$ increases when $|Z_{\rm X}|$ decreases as discussed above, 
the decrease in $|Z_{\rm X}|$ also accelerates the decrease in $R$ with decreasing $n_{\rm AB}$,
which is not captured by Eq.~\eqref{eq:intro-R}.
This is why the three color points with the lowest $R$ in Fig.~\ref{fig:results-R}(b) deviate from the linear interpolation, which is a major cause of the relatively low value of $f_{(2)}[R,\mathcal{V}_{1}] = 0.642$.
The MOD2 in Eq.~\eqref{eq:intro-R} may be combined with the results in Appendix~\ref{app:xfcl} to obtain a more accurate picture of the dependence of $R$ on $|Z_{\rm X}|$ and $n_{\rm AB}$.
}
\\

\paragraph*{$(7)$ Dependence of $R$ on $\mathcal{V}_{3}$ ---} 	
$R$ may be entirely determined by the subset $\{ \epsilon^{\rm X}_{p_x},\epsilon_{o},\epsilon_{e}\}$ of $\mathcal{V}_{3}$.
The ${\rm HDE}{[}R$,$\mathcal{V}_{3}$] yields $f_{\infty}$[$R$,$\mathcal{V}_{3}$] = 0.976, 
so that $\mathcal{D}_{}{[}R$,$\mathcal{V}_{3}$] is reasonably correct\jbmd{; also, the} dependence of $R$ on $x^{\rm opt}_{(15)}$ \jbmd{is almost linear [see Fig.~\ref{fig:results-R}(e)]}. 
The r${\rm HDE}{[}R$,$\mathcal{V}_{3}$] 
yields $f_{\infty}$[$R$,$\{ |\epsilon^{\rm X}_{p_x}|,|\epsilon_{o}|,\epsilon_{e}\}$] = 0.974, 
so that $\mathcal{D}_{}{[}R$,$\{ |\epsilon^{\rm X}_{p_x}|,|\epsilon_{o}|,\epsilon_{e}\}$] is reasonably correct as well.
Note that even though $R$ has complex dependencies on the whole band structure \textit{via} the polarization formula in Appendix~\ref{app:meth-mace}, Eq.~\eqref{eq:chiou},
the above result shows that $R$ can be described reasonably by only three characteristic energies in the band structure.

In the following, we discuss the MOD3 of $R$ on $\mathcal{V}_3$
instead of the MOD2 as usually done before. %
This is justified as follows:
In the case of $|t_1|$ and $v$, we have 
$f_{(2)}[|t_1|,\mathcal{V}_3] = 0.994$ 
and $f_{(2)}[v,\mathcal{V}_3] = 0.984$,
so that the MOD2 is accurate.
However, for $R$, we have 
$f_{(2)}[y,\mathcal{V}_3] = 0.846$ but 
$f_{(3)}[y,\mathcal{V}_3] = 0.923$,
so that the MOD3 is more accurate than the MOD2.
To obtain a compromise between accuracy and simplicity,
we choose to discuss the MOD3.
Although the MOD3 does not describe $R$ perfectly,
the \jbmd{dependence of $R$ on $x^{\rm opt}_{(3)}$ is almost linear} as seen in Fig.~\ref{fig:results-R}(d).
The MOD3 of $R$ on $\mathcal{V}_3$ is
\begin{widetext}
\begin{equation}
R_{\rm MOD3} = 0.309 + 0.000105 \Big[ |\epsilon^{\rm X}_{p_x}|^{3.94} - 93.1 |\epsilon_{o}|^{-1.01} - 184 \epsilon_{e}^{-0.93} \Big].
\label{eq:R-3var}
\end{equation}
\end{widetext}
The score analysis confirms that $|\epsilon^{\rm X}_{p_x}|$, $|\epsilon_{o}|$ and $\epsilon_{e}$ correspond respectively to $x_{i^{\rm opt}_{1}}$, $x_{i^{\rm opt}_{2}}$ and $x_{i^{\rm opt}_{3}}$ 
(see Appendix~\ref{app:HDEinterp}).

Qualitatively, $R$ increases (i.e. the screening decreases) when $|\epsilon^{\rm X}_{p_x}|$ increases (i.e. the occupied apical X$2p_x$ orbital becomes farther from the Fermi level), $|\epsilon_{o}|$ increases (the highest occupied energy band becomes farther from the Fermi level), or $\epsilon_{e}$ increases (the lowest empty energy band becomes farther from the Fermi level).
These three dependencies are intuitive, because the screening from a given band decreases when the band energy is farther from the Fermi level.
[See the polarization formula in Appendix~\ref{app:meth-mace}, Eq.~\eqref{eq:chiou}.]

\begin{figure*}[!htb]
\includegraphics[scale=0.13]{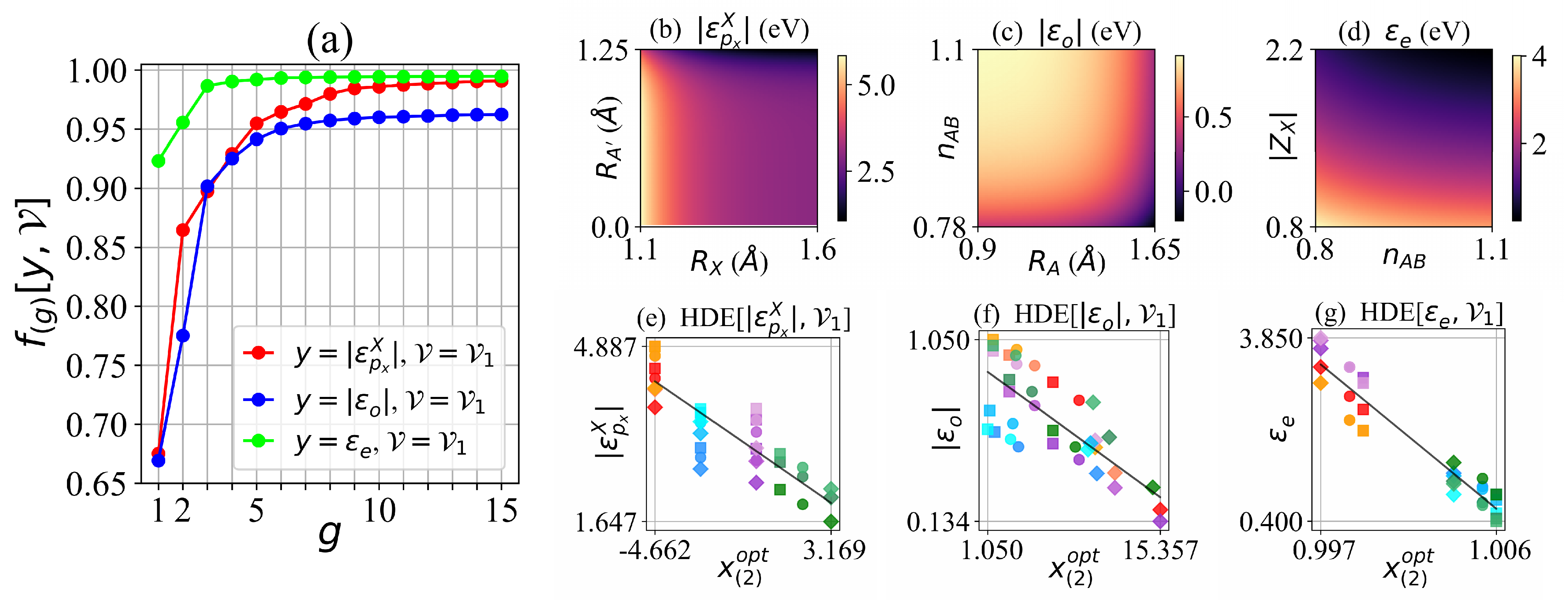}
\caption{\mycaption{
Dependence of $y$ = $|\epsilon^{\rm X}_{p_x}|$, $|\epsilon_o|$ and $\epsilon_e$ on $\mathcal{V}_{1}$.
Panel (a): Values of $f_{(g)}[y,\mathcal{V}]$
in the ${\rm HDE}{[}y$,$\mathcal{V}_{1}$].
Panels (b-d): Representation of the MOD2 of $y$ on $\mathcal{V}_{1}$
in Eqs.~\eqref{eq:R-g1},~\eqref{eq:R-g2},~\eqref{eq:R-g3}.
Panels (e-g): Dependence of $y$ on $x^{\rm opt}_{(2)}$ \jbmd{which corresponds to the panels (b-d), respectively}.
\jbmb{\jbmd{T}he CF that corresponds to each color point is shown in Fig.~\ref{fig:chart-comp},
and the solid black line shows the linear interpolation.}
}}
\label{fig:results-R-heatmap}
\end{figure*}

The three variables $\{ |\epsilon^{\rm X}_{p_x}|,|\epsilon_{o}|,\epsilon_{e}\}$ are entirely determined by $\mathcal{V}_{1}$
but not by $\mathcal{V}_{2}$.
For $y$ in $\{ |\epsilon^{\rm X}_{p_x}|,|\epsilon_{o}|,\epsilon_{e}\}$,
the ${\rm HDE}{[}y$,$\mathcal{V}_{2}$] yields 
$f_{\infty}$[$|\epsilon^{\rm X}_{p_x}|$,$\mathcal{V}_{2}$] = 0.892, 
$f_{\infty}$[$|\epsilon_{o}|$,$\mathcal{V}_{2}$] = 0.728, 
and $f_{\infty}$[$\epsilon_{e}$,$\mathcal{V}_{2}$] = 0.948,
so that $\mathcal{V}_{2}$ does not describe accurately $|\epsilon^{\rm X}_{p_x}|$, $|\epsilon_{o}|$ and $\epsilon_{e}$.
However, the ${\rm HDE}{[}y$,$\mathcal{V}_{1}$] yields 
$f_{\infty}$[$|\epsilon^{\rm X}_{p_x}|$,$\mathcal{V}_{1}$] = 0.991, 
$f_{\infty}$[$|\epsilon_{o}|$,$\mathcal{V}_{1}$] = 0.962, 
and $f_{\infty}$[$\epsilon_{e}$,$\mathcal{V}_{1}$] = 0.995
[see Fig.~\ref{fig:results-R-heatmap}(a)].
Here, we choose to express $|\epsilon^{\rm X}_{p_x}|$, $|\epsilon_{o}|$ and $\epsilon_{e}$
directly as a function of $\mathcal{V}_{1}$ instead of $\mathcal{V}_{2}$
in order to obtain a more accurate expression.
Thus, in the following, we discuss the dependence of $|\epsilon^{\rm X}_{p_x}|$, $|\epsilon_{o}|$ and $\epsilon_{e}$ on $\mathcal{V}_{1}$.

The MOD2s of $|\epsilon^{\rm X}_{p_x}|$, $|\epsilon_{o}|$ and $\epsilon_{e}$ on $\mathcal{V}_{1}$ are:
\begin{align}
&|\epsilon^{\rm X}_{p_x}|_{\rm MOD2}   =  2.902   -    0.287 \Big[ R_{\rm A'}^{8.49} - 26.5 R_{\rm X}^{-9.99} \Big] \label{eq:R-g1} \\
&|\epsilon_{o}|_{\rm MOD2}                            = 0.934   -    0.044\jbmb{3} \Big[ n_{\rm AB}^{-9.99} + 0.151 R_{\rm A}^{8.99} \Big] \label{eq:R-g2}\\
&\epsilon_{e {\rm MOD2}}                             = 298.06\jbmb{2} - 295.6\jbmb{36} \Big[ |Z_{\rm X}|^{0.01} - 0.000899 n_{\rm AB}^{-5.58} \Big]  \label{eq:R-g3}
\end{align}
and are represented in Fig.~\ref{fig:results-R-heatmap}.
[These correspond to the items (8), (9) and (10) in Fig.~\ref{fig:treeplot}, respectively.]
The score analysis confirms that $x_{i^{\rm opt}_{1}}$ and $x_{i^{\rm opt}_{2}}$ correspond to
$R_{\rm A'}$ and $R_{\rm X}$ in the dependence of $|\epsilon^{\rm X}_{p_x}|$,
$n_{\rm AB}$ and $R_{\rm A}$ in the dependence of $|\epsilon_{o}|$,
and $Z_{\rm X}$ and $n_{\rm AB}$ in the dependence of $\epsilon_{e}$.
(See Appendix~\ref{app:HDEinterp}.)
\\

On (III) and (7), let us discuss \jbmd{the} common dependencies of $R$, $|\epsilon^{\rm X}_{p_x}|$, $|\epsilon_{o}|$ and $\epsilon_{e}$ 
on $|Z_{\rm X}|$ and $n_{\rm AB}$. %
First, as for $|Z_{\rm X}|$, the MODs of $R$ and $|\epsilon^{\rm X}_{p_x}|$ on $|Z_{\rm X}|$ are consistent.
Indeed, both $R_{\rm MOD2}$ [Eq.~\eqref{eq:intro-R}] and $|\epsilon^{\rm X}_{p_x}|_{\rm MOD2}$ [Eq.~\eqref{eq:R-g1}] increase when $|Z_{\rm X}|$ decreases;
consistently, $R_{\rm MOD3}$ [Eq.~\eqref{eq:R-3var}] increases when $|\epsilon^{\rm X}_{p_x}|$ increases.

Second, as for $n_{\rm AB}$,
$R_{\rm MOD2}$ decreases with decreasing $n_{\rm AB}$ in Eq.~\eqref{eq:intro-R}:
This is the result of a competition between the MODs of $|\epsilon_{o}|$ and $\epsilon_{e}$ on $n_{\rm AB}$
[Eqs.~\eqref{eq:R-g2} and~\eqref{eq:R-g3}].
Indeed, when $n_{\rm AB}$ decreases, the two following mechanisms (i,ii) occur.
On one hand, (i) 
$|\epsilon_{o}|_{\rm MOD2}$ [Eq.~\eqref{eq:R-g2}] decreases,
\jbmd{which contributes to decrease $R_{\rm MOD3}$} according to Eq.~\eqref{eq:R-3var}.
On the other hand, (ii) 
$\epsilon_{e {\rm MOD2}}$ [Eq.~\eqref{eq:R-g3}] increases,
which contributes to increase $R_{\rm MOD3}$ according to Eq.~\eqref{eq:R-3var}.
The fact that $R_{\rm MOD2}$ decreases with decreasing $n_{\rm AB}$ in Eq.~\eqref{eq:intro-R}
suggests that (ii) dominates over (i).
\\

\paragraph*{$(8)$ Dependence of $|\epsilon^{\rm X}_{p_x}|$ on $\mathcal{V}_{1}$ ---} 	
The MOD of $|\epsilon^{\rm X}_{p_x}|$ on $R_{\rm A'}$ and $R_{\rm X}$ in Eq.~\eqref{eq:R-g1} is understood as follows.
First, $|\epsilon^{\rm X}_{p_x}|$ decreases when $R_{\rm A'}$ increases. 
\jbmd{Zero $R_{\rm A'}$ corresponds to A' = $\varnothing$,
whereas
nonzero $R_{\rm A'}$ corresponds to A' = Hg$_{1-\delta}$Au$_{\delta}$.}
The symmetry of the primitive cell changes from A' = $\varnothing$ to A' = Hg$_{1-\delta}$Au$_{\delta}$ (see Fig.~\ref{fig:nnmace}),
and the crystalline environment changes as well.
And, if we represent $|\epsilon^{\rm X}_{p_x}|$ as a function of $|\epsilon^{\rm X}_{p_z}|$
\jbmd{(in Appendix~\ref{app:int}),}
we see that $|\epsilon^{\rm X}_{p_x}| \simeq k_0 + k_1 |\epsilon^{\rm X}_{p_z}|$ has an affine dependence on $|\epsilon^{\rm X}_{p_z}|$,
but the coefficients $k_0$ are different for A' = $\varnothing$ and A' = Hg$_{1-\delta}$Au$_{\delta}$
whereas the coefficients $k_1$ are nearly identical.
Namely, the affine regression yields
\begin{align}
|\epsilon^{\rm X}_{p_x}| & = 0.181 + 0.940|\epsilon^{\rm X}_{p_z}| \ \ ({\rm A' = \varnothing}), \label{eq:affeap_1}\\
|\epsilon^{\rm X}_{p_x}| & = -1.436 + 0.965|\epsilon^{\rm X}_{p_z}| \ \ ({\rm A' = {\rm Hg}_{1-\delta}{\rm Au}_{\delta}}). \label{eq:affeap_2}
\end{align}
Thus, from A' = $\varnothing$ to A' = Hg$_{1-\delta}$Au$_{\delta}$, 
$|\epsilon^{\rm X}_{p_x}|$ is reduced by $1.62$ eV
for a given value of $|\epsilon^{\rm X}_{p_z}|$.
On the other hand, for A' = $\varnothing$, the value of $k_0$ is universal irrespective of A and X.
This suggests that the decrease in $|\epsilon^{\rm X}_{p_x}|$ from A' = $\varnothing$ to A' = Hg$_{1-\delta}$Au$_{\delta}$ does not depend on A or X,
but rather on the presence of the A' atom and the subsequent change in crystal structure and crystal electric field.
A possible explanation is the following:
For A' = Hg$_{1-\delta}$Au$_{\delta}$, there is a A' atom close to the apical X (see Fig.~\ref{fig:nnmace}),
and the apical X$2p$ orbital overlaps with the %
A'$5d$ orbitals.
This may cause the apical X$p_{x}$ MLWO to catch antibonding %
X$2p$/A'$5d$ character\footnote{\jbmd{Note that the X$p_{x}$ orbital that is considered here is a MLWO, whose character may be slightly different from the purely atomic $p_x$ character.}}, 
which may destabilize the apical X$p_{x}$ MLWO (i.e. increase its onsite energy and thus reduce $|\epsilon^{\rm X}_{p_x}|$).
This is consistent with the isosurface of the apical X$p_{x}$ MLWO in Fig.~\ref{fig:nnmace} for A' = Hg$_{1-\delta}$Au$_{\delta}$:
We see that the apical X$p_x$ MLWO has a slight A'$yz/zx$ character near the A' atom.

Second, $|\epsilon^{\rm X}_{p_x}|$ increases when $R_{\rm X}$ decreases.
This is consistent with ~\cite{Moree2023Hg1223}:
When $a$ decreases, the occupied bands in the M space (including apical X bands) become farther from the Fermi level [see~\cite{Moree2023Hg1223}, Fig.~3(h-k)]. 
This is because the MP$^{-}$ created by the in-plane O and felt by the Cu increases [as illustrated in Fig.~\ref{fig:cation_madelung}(c)], so that the energy of the Cu$3d_{x^2-y^2}$ increases (and this shifts the Fermi level upwards), whereas the apical X orbitals are less affected.
And, the decrease in $R_{\rm X}$ contributes to decrease $a$, as discussed previously.
\\

\paragraph*{$(9)$ Dependence of $|\epsilon_{o}|$ on $\mathcal{V}_{1}$ ---} 	
The MOD of $|\epsilon_{o}|$ on $n_{\rm AB}$ and $R_{\rm A}$ in Eq.~\eqref{eq:R-g2} is understood as follows.
First, $|\epsilon_{o}|$ decreases when $n_{\rm AB}$ decreases (i.e. the hole doping increases).
This is consistent with~\cite{Moree2022}, and \jbmd{the cause is interpreted as a rigid shift of the Fermi level upon hole doping.}
Because the partially filled AB band is the only band at the Fermi level,
increasing the hole doping reduces the number of electrons in the AB band,
which shifts the Fermi level downwards.
As a result, the energy of occupied bands relative to the Fermi level increases.
These include the highest occupied band outside the AB band, whose energy is $\epsilon_{o}$.
\jbmd{Thus, $\epsilon_{o}<0$ increases, so that $|\epsilon_{o}|$ decreases.}

Second, $|\epsilon_{o}|$ increases when $R_{\rm A}$ decreases.
\jbmd{
The microscopic mechanism is discussed in detail in Appendix~\ref{app:doseo}.
}
\\

\paragraph*{$(10)$ Dependence of $\epsilon_{e}$ on $\mathcal{V}_{1}$ ---} 
The MOD of $\epsilon_{e}$ on $|Z_{\rm X}|$ and $n_{\rm AB}$ in Eq.~\eqref{eq:R-g3} is understood as follows.
First, $\epsilon_{e}$ increases when $|Z_{\rm X}|$ decreases.
\jbmd{The mechanism was summarized in Sec.~\ref{sec:overview} [see (IV), item (i)], and is detailed here.}
Apical X anions with the negative charge $Z_{\rm X}$ emit a MP$^{-}$ that increases the energy of surrounding electrons,
in particular those in the M bands, because the Cu and in-plane O atoms are in the vicinity of apical X.
Reducing $|Z_{\rm X}|$ reduces the MP$^{-}$ from apical X felt by the Cu and in-plane O.
[See Fig.~\ref{fig:cation_madelung}(b) for an illustration.]
This reduces the energy of M bands.
The Fermi energy is also reduced, because it is determined by the partially filled AB band which is in the M space.
Thus, the empty bands become farther from the Fermi level.
This is consistent with results on Hg1223 in~\cite{Moree2023Hg1223} [see e.g. Appendix E1].

Second, $\epsilon_{e}$ increases when $n_{\rm AB}$ decreases.
This is because (i) the Fermi level is shifted downwards when $n_{\rm AB}$ decreases \jbmd{because of the hole doping of the AB band} [as discussed in (9)]:
As a consequence, the empty bands become farther from the Fermi level, and thus $\epsilon_{e}$ increases.
In addition, (ii) when $n_{\rm AB}$ decreases, the M bands are stabilized, which further shifts the M bands and thus the Fermi level downwards.
This is because the hole doping of in-plane O increases when $n_{\rm AB}$ decreases \jbmd{as mentioned before}: %
This reduces the negative charge of the in-plane O ions, and thus the Madelung potential created by the in-plane O ions.
[See Fig.~\ref{fig:cation_madelung}(c) for an illustration.]
This stabilizes the M bands, which increases the energy gap between M bands and empty bands.
\\

\section{Discussion}
\label{sec:disc}

Here, we discuss
the universality and accuracy of the obtained expressions of $|t_1|$, $v$ and $R$\jbmb{. Also,}
we propose prescriptions to optimize $T_{c}^{\rm opt}$ in future design of SC cuprates and akin materials.

\subsection{Universality and accuracy of the expressions of AB LEH parameters}

The CF dependencies of $|t_1|$, $v$ and $R$ obtained in this paper 
offer reliable guidelines for design \jbm{of SC} cuprates and akin compounds,
if we assume \jbmb{that
(A) the} training set considered in this paper is representative of the diversity in single-layer cuprates\jbmb{, and
(B) the} CF dependence of the AB LEH parameters $|t_1|$ and $u$ is correctly captured (at least qualitatively) by the GGA+cRPA version of MACE employed in this paper.
\\

(A) is supported in Appendix~\ref{app:sc}; below, we support (B).
\todounseen{Transfer this to appendix as well ?}
\jbmb{The detailed discussion on (B) is necessary, because t}he GGA+cRPA is the simplest level of the MACE scheme;
more sophisticated versions of MACE have been employed in previous works, such as the 
\jbm{constrained $GW$ (c$GW$) supplemented with self-interaction correction (SIC) at the c$GW$-SIC level}~\cite{Hirayama2018}
and \jbm{level renormalization feedback (LRFB) at the c$GW$-SIC+LRFB level}~\cite{Hirayama2019}.
Eq.~\eqref{eq:tcopt} was determined by solving AB LEHs at the c$GW$-SIC+LRFB level~\cite{Moree2022}.

The GGA+cRPA is expected to capture correctly the qualitative materials dependence of $|t_1|$ and $u$~\cite{Moree2022,Moree2023Hg1223} while avoiding the complexity of the c$GW$-SIC+LRFB calculation.
For instance, $u$ is smaller in Bi2201 compared to Bi2212 at the c$GW$-SIC+LRFB level, and this result is reproduced qualitatively by the GGA+cRPA~\cite{Moree2022}.
In addition, in Hg1223, the qualitative pressure dependence of $|t_1|$ and $u$ is captured by the GGA+cRPA~\cite{Moree2023Hg1223}.

Still, it should be noted that the accurate prediction of the materials dependence of $T_{c}^{\rm opt}$ cannot be done 
by considering the materials dependent $|t_1|$ and $u$ at the GGA+cRPA level (the values in Fig.~\ref{fig:chart-t1u}).
Indeed, the GGA+cRPA has a limitation at the quantitative level.
For instance, in Bi2201 and Bi2212, $u$ is underestimated at the GGA+cRPA level,
and the difference between the values of $u$ ($|t_1|$) at the GGA+cRPA and c$GW$-SIC+LRFB levels is around $10\%$ ($5\%$)~\cite{Moree2022}.
The uncertainty on $u$ may cause a significant uncertainty on $F_{\rm SC}$ and thus $T_{c}^{\rm opt}$,
because $T_{c}^{\rm opt}$ strongly depends on $u$ \textit{via} $F_{\rm SC}$ in Eq.~\eqref{eq:tcopt}, 
especially when $u$ is located in the weak-coupling region $u \simeq 6.5-8.0$~\cite{Schmid2023}.
Thus, the quantitative improvement from the GGA+cRPA level to the c$GW$-SIC+LRFB level is required to tackle the \jbmb{accurate} prediction of $T_{c}^{\rm opt} \simeq 0.16 |t_1| F_{\rm SC}$~\cite{Schmid2023} from the values of $|t_1|$ and $u$.

Nonetheless, we restrict to the simplest GGA+cRPA level in the present paper,
because 
(i) the c$GW$-SIC+LRFB calculation is complex and computationally expensive,
and (ii) the GGA+cRPA is expected to capture correctly the materials dependence of the AB LEH at least qualitatively~\cite{Moree2022,Moree2023Hg1223} as discussed above.
Quantitative prediction of the CF dependence of $T_{c}^{\rm opt}$ by using Eq.~\eqref{eq:tcopt} requires the CF dependence of the c$GW$-SIC+LRFB result, which is left for future studies.

Also, note that some of the results obtained in this paper are independent of the restriction to the GGA+cRPA level,
and are expected to remain valid for the AB LEH at the c$GW$-SIC+LRFB level.
This is the case of the items (3,5,6,7) in Fig.~\ref{fig:treeplot}.
For instance, the dependencies of $R$ and $v$ on $\mathcal{V}_{3}$ in Eqs.~\eqref{eq:R-3var} and~\eqref{eq:main_v_vs3} 
make complete abstraction of the level of the electronic structure from which the AB LEH is calculated (GGA in the case of GGA+cRPA, or $GW$+LRFB~\cite{Hirayama2019} in the case of c$GW$-SIC+LRFB).
Thus, the dependencies of $R$ and $v$ on $\mathcal{V}_{3}$ in Eqs.~\eqref{eq:R-3var} and~\eqref{eq:main_v_vs3} are expected to be rather universal.
(The detailed dependencies of \jbmb{$|t_{xp}|$, $\Delta E_{xp}$, $|\epsilon^{\rm X}_{p_x}|$, $|\epsilon_{o}|$ and $\epsilon_{e}$} at the $GW$+LRFB level on the CF are left for future studies.)
However, on $|t_1|$, the dependence on $\mathcal{V}_{3}$ in Eq.~\eqref{eq:t1s3}
does not take into account the removal of exchange-correlation double counting~\cite{Hirayama2013} that is done at the c$GW$-SIC+LRFB level.
This corrects $|t_1|$ by a term which is materials dependent~\cite{Moree2022,Moree2023Hg1223}.
(The detailed materials dependence of this term is left for future studies.)

\subsection{Prescriptions to optimize $T_{c}^{\rm opt}$}

By considering the MODs given in Sec.~\ref{sec:overview}, 
we propose prescriptions to optimize $T_{c}^{\rm opt}$ in future design of SC cuprates. 
The overall strategy is to maximize $|t_1|$ while keeping $u$ as close as possible to its optimal value $u_{\rm opt} \simeq \jbmd{8.0-}8.5$.
Indeed, $T_{c}^{\rm opt}$ increases with both $|t_1|$ and $F_{\rm SC}$ in Eq.~\eqref{eq:tcopt},
and the universal $u$ dependence of $F_{\rm SC}$ has a maximum at $u_{\rm opt}$ (see Sec.~\ref{sec:intro}). %
To realize $u=U/|t_1|=vR/|t_1|$ as close as possible to $u_{\rm opt}$, the criterion
\begin{equation}
vR/|t_1| \simeq 8.5
\label{eq:vRt1uopt}
\end{equation}
should be satisfied.

To satisfy Eq.~\eqref{eq:vRt1uopt}, the values of $|t_1|$, $v$ and $R$ may be tuned rather independently 
if we consider their distinct dependencies on the CF.
For instance, $Z_{\rm A}$ appears in the MOD of $v$ [Eq.~\eqref{eq:intro-v}] but not in that of $|t_1|$ [Eq.~\eqref{eq:intro-t1}],
so that tuning $Z_{\rm A}$ allows to tune $v$ without affecting $|t_1|$ substantially.

Note that the dependence of $|t_1|$ on the CF in Eq.~\eqref{eq:intro-t1} predicts an upper bound $T_{\rm c, max}^{\rm opt} \simeq 140$ K for $T_{c}^{\rm opt}$ in single-layer cuprates at ambient pressure.
Indeed, according to Eq.~\eqref{eq:intro-t1}, there is a maximal value of $|t_1|$, which is $|t_1|_{\rm max} \simeq 0.53$ eV ($|t_1|_{\rm max} \simeq 0.58$ eV if we consider $g=15$ instead of $g=2$).
And, in the universal $u$ dependence of $F_{\rm SC}$~\cite{Schmid2023}, the maximal value of $F_{\rm SC}$ at $u=u_{\rm opt}$ is $F_{\rm SC, max} \simeq 0.13$.
Thus, for $|t_1|_{\rm max} \simeq 0.58$ eV, the maximal value of $T_{c}^{\rm opt}$ is
$T_{\rm c, max}^{\rm opt} = 0.16 |t_1|_{\rm max} F_{\rm SC, max} \simeq 0.012$ eV ($140$ K)
according to Eq.~\eqref{eq:tcopt}.

The upper bound $|t_1|_{\rm max}$ corresponds to $R_{\rm X}=R_{\rm A}=0$, which cannot be reached in experiment.
However, reducing $R_{\rm X}$ or $R_{\rm A}$ may allow to increase $|t_1|$
and make $|t_1|$ as close as possible to $|t_1|_{\rm max}$.
Reducing $R_{\rm X}$ or $R_{\rm A}$ can be done by e.g. replacing A or X by an isovalent atom that has a smaller ionic radius.
This contributes to reduce the interatomic distances between atoms in the crystal and thus $a$ according to Eq.~\eqref{eq:main_a_vs1},
by mimicking the effects of physical $P_a$ at the chemical level.

Under pressure, $|t_1|$ may be further increased due to the application of physical $P_a$,
and may \jbmd{reach or} exceed the upper bound $|t_1|_{\rm max}$ at ambient pressure.\footnote{For instance, in Hg1223, $|t_1| \simeq 0.57-0.60$ eV at $P=30$ GPa, and $|t_1| \simeq 0.62-0.67$ eV at $P=60$ GPa~\cite{Moree2023Hg1223}.}
Note that the $P_a$-induced increase in $|t_1|$ eventually causes a rapid decrease in $u$ and thus $F_{\rm SC}$ by making $u$ fall into the weak-coupling region~\cite{Moree2023Hg1223};
this may be countered by tuning $Z_{\rm A}$ and $Z_{\rm X}$ to increase $v$ and $R$, so that the criterion in Eq.~\eqref{eq:vRt1uopt} is still satisfied at high $P_a$.

\jbmc{Finally, for completeness, we discuss the particular case of X = F, Cl in Appendix~\ref{app:xfcl}.
Indeed, for X = F, Cl, the decrease in $u$ with decreasing $n_{\rm AB}$ is sharper compared to X = O,
which is an effect beyond the MOD2 in Eq.~\eqref{eq:intro-u}.
The origin of this sharper decrease 
\jbmd{is the sharper decrease in $R$ for X = F, Cl compared to X = O, as mentioned before and detailed in Appendix~\ref{app:xfcl}. P}ossible implications on SC properties are discussed in Appendix~\ref{app:xfcl}.
}

\todounseen{X=Cl,F: strong-coupling or weak-coupling ? -> analyze $\delta$ dependence of $u$ for each mother compound ?}

\section{Conclusion}
\label{sec:concl}

We have proposed the universal CF dependence of the AB LEH parameters $|t_1|$ and $u$ in single-layer cuprates,
by proposing \jbm{the HDE} procedure and applying it to analyze the results of \textit{ab initio} calculations of the AB LEH
for various single-layer cuprates.
The results and especially the MODs given in Sec.~\ref{sec:overview} provide insights to optimize $T_{c}^{\rm opt}$ in future design of SC cuprates as proposed in Sec.~\ref{sec:disc}.

The qualitative insights obtained in this paper may also be useful for design of other SC materials whose crystal structure is similar to that of cuprates,
namely, other strongly correlated electron materials \jbmd{whose low-energy physics may be described} by the AB LEH on the two-dimensional square lattice.
These include nickelates~\cite{Li2019} and the recently proposed Ag- and Pd-based compounds~\cite{Hirayama2022silverarxiv,Kitatani2023}.

More generally, the gMACE+HDE procedure employed in this paper may be used for design of strongly correlated electron materials
including those which do not have SC properties.
Even more generally, the HDE procedure offers a general platform 
to extract dependencies between any physical quantities $y$ and $x_i$, regardless their physical meaning. 
These quantities may be either calculated within a theoretical framework or measured in an experiment.

\section*{Acknowledgements}
This work was done under the Special Postdoctoral Researcher Program at RIKEN.
Part of the figures were drawn by using the software \texttt{VESTA} \cite{Momma2011}.
We thank Masatoshi Imada, Youhei Yamaji and Shiro Sakai for discussions during the preliminary phase of the project.

\renewcommand{\thesection}{A\arabic{section}}
\renewcommand{\theequation}{A\arabic{equation}}
\appendix

\begin{appendices}

\section{Dependence of superconducting order parameter on $u=U/|t_1|$}
\label{app:ufsc}
\renewcommand{\theequation}{A\arabic{equation}}

\begin{figure}[!htb]
\includegraphics[scale=0.065]{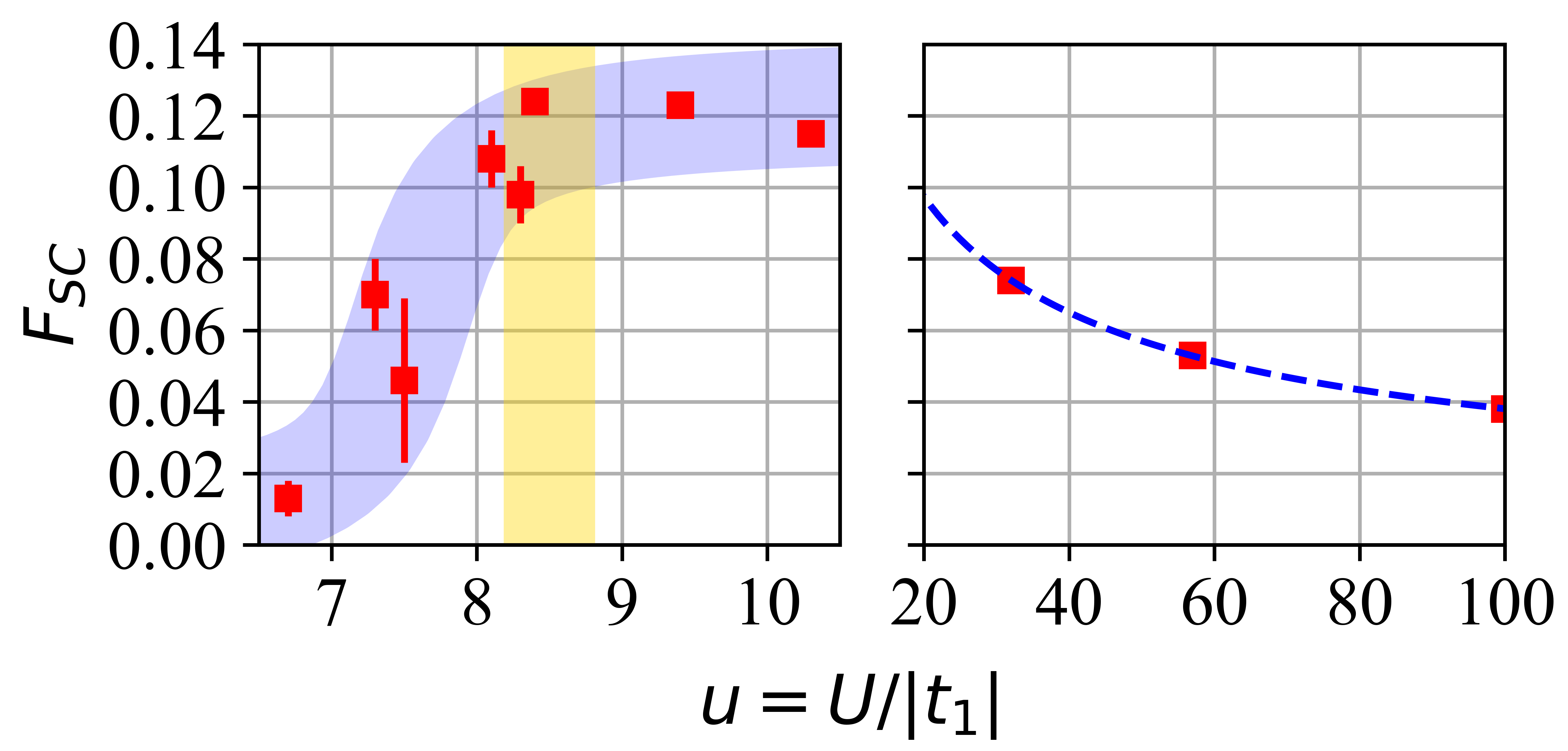}
\caption{
\jbmf{Dependence of the superconducting order parameter $F_{\rm SC}$ on $u=U/|t_1|$, from Ref.~\cite{Schmid2023}.
The values shown by the red squares and error bars are taken from Fig.~10 and Fig.~21 in Ref.~\cite{Schmid2023} (at hole doping $\delta = 0.167$).
The blue area shows the rough shape of the $u$ dependence of $F_{\rm SC}$ for $u \lesssim 10.5$.
The golden vertical bar shows the optimal regime ($u \simeq u_{\rm opt} \simeq 8.5$) in which $F_{\rm SC}$ reaches its maximum.
The dashed blue curve shows the scaling $F_{\rm SC} \simeq 0.56u^{-0.58}$ in the limit of large $u$.
}}
\label{fig:ufsc}
\end{figure}

In Sec.~\ref{sec:intro}, we mention the $u$ dependence of $F_{\rm SC}$ from Ref.~\cite{Schmid2023}.
Here, as a complement, we show this dependence in Fig.~\ref{fig:ufsc}.

\section{Choice of the training set of cuprates for the gMACE+HDE procedure}
\label{app:sc}
\renewcommand{\theequation}{B\arabic{equation}}

In Sec.~\ref{sec:disc}, we assume that
(A) the training set considered in this paper 
\jbmd{and represented in Fig.~\ref{fig:chart-comp}}
is representative of the diversity in single-layer cuprates.
Here, we support (A). %

First, the training \jbmd{set} includes CFs that correspond to experimentally confirmed SC cuprates
with a diverse distribution of $T_{c}^{\rm opt} \simeq 24-94$ K, 
including HgBa$_2$CuO$_4$ which has the highest known value of $T_{c}^{\rm opt} \simeq 94$ K 
among single-layer cuprates at ambient pressure.
These cuprates are listed in Table~\ref{tab:expsc} together with the \jbmd{experimental values of $T_{c}$.}

\begin{table}[!ht]
\caption{\mycaption{List of experimentally confirmed SC materials whose chemical formula is included in the training set.
$T_{c}^{\rm exp}$ is the experimental value of $T_{c}^{}$ \jbmd{from} the reference given in the third column.}
\todounseen{complete}
}
\label{tab:expsc}
\begin{ruledtabular}
\begin{tabular}{lll}
Chemical formula & $T_{c}^{\rm exp}$ (K)  & Reference\\
\hline
HgBa$_2$CuO$_4$ & $\simeq 94$ & \cite{Putilin1993} \\
HgSr$_2$CuO$_4$ & $\simeq 78$ \jbmd{(with Mo substitution)} & \cite{Singh1994} \\
La$_2$CuO$_4$ & $\simeq 40$ & \cite{Grant1987}. \\ %
Sr$_2$CuO$_2$F$_2$ & $\simeq 46$ & \cite{AlMamouri1994}\\
Ca$_{2-x}$K$_x$CuO$_2$Cl$_2$ & $\simeq 24$ & \cite{Kim2006} \\
\end{tabular}
\end{ruledtabular}
\end{table}

Second, the \textit{ab initio} values of $|t_1|$ and $u$ obtained within the training set reproduce the diverse distribution of $|t_1|$ and $u$ observed in realistic cuprates.
We have $|t_1| \simeq 0.40-0.55$ eV and $u \simeq 7-10.5$ in Fig.~\ref{fig:chart-t1u}(a).
The range of $u$ corresponds to that observed not only in single-layer cuprates but also in multi-layer cuprates~\cite{Moree2022,Schmid2023,Moree2023Hg1223}.
In particular, the values of $u$ correspond to the range $u \simeq 6.5-10.5$ in which $F_{\rm SC}$ is nonzero \jbmd{(}see~\cite{Schmid2023}, Fig.~10\jbmd{)},
and thus $T_{c}^{\rm opt}$ is nonzero according to Eq.~\eqref{eq:tcopt}.
This suggests the training set is a good platform to study the microscopic \jbmd{mechanism of the materials dependence of} $|t_1|$ and $u$ and thus $T_{c}^{\rm opt}$.

\jbmd{N}ote that for the CFs in the training set, the M space has a similar structure, which facilitates the comparison between the \jbmd{variables in $\mathcal{V}_{3}$}. %
Namely, the number of bands in the M space is $N_{\rm M}=17$ for all compounds in the training set.
\jbmb{(See the band structures in Sec.~S2 of SM~\cite{Moree2024Supplemental_PRX}.)}
In other single-layer compounds such as Bi2201 with $T_{c}^{\rm opt} \simeq 10-40$ K~\cite{Torrance1988,ARAO2005351}, we have $N_{\rm M}=23$ due to the presence of $6$ additional bands from the BiO block layer.
However, we do not consider cuprates with a Bi2201-like CF and crystal structure,
in order to simplify the comparison between compounds.
This is not expected to weaken the generality of the result,
because the diverse distribution in experimental $T_{c}^{\rm opt} \simeq 24-94$ K is already reproduced by the CFs in the training set,
as mentioned above.
Possible extensions of the training set to other cuprates including multi-layer cuprates and Bi2201-like single-layer cuprates are left for future studies.

\section{Details on the HDE procedure}
\label{app:meth-sr}
\renewcommand{\theequation}{C\arabic{equation}}

\subsection{Motivation of the HDE}
\label{app:meth-sr-motiv}

The essence of the HDE was presented in Sec.~\ref{sec:meth-algo}.
Here, as a complement, we discuss in more detail the motivation of the HDE procedure.
\\

The complete elucidation of the dependence of $y$ on $\mathcal{V}=\{x_i\}$ requires several conditions:
(a) Probe the completeness of the dependence of $y$ on $\mathcal{V}$;
(b) Extract the hierarchy in the dependencies of $y$ on $x_i$; 
(c) Capture the nonlinear dependence of $y$ on $x_i$;
(d) Obtain an explicit expression of $y$ as a function of $x_i$.
In addition, it is desirable to (e) keep the procedure as simple as possible\jbmd{, and (f) preserve the sparsity in the approached expression of $y$ (namely, the approached expression of $y$ should depend on as few $x_i$ as possible).}

On (a), probing the completeness of the dependence of $y$ on $\mathcal{V}$ 
allows to clarify whether or not it is possible to construct a perfect descriptor for $y$ as a function of $x_i$\jbmb{. This}
provides useful information prior to (b,c).

On (b), clarifying the hierarchy in the dependencies of $y$ on $x_i$ 
is necessary to pinpoint the MOD of $y$,
namely, the variables $x_i$ on which $y$ depends the most.
This allows to construct an approximation of $y$ as a function of these $x_i$
by combining (b) and (d), which consists in $y_{{\rm MOD}g}$ \jbmd{which is defined and discussed} in the main text.
The MOD$g$ of $y$ contains the principal microscopic mechanism of the dependence of $y$,
and thus, provides useful clues in the context of materials design.

On (c), $y$ has a nonlinear dependence on $x_i$ in the general case. 
For instance, in previous works on cuprates~\cite{Moree2022,Moree2023Hg1223},
it has been shown that the AB LEH \jbmd{parameters have} a nonlinear dependence on band structure variables and crystal structure \jbmd{variables.}

On (d), the explicit expression of $y$ as a function of $x_i$ 
is necessary to understand whether $y$ increases or decreases with increasing $x_i$.
[Such information is not captured by (b) alone.]
As mentioned above, the combination of (b) and (d) allows to obtain $y_{{\rm MOD}g}$.

\jbmd{On (f), the sparsity allows to facilitate the physical interpretation of the approached expression of $y$.
Indeed, if an accurate approximation of $y$ can be constructed from only a few $x_i$, the distinct contributions of the $x_i$ to the dependence of $y$ may be analyzed and discussed more easily.
}
\\

Already existing procedures
such as 
linear regression,
polynomial regression,
and also symbolic regression
and sparse regression
fulfill part of the above conditions (a-e),
but not all of them.
\jbmd{This is discussed below.}

\jbmd{Linear regression fulfills (d) and (e), but not other points.
Although it is the simplest approach (e) to obtain an explicit expression of $y$ as a function of the $x_i$ (d),
the dependence of $y$ on $x_i$ is nonlinear in the general case as discussed before,
so that (c) is not fulfilled.
}

Polynomial regression \jbmd{and symbolic regression fulfill (c) in addition to (d) and (e),} but not other points.
\jbmd{Polynomial regression} has good approximation properties provided that $y=h(\{x_i\})$ is a continuous function of the $x_i$:
In that case, $y$ can be approached uniformly by a polynomial of the $x_i$ according to the Stone-Weierstrass theorem. \todounseen{~\cite{?}}
\jbmd{Symbolic regression allows to go beyond the polynomial regression without introducing assumptions on the expression of $y$ as a function of the $x_i$.}
However, \jbmd{these techniques do not allow to fulfill (a), (b) and (f)} in the general case.

Sparse regression allows to perform polynomial and symbolic regression by fulfilling \jbmd{(f)}.
Regularization techniques \todounseen{\cite{?}} allow to penalize expressions of $y$ that depend on many $x_i$,
allowing to construct an expression of $y$ as a function of a reduced number of $x_i$ (f).
However, \jbmd{they do not offer a clear framework to fulfill (a) or (b).}
\\

The HDE is designed to fulfill all conditions (a-f).
First, the HDE allows to probe the completeness of the dependence (a)
by examining the value of $f_{\infty}[y,\mathcal{V}]$ in Eq.~\eqref{eq:finfty}.
Second, the HDE allows to extract the hierarchy in dependencies of $y$ on $x_i$ (b)
thanks to the recurrent expression of the candidate descriptor \jbmd{$x_{(g)}$} in Eq.~\eqref{eq:editer}.
In the HDE${[}y,\mathcal{V}{]}$, 
when we increment $g$,
we successively add terms to $x^{\rm opt}$ in Eq.~\eqref{eq:xN}
that correspond to the higher-order dependencies of $y$ on $\{x_i\}$
(the order increases with $g$).
The competition between dependencies can also be analyzed by calculating the score of each variable in Eq.~\eqref{eq:score}.
Third, the wildcard operator in Eq.~\eqref{eq:op} encompasses polynomials of variables,
so that the nonlinear dependence of $y$ may be captured at an acceptable level of accuracy (c).
(Still, note that its approximation properties are not perfect, as discussed below.)
Fourth, the HDE allows to obtain an explicit expression of $y$ as a function of $x_i$ (d).
Fifth, the HDE procedure is simple and deterministic (e), %
and can be applied even if the size of the training set is relatively small (we should have $N_{t} \geq 3$).
\jbmd{Sixth, the HDE procedure allows to preserve the sparsity
by introducing no more than one variable $x_i$ in Eq.~\eqref{eq:xN} when $g$ is incremented.}

Note that, in the HDE[$y,\mathcal{V}$], the ratio 
\begin{equation}
\Delta f_{(g)} = \frac{f_{(g+1)}[y,\mathcal{V}]-f_{(g)}[y,\mathcal{V}]}{f_{(g)}[y,\mathcal{V}]}
\end{equation}
usually decreases with increasing $g$,
but the amplitude of the ratio
\begin{equation}
\Delta x^{\rm opt}_{(g)} = \frac{x^{\rm opt}_{(g+1)}-x^{\rm opt}_{(g)}}{x^{\rm opt}_{(g)}} = \zeta^{\rm opt}_{g+1} \frac{[x_{i^{\rm opt}_{g+1}}]^{\alpha^{\rm opt}_{g+1}}}{[x^{\rm opt}_{(g)}]^{\beta^{\rm opt}_{g+1}}}
\end{equation}
does not necessarily decrease with increasing $g$.
Namely, when incrementing $g$, the effect on the \textit{dependence} of $y$ on $x_i$ may be corrective
(i.e. $f_{(g)}[y,\mathcal{V}]$ increases by a small amount so that $\Delta f_{(g)}$ is small),
but the effect on the \textit{amplitude} of $x^{\rm opt}$ may be significant:
The amplitude of the additional term (which is encoded in $|\Delta x^{\rm opt}_{(g)}|$) is not necessarily small.
\\

Note that the approximation properties of the HDE are not perfect
due to the compromise between approximability and simplicity in the present framework.
Namely, we enforce sparsity by allowing only one variable \jbmd{$x_i$ in $\mathcal{V}$} to be introduced in $x^{\rm opt}_{(g)}$ when $g$ is incremented, %
and the expression of $\Delta x^{\rm opt}_{(g)}$ is kept as simple as possible to facilitate its interpretation.
On one hand, this choice does not allow to represent all polynomials of $x_i$,
which requires to introduce several variables at once in $x^{\rm opt}_{(g)}$ when $g$ is incremented.
On the other hand, this allows to obtain the hierarchy between the dependencies of $y$ on $x_i$ in a simple manner.
In the scope of this paper, the consistency between results provided by the HDE and previous works 
suggests the HDE in its present form is reliable.
Possible extensions of the HDE that involve more sophisticated expressions of $\Delta x^{\rm opt}_{(g)}$ \jbmb{are} left for future studies.
\\

\jbmd{
Finally, on (f), it should be noted that sparsity is distinct from dimensionality reduction.
Dimensionality reduction techniques such as principal component analysis (PCA)~\cite{Pearson1901}
have been used in e.g. recent studies on iron-based SC materials~\cite{Ido2023}.
The PCA allows to reduce the size of the variable space $\mathcal{V} = \{ x_i, i=1..N_{\mathcal{V}} \}$,
by extracting principal components that are linear combinations of the $x_i$.
(Details can be found in e.g. Ref.~\cite{Bishop2006}.)
The $m^{\rm th}$ principal component is denoted as
\begin{equation}
z^{(m)}[j] = \sum_{i=1}^{N_{\mathcal{V}}} v^{(m)}_{i} x_i[j].
\label{eq:pca}
\end{equation}
The principal components are sufficient to reproduce the diverse distribution of values of $x_i$ in the training set,
so that an approached expression of $y$ may be constructed as a function of a few $z^{(m)}$.
However, in Eq.~\eqref{eq:pca}, the weights $v^{(m)}_{i}$ may be nonzero for many variables $x_i$ in the general case.
Thus, if we try to express $y$ as a function of $\{ z^{(m)}, m \}$ instead of $\{ x_i, i=1..N_{\mathcal{V}} \}$,
the expression of $y$ will depend on a few $z^{(m)}$ but may depend on many $x_i$, so that the sparsity is not preserved.
To interpret the expression of $y$, we discuss the dependence of $y$ on the $x_i$ rather than on the $z^{(m)}$,
because the variables $x_i$ have a physical meaning (see Sec.~\ref{sec:gMACE+HDE}).
}

\subsection{Comments on the choice of fitness function}
\label{app:meth-sr-fit}

Here, we discuss in further detail the choice of the fitness function in Eq.~\eqref{eq:deffit}.
\jbmb{Namely, we remind the definition of the Pearson correlation coefficient that is used in Eq.~\eqref{eq:deffit}, and we discuss the invariance property of the fitness function [Eq.~\eqref{eq:finv}].}
\\

\paragraph*{Pearson correlation coefficient ---}
The Pearson correlation coefficient $\rho(y,x)$ between two variables $y$ and $x$ is defined as follows.
The variables $y$ and $x$ are represented by two data samples $\{ y[j], \ j=1..N_t \}$ and $ \{ x[j], \ j=1..N_t \}$, where $N_t$ is the size of the training set.
The sample mean of $x$, sample covariance of $y$ and $x$ and sample variance of $x$ are defined as, respectively:
\begin{align}
m(x)_{}   & = \frac{1}{N_t}\sum_{j=1}^{N_t} x[j],\\
c(y,x)  & = \frac{1}{N_t \jbmd{-1}} \sum_{j=1}^{N_t} \Big[ y[j] - m(y) \Big] \Big[ x[j] - m(x) \Big], \label{eq:cov}\\
v(x)_{} & = c(x,x),
\end{align}
and we calculate $\rho(y,x)$ as
\begin{equation}
\rho(y,x) = \frac{c(y,x)}{\sqrt{v(y)} \sqrt{v(x)}}.
\label{eq:pearson}
\end{equation}
The value of $\rho(y,x)$ is a real number between $-1$ and $+1$.
If $\rho(y,x)=1$ [$\rho(y,x)=-1$], then $x$ and $y$ are perfectly correlated (anticorrelated),
and there exist $k_0$ and $k_1$ such that the equation $y = k_0 + k_1 x$ is rigorously satisfied,
with $k_1 > 0$ ($k_1 < 0$) if $\rho(y,x)=1$ [$\rho(y,x)=-1$].
\\

\paragraph*{Invariance property of the fitness function ---}
The invariance property [Eq.~\eqref{eq:finv}] of the fitness function %
comes from the property 
\begin{equation}
\rho[y,k_0 + k_1x] = {\rm sgn}(k_1) \rho[y,x]
\label{eq:rhoaxb}
\end{equation}
of the Pearson correlation coefficient. 
[In Eq.~\eqref{eq:rhoaxb}, we have $k_1 \neq 0$, ${\rm sgn}(k_1)=1$ if $k_1 >0$ and ${\rm sgn}(k_1)=-1$ if $k_1<0$].

The invariance property simplifies the search of the best candidate descriptor:
The fitness function [Eq.~\eqref{eq:deffit}] encodes only the linear dependence of $y$ on $x({\bf p})$, which is sufficient to judge how well $y$ is described by $x({\bf p})$.\footnote{Note that, as mentioned before, $x({\bf p})$ depends nonlinearly on $x_i$, so that $x({\bf p})$ allows to represent the nonlinear dependence of $y$ on $x_i$.}
From the viewpoint of the fitness function, 
the candidate descriptor $x({\bf p})$ is equivalent to any other candidate descriptor in the equivalence class
\begin{equation}
\mathcal{E}[x({\bf p})] = \{ k_0 + k_1 x({\bf p}), (k_1 \neq 0,k_0)\},
\label{eq:eqclass}
\end{equation}
and we do not need to distinguish between elements of $\mathcal{E}[x({\bf p})]$ during the optimization in Eq.~\eqref{eq:ftilde}.
The values of $k_0$ and $k_1$ are determined after $x^{\rm opt}$ has been determined, 
by performing the affine regression in Eq.~\eqref{eq:ff}.

In addition, the value of the fitness function has a rather universal meaning \jbmd{(at least at a fixed value of $N_t$)}, 
which facilitates the judgment of the quality of the candidate descriptor $x({\bf p})$.
This is a consequence of the invariance property [Eq.~\eqref{eq:finv}].
For a given $y$ and $x({\bf p})$,
the value of $f[y,x({\bf p})]$ allows to deduce how well $y$ is described by $x({\bf p})$,
irrespective of the scale or order of magnitude of $y$ and $x({\bf p})$.
The description is perfect for $f[y,x({\bf p})] = 1.0$.
Empirically (but rather universally \jbmd{at least at $N_t=36$}), we observe in Sec.~\ref{sec:results} that the description is almost perfect for $f[y,x({\bf p})] \gtrsim 0.98$.
[In this paper, we assume that $x({\bf p})$ describes $y$ entirely if $f_{\infty}[y,\mathcal{V}] \gtrsim 0.98$.]
If $0.95 \lesssim f[y,x({\bf p})] \lesssim 0.97$, the description is good, but corrective higher-order dependencies of $y$ on $x_i$ may be missing.
If $f[y,x({\bf p})] \simeq 0.6-0.9$, the description is rough or very rough, but \jbmd{$x({\bf p})$} captures correctly the MOD.
Typically, for the MOD2s discussed in Sec.~\ref{sec:overview} and Sec.~\ref{sec:results},
the values of $f_{(2)}[y,x^{\rm opt}_{\jbmd{(2)}}]$ are above $0.6$,
and most of the time above $0.8-0.9$.
There are also particular cases in which $f_{(2)}[y,x^{\rm opt}_{\jbmd{(2)}}] \gtrsim 0.98$, so that the MOD2 is sufficient to describe $y$ entirely.

\subsection{Comments on the wildcard operator}
\label{app:meth-sr-gen}

\begin{figure}
\includegraphics[scale=0.53]{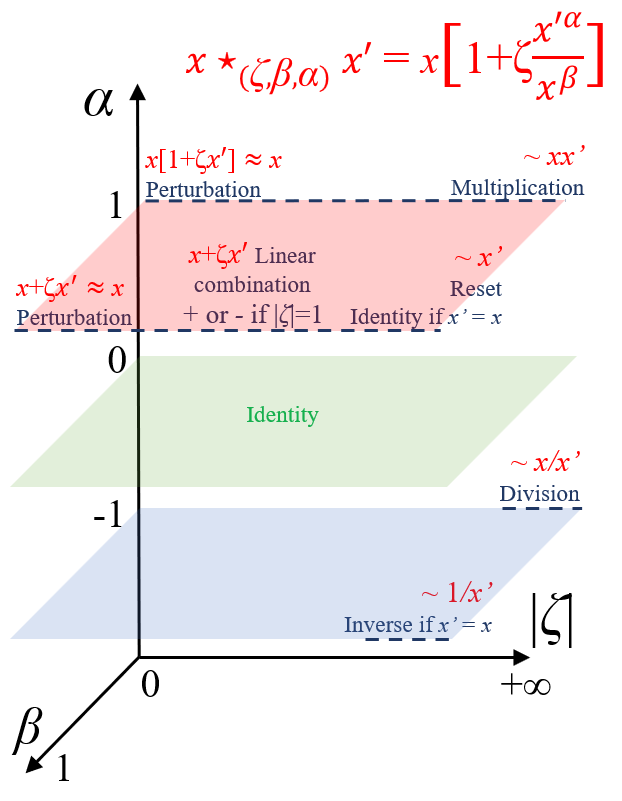}
\caption{\mycaption{Character of the wildcard operator $\star_{(\zeta,\beta,\alpha)}$ depending on the values of ($\zeta,\beta,\alpha$),
when assuming the fitness function in Eq.~\eqref{eq:deffit}.
In the general case, we have $x \star_{(\zeta,\beta,\alpha)} x' = x[1 + \zeta x^{' \alpha}/x^{\beta}]$.
\jbmb{The red, green and blue planes correspond to $\alpha=1$, $\alpha=0$ and $\alpha=-1$, respectively.
In Appendix~\ref{app:meth-sr-gen}, we comment the character that is acquired by the wildcard operator 
for the values of $(\zeta,\beta,\alpha)$ that correspond to the blue dashed lines.
(For these values, the character is shown explicitly in the figure.)}
At $|\zeta| \rightarrow \infty$, we have $x \star_{(\zeta,\beta,\alpha)} x' \simeq \zeta x^{' \alpha} x^{1-\beta}$, and the factor $\zeta$ can be traced out by using the invariance property of the fitness function [Eq.~\eqref{eq:finv}].
}}
\label{fig:starop}
\end{figure}

Here, we detail the properties of the wildcard operator $\star_{(\zeta,\beta,\alpha)}$ [Eq.~\eqref{eq:op}].
\jbmb{These properties are illustrated in Fig.~\ref{fig:starop}.}
\\

(i) The wildcard operator can represent basic algebraic operations.
First, multiplication and division are accounted for by:
\begin{equation}
x \star_{(\zeta \rightarrow +\infty,\beta=0,\alpha=\pm 1)} x' \propto xx'^{\pm 1}.
\label{eq:op1}
\end{equation}
Indeed, for $\zeta \rightarrow +\infty$, we have
\begin{equation}
x \star_{(\zeta,\beta=0,\alpha)} x' = x [1 + \zeta x'^{\alpha}] \sim_{\zeta \rightarrow +\infty}  \zeta  x x'^{\alpha} \propto x x'^{\alpha},
\label{eq:mult}
\end{equation}
in which the arbitrarily large yet finite factor $\zeta$ is traced out by using the invariance property of the fitness function.
Furthermore, Eq.~\eqref{eq:mult} encompasses products and ratios between $x$ and exponents of $x'$.
Second, addition and subtraction are accounted for by:
\begin{equation}
x \star_{(\zeta,\beta=1,\alpha=1)} x' = x + \zeta x'
\label{eq:op2}
\end{equation}
for $\zeta = \pm 1$.
Furthermore, Eq.~\eqref{eq:op2} encompasses any linear combination of $x$ and $x'$ if $|\zeta| \neq 1$.
Third, exponents and polynomials of any variable $x$ are accounted for by:
\begin{eqnarray}
x \star_{(\zeta,\beta = 1,\alpha)} x & = & x + \zeta x^{\alpha}, \label{eq:op3} \\
x \star_{(\zeta \rightarrow +\infty, \beta =1,\alpha)} x & = & x^{\alpha + 1}. \label{eq:op4}
\end{eqnarray}

(ii) The wildcard operator can represent the reset and identity operators.
The reset operator is defined as
\begin{equation}
x \star_{(\zeta \rightarrow +\infty,\beta=1,\alpha=1)} x' \propto x'
\label{eq:opreset}
\end{equation}
and allows to replace $x$ by $x'$. %
If $x'=x$, Eq.~\eqref{eq:opreset} becomes the identity operator.
[The possible values of $(\zeta^{\rm opt},\beta^{\rm opt},\alpha^{\rm opt})$ which represent the identity operator
are $(\zeta \rightarrow +\infty,\beta=1,\alpha=1)$ and also $(\zeta >0,\beta,\alpha=0)$.]
The fact that the wildcard operator is able to mimic the identity operator 
guarantees that the best candidate descriptor from the generation $g$ is included in the candidate descriptor space for the generation $g+1$.
In particular, this implies Eq.~\eqref{eq:fincreases},
which is a desired property of the fitness function:
Given the best candidate descriptor at $g$, 
the iteration from $g$ to $g+1$ adds the $g+1^{\rm th}$ order dependence, 
and taking into account the $g+1^{\rm th}$ order dependence necessarily improves the descriptor.

(iii) The wildcard operator can represent a perturbation of $x$ by $x'$.
For small $|\zeta|$ \jbmd{and $\alpha=1$},
we have
\begin{equation}
\Big|\zeta \frac{x'^{}} {x^{\beta}}\Big| \ll 1
\end{equation}
in Eq.~\eqref{eq:op}, so that $x \star_{(\zeta,\beta,\alpha=1)} x'$ yields $x$ corrected by a small term that depends on $x'$.

\subsection{Computational details of the HDE procedure}
\label{app:meth-sr-cd}

Here, we give computational details of the $\mathcal{P}{[}y,\mathcal{V}{]}$ procedure
that is employed to obtain $x^{\rm opt}_{(N)}$ in Eq.~\eqref{eq:xN}.
\jbmb{First}, we use a finite number $N$ of generations.
In this paper, we use $N=15$ in Sec.~\ref{sec:results-R}, 
and we check that the fitness function varies slowly \jbmd{with increasing $N$ at $N \simeq 15$},
so that it is reasonable to stop at $N=15$.

Second, at fixed $g$, we determine ${\bf p}^{\rm opt}$ that maximizes $f[y,x({\bf p})]$ in Eq.~\eqref{eq:ftilde} as follows.
We scan $\mathcal{C}_{\mathcal{V}}^{}$ by computing $f[y,x({\bf p})]$ for each $x({\bf p})$ in $\mathcal{C}_{\mathcal{V}}^{}$.
Note that this is computationally expensive as discussed below;
however, this allows to avoid falling into any local extremum of $f[y,x(\jbmb{{\bf p}})]$.
Another possibility to reduce the computational cost would be to 
prepare an initial guess for $\jbmb{{\bf p}}$
and then optimize $\jbmb{{\bf p}}$ by using e.g. a gradient descent algorithm.
Such extensions and the detailed study of the dependence of the result on the initial guess for $\jbmb{{\bf p}}$
are left for future studies.

At fixed $g$, the variational parameters are $\jbmd{{\bf p}_{g}} = (\alpha_{\jbm{g}},\beta_{\jbm{g}},\zeta_{\jbm{g}},i_{\jbm{g}})$ as mentioned in Sec.~\ref{sec:meth-algo}.
The number of indices $i_{\jbm{g}}$ is given by the total number of variables $N_{\mathcal{V}}$ in $\mathcal{V}= \{ x_i \}$.
As for $\beta$, we consider only $N_{\beta} = 2$ values, which are $\beta=0,1$;
this choice preserves the polyvalence of the wildcard operator as illustrated in Fig.~\ref{fig:starop}.
As for $\alpha$ and $\zeta$, we use a finite grid of values for the optimization of the fitness function.
The numbers of values of $\alpha$ and $\zeta$ in the grid are denoted as $N_{\alpha}$ and $N_{\zeta}$, respectively.
Below, we detail the choice of this grid.

The computational cost increases rapidly with $N_{\alpha}$ and $N_{\zeta}$:
At fixed $g$, the number of candidate descriptors to be evaluated is 
$N_{\mathcal{V}} N_{\alpha}$ if $g=1$ and $N_{\mathcal{V}} N_{\alpha} N_{\beta}N_{\zeta}$ if $g \geq 2$.
In practice, we reduce the computational cost by employing a multi-step optimization, as detailed in (i,ii) below.

(i) First, we optimize $(\alpha_{\jbm{g}},\zeta_{\jbm{g}})$ on a coarse grid 
together with $(\beta_{\jbm{g}},i_{\jbm{g}})$.
The coarse grid is the following:
\begin{align}
\alpha_{\jbm{g}}^{(1)} & = n \ \ \ \ \ \ \ \ \ \ n=-9...9, \label{eq:gridalpha1}\\
\zeta_{\jbm{g}}^{(1)}  & = m 10^{m'} \ \ \ m,m'=-9...9 \ (m \neq 0).
\end{align}
The optimized values are denoted as $(\alpha_{\jbm{g}}^{\rm opt(1)},\beta_{\jbm{g}}^{\rm opt},\zeta_{\jbm{g}}^{\rm opt(1)},i_{\jbm{g}}^{\rm opt})$.

(ii) Then, we keep the values of $\beta_{\jbm{g}}^{\rm opt}$ and $i_{\jbm{g}}^{\rm opt}$ that were determined in (i),
and we refine the optimization of $(\alpha_{\jbm{g}},\zeta_{\jbm{g}})$ on a finer grid, %
which is constructed iteratively as follows.
We introduce the iteration index $j$, starting from $j=2$.
The fine grid is built around $\alpha_{\jbm{g}}^{{\rm opt}(j-1)},\zeta_{\jbm{g}}^{{\rm opt}(j-1)}$:
\begin{align}
\alpha_{\jbm{g}}^{(j)} & = \alpha_{\jbm{g}}^{{\rm opt}(j-1)} [ 1 + 0.1n] \ \ \ n=-9...9, \label{eq:gridalpha2} \\
\zeta_{\jbm{g}}^{(j)}  & = \zeta_{\jbm{g}}^{{\rm opt}(j-1)} [ 1 + 0.1m] \ \ \ m=-9...9.
\end{align}
We optimize the fitness function to obtain $\alpha_{\jbm{g}}^{{\rm opt}(j)},\zeta_{\jbm{g}}^{{\rm opt}(j)}$.
We iterate up to $j=3$.

Note that the grid includes values of $|\alpha|$ from $|\alpha|_{\rm min}=0.01$ to $|\alpha|_{\rm max}=9.99$
and values of $|\zeta|$ from $|\zeta|_{\rm min}=0.01 \jbmb{\times} 10^{-9}$ to $|\zeta|_{\rm max}=9.99 \jbmb{\times} 10^{9}$
if we iterate up to $j=3$.
Also, note that the score [Eq.~\eqref{eq:score}] is calculated on the coarse grid, because $i_{\jbm{g}}$ is optimized on the coarse grid.

\subsection{Numerical aspects and pitfalls}
\label{app:meth-sr-num}

Here, we discuss a few limitations and numerical pitfalls in the HDE procedure.
\jbmb{First}, note that if $\beta \jbmd{>} 0$, the values of $x$ [or $x^{\rm opt}_{(g-1)}$ in Eq.~\eqref{eq:editer}]
must be nonzero when calculating $x \star_{(\zeta,\beta,\alpha)} x' = x[1 + \zeta x'^{\alpha}/x^{\beta}]$.
Namely, if the variable $x$ is represented by the sample $ \{ x[j], \ j=1..N_t \}$ introduced in Appendix~\ref{app:meth-sr-fit},
we should have $x[j] \neq 0$ for all $j$.
This implies $x_i[j] \neq 0$ for all $x_i$ in $\mathcal{V}$,
because we have $x^{\rm opt}_{(1)} = x_{i_1}^{\alpha_1}$:
If $x_{i_1}$ has zero values, then $x^{\rm opt}_{(1)}$ will have zero values if $\alpha_1 > 0$,
or $x^{\rm opt}_{(1)}$ will diverge if $\alpha_1 < 0$.
In our calculations, some of the variables $x_i$ in $\mathcal{V}$ have zero values:
For instance, in $\mathcal{V}_1$, the variable $R_{\rm A'}$ has zero values if A'$=\varnothing$.
Thus, we introduce an infinitesimal offset $\delta_{\rm off} \jbmd{> 0}$ in all variables.
(Namely, we replace $x_i[j]$ by $x_i[j] + \delta_{\rm off}$.)
We choose \jbmd{the value} $\delta_{\rm off} = \jbmd{10^{-4}}$ \jbmd{which} is small enough to be negligible with respect to the difference between values of $x_i[j]$,
but large enough so that $|\zeta|_{\rm max} \delta_{\rm off}  \gg 1$ and $|\zeta|_{\rm min} / \delta_{\rm off}  \ll 1$.
(The condition $|\zeta|_{\rm max}\delta_{\rm off} \gg 1$ is necessary \jbmd{to have} $1 + \zeta x^{\alpha} \sim \zeta x^{\alpha}$ when $|\zeta| \rightarrow |\zeta|_{\rm max}$,
so that the character of the wildcard operator at $|\zeta| \rightarrow + \infty$ shown in Fig.~\ref{fig:starop} is valid.)
\\

Second, in some particular cases, 
there may be no global maximum in the $\alpha$ dependence of \jbmd{$f_{}[y,x_{(g)}]$, e.g.} \jbmb{$f_{}[y,x_{i^{\rm opt}_{1}}^{\alpha}]$ at $g=1$}.
\todounseen{There IS a theoretical maximal value when $\alpha \rightarrow +\infty$, but this value is never reached for finite $\alpha$ (it is approached asymptotically).}
Namely, the value of $|\alpha|$ becomes arbitrary large when attempting to maximize $f[y,x^{\alpha}]$.
\jbmb{This is discussed in Sec.~S3 of SM~\cite{Moree2024Supplemental_PRX}.}
To avoid the divergence of $|\alpha|$, we consider a maximal value $|\alpha|_{\rm max}$ for $|\alpha|$ in practice,
and if $f[y,x^{\alpha}]$ is maximal for $\alpha^{\rm opt}$ such that $|\alpha^{\rm opt}|=|\alpha|_{\rm max}$,
\jbmd{then} we assume $|\alpha^{\rm opt}|=|\alpha|_{\rm max}$.
In this paper, we have $|\alpha|_{\rm max}=9.99$ by employing the grids in Eqs.~\eqref{eq:gridalpha1} and~\eqref{eq:gridalpha2} up to $j=3$.
Note that $|\alpha|=|\alpha|_{\rm max}$ is sometimes reached in the MODs discussed in this paper
[see e.g. the dependence of $|t_1|$ on $R_{\rm X}$ in Eq.~\eqref{eq:intro-t1}].
\\

Third, the maximal value of $f_{}[y,x_{\jbmd{(g)}}]$ may be obtained for several candidate descriptors \jbmd{with the same $i_{g}$ but different $\alpha_{g}$}.
\jbmd{These} are in the same equivalence class [Eq.~\eqref{eq:eqclass}], but this is nontrivial.
\jbmb{This is discussed in Sec.~S4 of SM~\cite{Moree2024Supplemental_PRX}.}
In this case, we choose to apply the following convention:
We choose $\alpha$ such that $|\alpha|$ is minimal and $\alpha >0$.
(Then, if $g=2$ and if there are several values of $\zeta$ for which the fitness function has the same value, we choose $\zeta$ such that $|\zeta|$ is minimal and $\zeta >0$.)
In practice, if we use this convention, the optimization of $\alpha$ on the coarse grid then on the fine grid yields $\alpha = |\alpha|_{\rm min} = 0.01$.

A concrete example is $|Z_{\rm X}|$, whose value is either $1$ and $2$.
In this case, $f[R,|Z_{\rm X}|^{\alpha}] = 0.503$ for all values of $\alpha \neq 0$.
This is why we have e.g. $\alpha_1 = 0.01$ in Eq.~\eqref{eq:intro-R}.
Note that this choice should not affect the physical interpretation of the MOD of $R$ on $|Z_{\rm X}|$ in Eq.~\eqref{eq:intro-R}.
The MOD1 of $R$ on $\mathcal{V}_{1}$ is
\begin{equation}
R =4.22 - 3.91 |Z_{\rm X}|^{0.01}, 
\end{equation}
and if we choose another convention,
e.g. we choose $\alpha$ such that $|\alpha|$ is minimal and $\alpha < 0$,
then we obtain
\begin{equation}
R = -3.63 + 3.94 |Z_{\rm X}|^{-0.01},
\end{equation}
so that the physical interpretation of the MOD1 ($R$ increases when $|Z_{\rm X}|$ decreases) remains the same irrespective of the selected convention.
The relatively large values of $|k_0|$ and $|k_1|$ in the above MOD1
and in the MOD2 [Eq.~\eqref{eq:intro-R}]
with respect to the \textit{ab initio} values of $R \simeq 0.22-0.34$
are a consequence of the small value of $|\alpha|$. 
Still, the range of \textit{ab initio} values of $R$ is correctly reproduced by the MOD2, as seen in Fig.~\ref{fig:summary}.
\\

Finally, note that
the complexity of the expression $y_{N}$ of $y$ in Eq.~\eqref{eq:ff}
increases with increasing $g$.
However, incrementing $g$ up to a large value of $N$ allows to probe the completeness of the dependence of $y$ on $\mathcal{V}$,
as discussed in Sec.~\ref{sec:meth-algo}.
The principal microscopic mechanism of the dependence of $y$ on $\mathcal{V}$
is usually contained in the MOD$g$ for $g \lesssim 3$.

\section{Details on the gMACE procedure}
\label{app:meth-mace}
\renewcommand{\theequation}{D\arabic{equation}}

\subsection{Computational details of the gMACE procedure}
\label{app:meth-mace-cd}

Here, we give computational details of the \textit{ab initio} calculations in the gMACE procedure.
We also detail \jbm{the procedure} that is employed to construct the AB MLWO.
\\

\paragraph{Structural optimization and DFT calculation ---}
We use \texttt{Quantum ESPRESSO} \cite{QE-2009,QE-2017}
and optimized norm-conserving Vanderbilt pseudopotentials~\cite{Hamann2013,Schlipf2015} with the GGA-PBE functional~\cite{Perdew1996}.
To model hole doping, 
\jbmd{we use the virtual crystal approximation~\cite{nordheim1931electron} as done in~\cite{Moree2022,Moree2023Hg1223} and as mentioned in the main text:
The pseudopotential of A or A' cation is interpolated with that of the chemical element whose atomic number is that of A or A' minus one.}
We use a planewave cutoff of 100 Ry for the wavefunctions.
The full Brillouin zone is sampled by using a $k$-point grid of size
$8 \times 8 \times 8$ in the structural optimization, 
$12 \times 12 \times 12 $ in the self-consistent DFT calculation,
and $6 \times 6 \times 6$ in the non-self-consistent DFT calculation ($8 \times 8 \times 4$ if A'$=$Hg$_{1-\delta}$Au$_{\delta}$).
We use a Fermi-Dirac smearing of 0.002 Ry.
\\

\paragraph{Crystal parameters and symmetry ---}
The primitive vectors of the Bravais lattice 
and the positions of atoms are given in Table~\ref{tab:atompos}.
During the structural optimization, we optimize the values of $a$, $c$, $d^{z}_{A}$ and $d^{z}_{X}$ altogether.
Then, we use the optimized values in the self-consistent and non-self-consistent DFT calculations.
 
For completeness, note that the high-symmetry structure that is considered in Table~\ref{tab:atompos}
may be lowered in experiment.
Namely, atoms may undergo displacements around their ideal positions listed in Table~\ref{tab:atompos},
which creates incommensurate modulations in the crystal structure.
The origin of these displacements is the nonideal Goldschmidt tolerance factor of the perovskite-like structure of the cuprate
(i.e. the ratio between the radii of atoms in the crystal is not ideal),
which causes a geometric mismatch between the block layer and the CuO$_2$ plane.
This happens e.g. in the case of Bi2201~\cite{Petricek1990,Shamray2009}.
Also, in the case of Sr$_2$CuO$_2$F$_2$, the F atoms are distorted,
and the doping may introduce excess F at different positions~\cite{AlMamouri1994}.

It is possible to account for the structural distortion in a simplified manner,
by considering a distortion that is restricted to the unit cell
and ignores the incommensurate character of the distortion,
as done in~\cite{Moree2022} in the case of Bi2201. (See Appendix~C of~\cite{Moree2022}.)
However, in Bi2201, the effect of the distortion $|t_1|$ and $u$ is small~\cite{Moree2022}:
If we compare the AB LEH obtained from the crystal structure with and without distortion,
 $|t_1|$ does not vary and $U$ varies by no more than $3\%$.
In the present paper, for simplicity, we do not consider the distortion and always assume the ideal atomic positions in Table~\ref{tab:atompos}.
\\

\begin{table}[!ht]
\caption{\mycaption{Primitive vectors \jbmd{${\bf a}$, ${\bf b}$, ${\bf c}$ of the Bravais lattice} 
and atomic positions in Cartesian coordinates.
We \jbmd{consider} $c_\perp=0$ if A' = Hg$_{1-\delta}$Au$_{\delta}$ and $c_\perp=a/2$ if A' = $\varnothing$.
The atomic positions are entirely determined by $a$, $c$, $c_{\perp}$, $d^{z}_{A}$ and $d^{z}_{X}$.
Note that there are \jbmd{two O atoms,} two A atoms and two X atoms in the unit cell.
\jbmd{The first and second O atoms in the unit cell are denoted as O and O', respectively.}
}}
\label{tab:atompos}
\begin{ruledtabular}
\begin{tabular}{llll}
 & $x$ & $y$ & $z$\\
 \hline
${\bf a}$ & $a$ & 0 & 0 \\
${\bf b}$ & 0 & $a$ & 0 \\
${\bf c}$ & $c_\perp$ & $c_\perp$ & $c$ \\ 
 \hline 
Cu & 0 & 0 & 0\\
O & $a/2$ & 0 & 0\\
O' & 0 & $a/2$ & 0\\
A & $a/2$ & $a/2$ & $\pm d^{z}_{A}$\\
X & 0 & 0 & $\pm d^{z}_{X}$\\
A' & 0 & 0 & $c/2$\\
\end{tabular}
\end{ruledtabular}
\end{table}

\paragraph{Wannier orbital ---}
Here, we detail the procedure to construct the AB orbital.
We use the \texttt{RESPACK} code \cite{Nakamura2021,Moree2022}.
The initial guess consists in an atomic Cu$3d_{x^2-y^2}$ orbital centered on the Cu atom in the unit cell.
In the outer window, the spillage functional~\cite{Souza2001} is minimized to obtain the AB subspace and band dispersion,
then the spread functional~\cite{Marzari1997} is minimized to obtain the AB MLWO.
In previous works~\cite{Hirayama2018,Hirayama2019,Moree2022,Moree2023Hg1223},
the outer window that is used to construct the AB orbital
consists in the M space, from which the $N_{B}$ lowest bands are excluded to avoid catching the Cu$3d_{x^2-y^2}$/O$2p_{\sigma}$ bonding (B) character.
However, the characteristics of the AB \jbmd{MLWO} and values of AB LEH parameters depend slightly on $N_{B}$: 
For instance, the AB LEH parameters may vary by a few percent if $N_{B}$ varies by 2 or 3~\cite{Hirayama2018,Moree2022}.
Although this small dependence \jbmd{of the AB MLWO} on $N_{B}$ does not change the physics of the AB LEH,
\jbmd{it} may prevent the very accurate comparison between AB LEHs that are derived from different CFs.

To estimate accurately the materials dependence of the AB LEH, 
we employ a \jbm{modified} procedure to construct the AB orbital,
in which we remove the B character without excluding the B bands from the outer window.
This procedure is based on the \jbmb{antibonding-bonding (ABB)} transformation~\cite{Hirayama2022silverarxiv},
and is described below.
\jbmd{W}e use the whole M space as the outer window, and we consider three initial guesses for the MLWOs:
The Cu$3d_{x^2-y^2}$ atomic orbital centered on Cu,
and the O$2p_{\sigma}$ (O'$2p_{\sigma}$) orbital centered on O (O').
Then, we minimize the spillage functional~\cite{Souza2001} 
to obtain the band dispersion that corresponds to the three MLWOs with Cu$3d_{x^2-y^2}$ and O/O'$2p_{\sigma}$ character.
This band dispersion consists in the partly filled AB band plus the two fully occupied \jbmd{B} bands;
see e.g. Fig.~8 in~\cite{Moree2022} for an illustration.
Then, we \jbmd{discard the B band dispersion and we keep only the AB band, which} contains the AB subspace and has been determined without introducing the dependence on $N_{B}$.
Finally, we obtain the AB MLWO from the AB band
by considering the AB band as the outer window,
reinitializing the Cu$3d_{x^2-y^2}$ initial guess 
and minimizing the spread functional~\cite{Marzari1997}.
(Note that the ABB transformation allows to derive a three-orbital Hamiltonian that includes the AB orbital and two \jbmd{B} orbitals; here, contrary to the ABB transformation, we discard the \jbmd{B} subspace.)
\\

\paragraph{Polarization}
We use the \texttt{RESPACK} code \cite{Nakamura2021,Moree2022}.
The cRPA polarization at zero frequency is expressed as \cite{Nakamura2021}:
\begin{widetext}
\begin{equation}
[\chi_{\rm H}]_{GG'}^{}(q)= -\frac{4}{N_k} \sum_{k} \sum_{n_u}^{\rm empty} \sum_{n_o}^{\rm occupied} (1-T_{n_o k} T_{n_u k+q}) 
\frac{M^{G}_{n_o, n_u}(k+q,k)  [M^{G'}_{n_o, n_u}(k+q,k) ]^{*}}{\Delta_{n_o, n_u}(k,q)-i\eta},
\label{eq:chiou}
\end{equation}
\end{widetext}
with
\begin{equation}
\Delta_{n_o, n_u}(k,q) = \epsilon_{n_u k+q} - \epsilon_{n_o k}
\label{eq:cte}
\end{equation}
and
\begin{equation}
M^{G}_{n_o, n_u}(k+q,k) = \int_{\Omega} dr \psi^{*}_{n_u k +q}(r) e^{i(q+G)r} \psi_{n_o k}(r);
\label{eq:m}
\end{equation} 
in the above equations, $\eta$ is an infinitesimal positive number,
$q$ is a wavevector in the Brillouin zone, $G,G'$ are reciprocal lattice vectors, $nk$ is the Kohn-Sham one-particle state with energy $\epsilon_{nk}$ and wavefunction $\psi_{nk}$, $T_{nk}=1$ if $nk$ belongs to the AB band and $T_{nk}=0$ else. 
The cRPA effective interaction is deduced as
\begin{equation}
W_{\rm H} = \Big(1 - v\chi_{\rm H}\Big)^{-1} v,
\label{eq:wh}
\end{equation}
in which $v$ is the bare Coulomb interaction.
We use a planewave cutoff of 8 Ry for the calculation of the cRPA polarization and dielectric function.

\subsection{Choice of the variables in $\mathcal{V}_{3}$}
\label{app:v3}

\jbmd{As a complement to Sec.~\ref{sec:gMACE+HDE}, t}he choice of the variables in $\mathcal{V}_{3}$ is discussed below.
The DFT band structure is entirely described by the Kohn-Sham energies $\epsilon_{nk}$ and orbitals $\psi_{nk}$.
However, it is too complex to consider the whole set $\{ \epsilon_{nk}, \psi_{nk} \}$ as the variable space $\mathcal{V}_{3}$.
Instead, we consider variables that capture the essential characteristic energies in the band structure.
To choose the variables, we use as a guide the M space.
Prioritizing the M space for the choice of variables is natural, for two reasons.
First, the M space determines the characteristics of the AB Wannier orbital including $|t_1|$ and $v$~\cite{Moree2023Hg1223}, 
because the M space contains the Cu$3d_{x^2-y^2}$-like and O$2p_{\sigma}$-like bands and the AB band.
Second, the M space plays a prominent role on the cRPA screening and thus on $R=U/v$.
Indeed, the screening increases when the charge transfer energies between occupied and empty bands decrease [see Eqs.~\eqref{eq:chiou} and~\eqref{eq:cte}], and the charge transfer energies are the smallest for the occupied M bands.

\subsection{Values of intermediate quantities within the gMACE procedure}
\label{app:int}

\begin{figure}[!htb]
\includegraphics[scale=0.40]{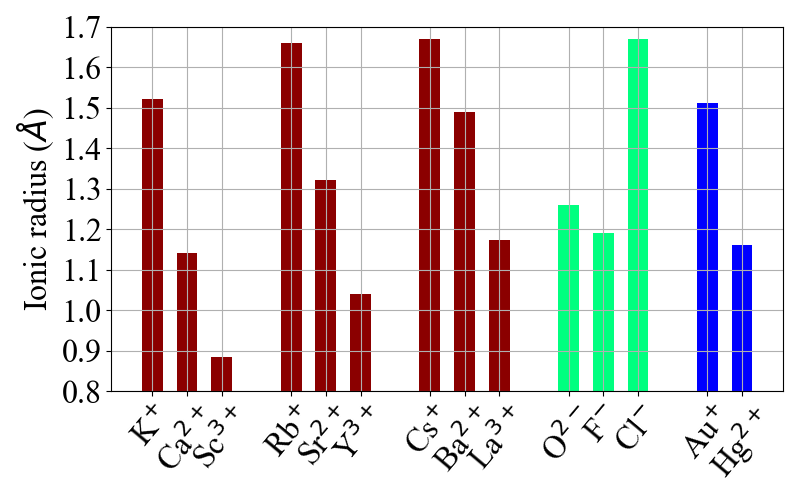}
\caption{\mycaption{Values of ionic radii and charges
that are considered in this paper. 
\jbmb{The \jbmd{values of} ionic charges are $Z_{\rm X} = -2$ or $-1$
and $Z_{\rm A} = +1$, $+2$ or $+3$.
The values of $R_{\rm A}$, $R_{\rm X}$ and $R_{\rm A'}$ are represented in red, green and blue color, respectively.
In the case of the hole-doped compounds, the values of ionic radii and charges are interpolated according to Eqs.~\eqref{eq:rzinterp1} and~\eqref{eq:rzinterp2}.
}
}}
\label{fig:chart-ionic}
\end{figure}

Here, we give the values of intermediate quantities in the variables spaces $\mathcal{V}_{s}$ with $s=1..4$.
\jbmb{On} $\mathcal{V}_1$, 
the values of $R_{\rm A}$, $R_{\rm X}$, $R_{\rm A'}$, $|Z_{\rm X}|$ and $Z_{\rm A}$ that are considered in this paper are represented in Fig.~\ref{fig:chart-ionic}.
\jbmd{(The numerical values are given in Sec.~S1 of SM~\cite{Moree2024Supplemental_PRX}.)}
The values of $R_{\rm A}$, $R_{\rm X}$ and $R_{\rm A'}$ are the crystal ionic radii values from~\cite{Shannon1976}.
(For completeness, we mention that ions are assumed to be $6$-coordinate in the calculation of these values.)
For hole-doped compounds, the partial ion substitution is accounted for as follows.
If A'$=\varnothing$, the ion A becomes ${\rm A}_{2-\delta}\tilde{\rm A}_{\delta}$
where the atomic number of $\tilde{\rm A}$ is that of A minus one;
otherwise, the ion A' is ${\rm Hg}_{1-\delta}{\rm Au}_{\delta}$.
In this case, we interpolate linearly the values of ionic radii;
namely, we consider
\begin{align}
R_{{\rm A}_{2-\delta}\tilde{\rm A}_{\delta}} & = (1-0.5\delta)R_{{\rm A}} + 0.5\delta R_{\tilde{\rm A}}, \label{eq:rzinterp1}\\
R_{{\rm Hg}_{1-\delta}{\rm Au}_{\delta}} & = (1-\delta)R_{{\rm Hg}} + \delta R_{{\rm Au}}. \label{eq:rzinterp2}
\end{align}
\jbmb{The values of \jbmd{$Z_{\rm A}$} are interpolated similarly.}

\begin{figure*}[!htb]
\includegraphics[scale=0.105]{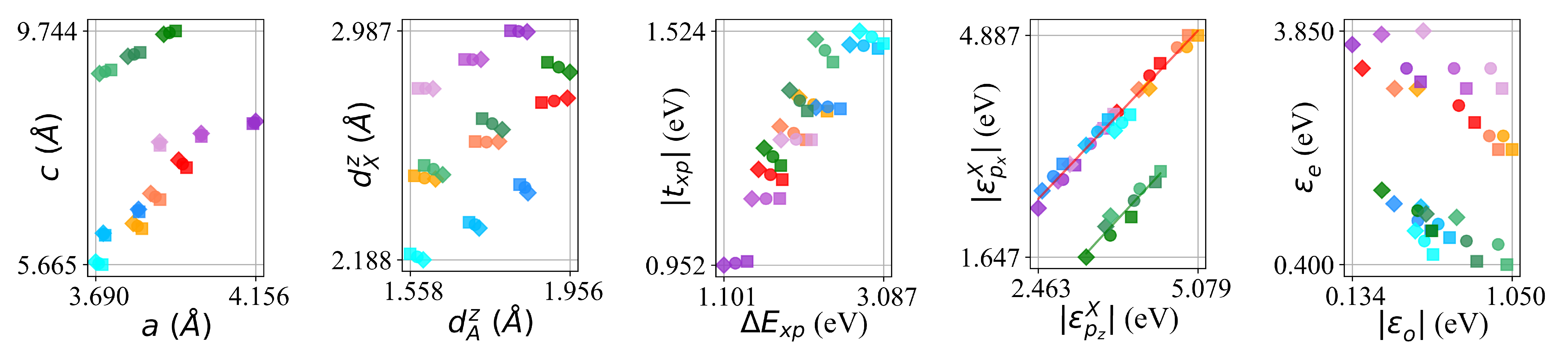}
\caption{\mycaption{Values of crystal parameters $a$, $c$, $d^{z}_{A}$ and $d^{z}_{X}$ obtained by the structural optimization,
and values of $\Delta E_{xp}$ and $|t_{xp}|$, $|\epsilon^{\rm X}_{p_z}|$ and $|\epsilon^{\rm X}_{p_x}|$, and $|\epsilon_{o}|$ and $\epsilon_{e}$ obtained in the MACE calculation.
On $|\epsilon^{\rm X}_{p_z}|$ and $|\epsilon^{\rm X}_{p_x}|$, the red and green solid lines correspond to the affine interpolations in Eq.~\eqref{eq:affeap_1} and Eq.~\eqref{eq:affeap_2},
respectively.
\jbmb{Values of other variables in $\mathcal{V}_{3}$ are given in Sec.~S1 of SM~\cite{Moree2024Supplemental_PRX}.}
\jbmb{The CF that corresponds to each color point is shown in Fig.~\ref{fig:chart-comp}.}
}}
\label{fig:chart-cryst}
\end{figure*}

On $\mathcal{V}_2$, 
the values of crystal parameters $a$, $c$, $d^{z}_{A}$ and $d^{z}_{X}$ obtained by the structural optimization
are represented in Fig.~\ref{fig:chart-cryst}.
(\jbmd{These parameters are} defined in Table~\ref{tab:atompos}).
\jbmd{We consider $c_{\perp}=0$} if A' = Hg$_{1-\delta}$Au$_{\delta}$ and \jbmd{$c_{\perp}=a/2$} if A' = $\varnothing$.

On $\mathcal{V}_3$, the values of variables that are discussed in the main text
are represented in Fig.~\ref{fig:chart-cryst}.
\jbmb{The \jbmd{numerical values of all variables in $\mathcal{V}_{3}$} and the band structures are given in Sec.~S1 and Sec.~S2 of SM~\cite{Moree2024Supplemental_PRX}, respectively.}

\section{Score analysis of the results of the HDE procedure}
\label{app:HDEinterp}
\renewcommand{\theequation}{E\arabic{equation}}

Here, we discuss how the score analysis allows to construct the physical interpretation of the dependence of $y$ on $\mathcal{V}$
based on the results of the HDE[$y,\mathcal{V}$].
Then, as a complement to Sec.~\ref{sec:results},
we \jbmd{detail} the analysis of the score in the items (I), (III), (IV) and (1-10) in Sec.~\ref{sec:results}.
\\

The fitness function $f_{}[y,x^{\rm opt}_{(g)}]$ \jbmd{as defined in Eq.~\eqref{eq:deffit}}
measures the linear correlation between $y$ and $x^{\rm opt}_{(g)}$.
However, the correlation does not always imply a causation, i.e. a physical dependence of $y$ on $x^{\rm opt}_{(g)}$.
Namely, when $y$ and $x^{\rm opt}_{(g)}$ are correlated,
there are two possible scenarios:
(i) Causative: There is a physical dependence of $y$ on $x^{}_{\jbmd{i^{\rm opt}_{g}}}$.
(ii) Noncausative: The correlation between $y$ and $x^{}_{\jbmd{i^{\rm opt}_{g}}}$ is coincidental, and there is no physical dependence of $y$ on $x^{}_{\jbmd{i^{\rm opt}_{g}}}$.

In practice, we need to judge whether the correlation is causative or not, especially in the MOD$g$.
Below, we discuss how to do so at $g=1$.
Two things should be examined.
First, whether \jbmd{there is a competition between} $x_{i^{\rm opt}_{1}}$ \jbmd{and} other variables, 
namely, \jbmd{whether some} variables other than $x_{i^{\rm opt}_{1}}$ have a score that is close to 1.
Second, whether the value of \jbmd{the maximal fitness} \jbmd{$\tilde{f}[y,x_{i^{\rm opt}_{g}}]$} \jbmd{[Eq.~\eqref{eq:mfit1}]} is close to 1 or not.
There are three possible scenarios (S1), (S2) and (S3):
\begin{description}

\item[${\rm (S1)}$ If $x_{i^{\rm opt}_1}$ is not in competition with other variables] 
The dominance of $x_{i^{\rm opt}_1}$ in the dependence is unambiguous,
and we may identify a physical dependence of $y$ on $x_{i^{\rm opt}_1}$.

\item[${\rm (S2)}$ If competition exists and $\tilde{f}{[}y,x_{i^{\rm opt}_1}{]}$ is not close to 1] 
then $y$ has a physical dependence on several variables.
If $y$ depends mainly on two variables $x_i$ and $x_{i'}$,
then each of these variables corresponds to either $x_{i^{\rm opt}_{1}}$ and $x_{i^{\rm opt}_{2}}$,
and $f_{}[y,x^{\rm opt}_{(2)}]$ has a rather high value, typically $\gtrsim 0.9$.

\item[${\rm (S3)}$ If competition exists and $\tilde{f}{[}y,x_{i^{\rm opt}_1}{]}$ is close to 1] 
The physical dependence is hidden in one of the variables
that is highly correlated with $y$,
and for other variables, the correlation with $y$ is coincidental and not physical.
Indeed, if \jbmd{$\tilde{f}{[}y,x_{i^{\rm opt}_1}{]}$} is close to 1,
then $y$ cannot have a strong dependence on more than one variable.
If $y$ has a strong dependence on two or more variables,
then each of these variables is necessary but not sufficient to describe $y$ accurately.
In this case, \jbmd{$\tilde{f}{[}y,x_{i^{\rm opt}_1}{]}$} cannot be close to 1, which contradicts (S3).
(Typically, we obtain \jbmd{$\tilde{f}{[}y,x_{i^{\rm opt}_1}{]} \lesssim 0.7$}.)
\end{description}

\begin{figure*}[!htb]
\includegraphics[scale=0.15]{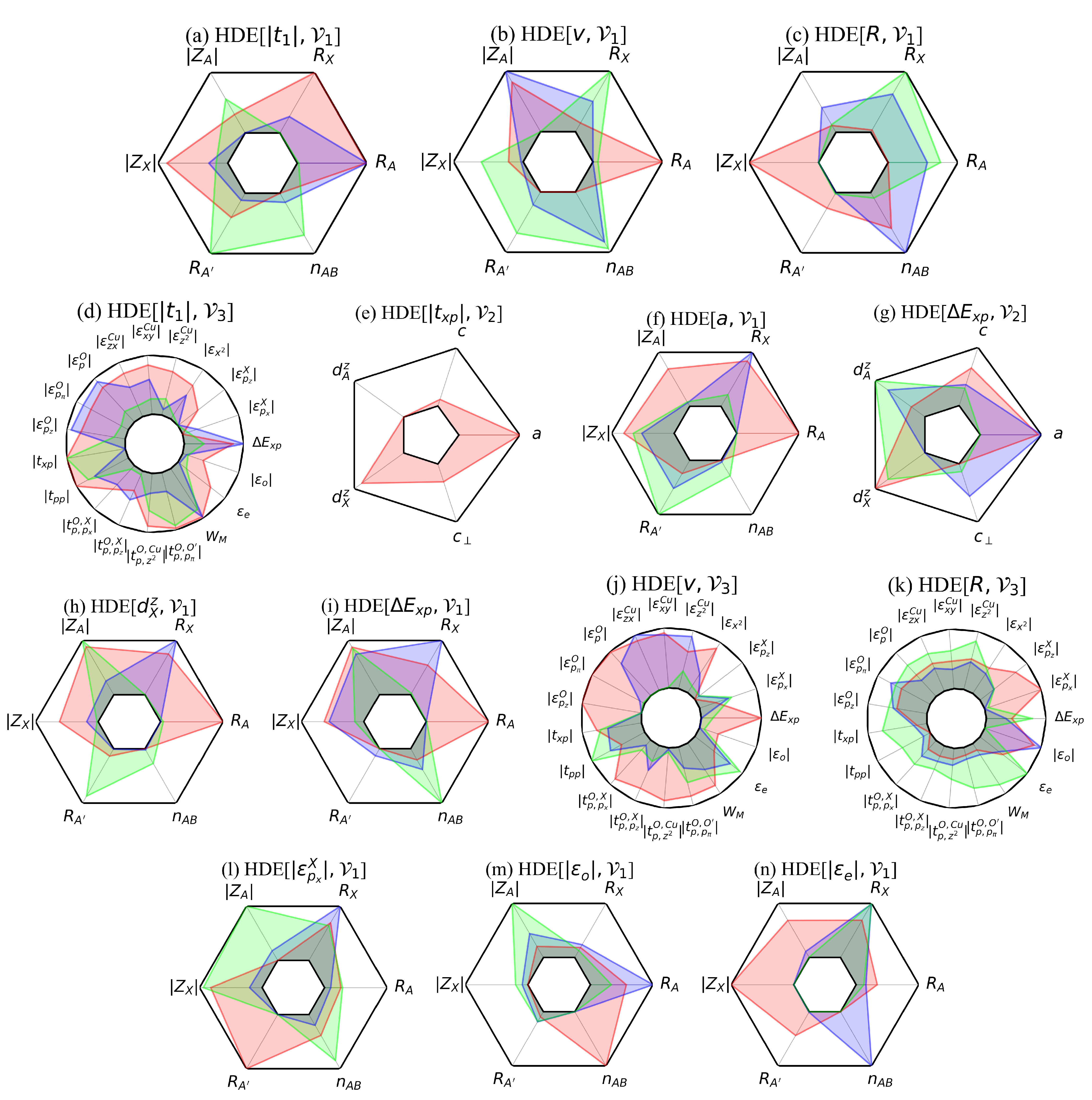}
\caption{\mycaption{
Score of each variable $x_i$ [Eq.~\eqref{eq:score}] in the dependence of $y$ at $g=1$ (red), $g=2$ (blue) and $g=3$ (green)
in the HDE[$y,\mathcal{V}=\{ x_i \}$], for the items (I), (III), (IV) and (1-10) in Sec.~\ref{sec:results}.
The inner and upper bounds of the histogram correspond to $s_{(g)}[y,x_i] = 0$ and $s_{(g)}[y,x_i] = 1$, respectively.
\jbmd{The numerical values of the score are given in Sec.~S1 of SM~\cite{Moree2024Supplemental_PRX}.
The panel (e) is restricted to $g=1$.
}
}}
\label{fig:score}
\end{figure*}

\paragraph*{$({\rm I})$ Dependence of $|t_1|$ on $\mathcal{V}_1$ ---}
See Fig.~\ref{fig:score}(a):
At $g=1$, $s_{(1)}[|t_1|,R_{\rm X}] = 1$,
but $s_{(1)}[|t_1|,R_{\rm A}]$ is very close to 1,
and $\tilde{f}_{(1)}[|t_1|,R_{\rm X}] = 0.68$  is not close to 1.
Then, at $g=2$ \jbmb{[Eq.~\eqref{eq:intro-t1}]}, $s_{(2)}[|t_1|,R_{\rm A}] = 1$,
$\tilde{f}_{(2)}[|t_1|,R_{\rm A}] = 0.88$,
and $R_{\rm A}$ is not in competition with other variables.
This corresponds to the scenario (S2),
and $R_{\rm X}$ and $R_{\rm A}$ have equal importance in the MOD2 of $|t_1|$.
\\

\paragraph*{$({\rm III})$ Dependence of $v$ on $\mathcal{V}_1$ ---}
See Fig.~\ref{fig:score}(b):
At $g=1$, $s_{(1)}[v,R_{\rm A}] = 1$ and $R_{\rm A}$ is not in close competition with other variables.
At $g=2$ \jbmb{[Eq.~\eqref{eq:intro-v}]}, $s_{(2)}[v,Z_{\rm A}] = 1$ and $Z_{\rm A}$ is not in close competition with other variables.
This corresponds to the scenario (S1).
Thus, $R_{\rm A}$ and $Z_{\rm A}$ correspond respectively to $x_{i^{\rm opt}_{1}}$ and $x_{i^{\rm opt}_{2}}$ unambiguously.
\\

\paragraph*{$({\rm IV})$ Dependence of $R$ on $\mathcal{V}_1$ ---}
See Fig.~\ref{fig:score}(c):
At $g=1$, $s_{(1)}[R,|Z_{\rm X}|] = 1$ and $|Z_{\rm X}|$ is not in competition with other variables.
At $g=2$ \jbmb{[Eq.~\eqref{eq:intro-R}]}, $s_{(2)}[R,n_{\rm AB}] = 1$ and $n_{\rm AB}$ is not in competition with other variables.
This corresponds to the scenario (S1).
Thus, $|Z_{\rm X}|$ and $n_{\rm AB}$ correspond respectively to $x_{i^{\rm opt}_{1}}$ and $x_{i^{\rm opt}_{2}}$ unambiguously.
\\

\paragraph*{$({\rm 1})$ Dependence of $|t_1|$ on $\mathcal{V}_3$ ---}
See Fig.~\ref{fig:score}(d):
At $g=1$, $s_{(1)}[|t_{1}|,|t_{xp}|] = 1$, 
but $s_{(1)}[|t_{1}|,|t_{pp}|]$, $s_{(1)}[|t_{1}|,|t^{O,O'}_{p,p_\pi}|]$ and $s_{(1)}[|t_{1}|,W_{\rm M}]$ are close to 1,
so that $|t_{xp}|$ is in competition with $|t_{pp}|$, $|t^{O,O'}_{p,p_\pi}|$ and $W_{\rm M}$.
$\tilde{f}_{(1)}[|t_{1}|,|t_{xp}|] = 0.95$ is rather close to 1.
This corresponds to the scenario (S3).
We identify the physical dependence of $|t_{1}|$ as that on $|t_{xp}|$.
\jbmd{(The intuitive dependence of $|t_1|$ on $|t_{xp}|$ is discussed in Sec.~\ref{sec:results}.)}
At $g=2$ \jbmb{[Eq.~\eqref{eq:t1s3}]}, $s_{(2)}[|t_1|,\Delta E_{xp}] = 1$,
but $s_{(1)}[|t_{1}|,W_{\rm M}]$, $s_{(1)}[|t_{1}|,|\jbmd{\epsilon}^{\rm O}_{p}|]$, $s_{(1)}[|t_{1}|,|\jbmd{\epsilon}^{\rm O}_{p_{\pi}}|]$ and $s_{(1)}[|t_{1}|,|\jbmd{\epsilon}^{\rm O}_{p_z}|]$ are also close to 1, because $\Delta E_{xp}$ is physically correlated with these variables. %
We identify the physical dependence as that on $\Delta E_{xp}$.
\jbmd{(The intuitive dependence of $|t_1|$ on $\Delta E_{xp}$ is discussed in Sec.~\ref{sec:results}.)}
\\

\paragraph*{$({\rm 2})$ Dependence of $|t_{xp}|$ on $\mathcal{V}_2$ ---}
See Fig.~\ref{fig:score}(e):
At $g=1$ \jbmb{[Eq.~\eqref{eq:main_txp_a}]}, $s_{(1)}[|t_{xp}|,a] = 1$,
and $a$ is not in very close competition with other variables.
(We have $s_{(1)}[|t_{xp}|,d^{z}_{\rm X}] = 0.93$, which is not so far from 1 but not very close either.)
Also, $\tilde{f}_{(1)}[|t_{xp}|,a] = 1.000$.
This corresponds to the scenario (S1),
and the dominance of $a$ in the MOD1 of $|t_{xp}|$ is unambiguous.
\\

\paragraph*{$({\rm 3})$ Dependence of $a$ on $\mathcal{V}_1$ ---}
See Fig.~\ref{fig:score}(f):
At $g=1$, $s_{(1)}[a,R_{\rm A}] = 1$,
$\tilde{f}_{(1)}[a,R_{\rm A}] = 0.80$,
 and $R_{\rm A}$ is not in close competition with other variables.
At $g=2$ \jbmb{[Eq.~\eqref{eq:main_a_vs1}]}, $s_{(2)}[a,R_{\rm X}] = 1$,
$\tilde{f}_{(2)}[a,R_{\rm X}] = 0.97$,
and $R_{\rm X}$ is not in competition with other variables.
This corresponds to the scenario (S1).
Thus, $R_{\rm A}$ and $R_{\rm X}$ correspond respectively to $x_{i^{\rm opt}_{1}}$ and $x_{i^{\rm opt}_{2}}$ unambiguously.
\\

\paragraph*{$({\rm 4})$ Dependence of $\Delta E_{xp}$ on $\mathcal{V}_2$ ---}
See Fig.~\ref{fig:score}(g):
At $g=1$, $s_{(1)}[\Delta E_{xp},d^{z}_{X}] = 1$, but $s_{(1)}[\Delta E_{xp},a] = 0.97$, so that $a$ is in close competition with $d^{z}_{X}$.
We have $\tilde{f}_{(1)}[\Delta E_{xp},d^{z}_{X}] = 0.93$.
At $g=2$ \jbmb{[Eq.~\eqref{eq:main_dexp_vs2_0}]}, $s_{(2)}[\Delta E_{xp},a] = 1$ and $\tilde{f}_{(2)}[\Delta E_{xp},a] = 0.94$, \jbmd{and $a$ is not in close competition with other variables}.
This corresponds to the scenario (S2):
\jbmd{the equal importance of $d^{z}_{X}$ and $a$ in the dependence of $|t_{xp}|$ is discussed in the main text.}
Consistently, in Fig.~\ref{fig:results-t1-interm}(c), 
the amplitude of the variation in $\Delta E_{xp}$ with either $d^{z}_{X}$ or $a$ looks similar:
The color map has a diagonal-like pattern.
\\

\paragraph*{$({\rm 5})$ Dependence of $d^{z}_{X}$ on $\mathcal{V}_1$ ---}
See Fig.~\ref{fig:score}(h):
At $g=1$, $s_{(1)}[d^{z}_{X},R_{\rm A}] = 1$ and $R_{\rm A}$ is not in \jbmd{close} competition with other variables.
At $g=2$ \jbmb{[Eq.~\eqref{eq:main_dza_vs1_0}]}, $s_{(2)}[d^{z}_{X},R_{\rm X}] = 1$ and $R_{\rm X}$ is not in competition with other variables.
This corresponds to the scenario (S1).
Thus, $R_{\rm A}$ and $R_{\rm X}$ correspond respectively to $x_{i^{\rm opt}_{1}}$ and $x_{i^{\rm opt}_{2}}$ unambiguously.
\\

\paragraph*{$({\rm 4,5})$ Dependence of $\Delta E_{xp}$ on $\mathcal{V}_1$ ---}
See Fig.~\ref{fig:score}(i):
At $g=1$, $s_{(1)}[\Delta E_{xp},R_{\rm A}] = 1$ and $R_{\rm A}$ is not in close competition with other variables.
We have $\tilde{f}_{(1)}[\Delta E_{xp},R_{\rm A}] = 0.90$.
At $g=2$, $s_{(2)}[\Delta E_{xp},R_{\rm X}] = 1$ and $R_{\rm X}$ is not in \jbmd{close} competition with other variables.
We have $\tilde{f}_{(2)}[\Delta E_{xp},R_{\rm X}] = 0.96$.
This corresponds to the scenario (S1).
Thus, $R_{\rm A}$ and $R_{\rm X}$ correspond respectively to $x_{i^{\rm opt}_{1}}$ and $x_{i^{\rm opt}_{2}}$ unambiguously \jbmb{in Eq.~\eqref{eq:main_dexp_vs1}.}
\\

\paragraph*{$({\rm 6})$ Dependence of $v$ on $\mathcal{V}_3$ ---}
See Fig.~\ref{fig:score}(j):
At $g=1$, $s_{(1)}[v,|\jbmd{\epsilon}^{\rm O}_{p_z}|] = 1$, 
but $s_{(1)}[v,|\jbmd{\epsilon}^{\rm O}_{p_\sigma}|]$,  $s_{(1)}[v,|\jbmd{\epsilon}^{\rm O}_{p_\pi}|]$ and $s_{(1)}[v,\Delta E_{xp}]$ are all close to 1,
so that $|\jbmd{\epsilon}^{\rm O}_{p_z}|$ is in close competition with $|\jbmd{\epsilon}^{\rm O}_{p_\sigma}|$, $|\jbmd{\epsilon}^{\rm O}_{p_\pi}|$ and $\Delta E_{xp}$.
Also, $\tilde{f}_{(1)}[v,|\jbmd{\epsilon}^{\rm O}_{p_z}|] = 0.97$ is close to 1.
This corresponds to the scenario (S3).
We identify the physical dependence of $v$ as that on $\Delta E_{xp}$ in Sec.~\ref{sec:results}.
At $g=2$, $s_{(2)}[v, |\jbmd{\epsilon}^{\rm Cu}_{z^2}|] = 1$ and $|\jbmd{\epsilon}^{\rm Cu}_{z^2}|$ is not in very close competition with other variables.
However, the physical meaning of this result is biased by the fact that $|\jbmd{\epsilon}^{\rm O}_{p_z}|$ corresponds to $x_{i^{\rm opt}_{1}}$, which is not physical as discussed above.
If we consider the HDE[$v,\{ \Delta E_{xp}, |t_{xp}|\}$], we obtain 
$s_{(1)}[v, \Delta E_{xp}] = 1$ 
then $s_{(2)}[v, |t_{xp}|] = 1$
\jbmb{in Eq.~\eqref{eq:main_v_vs3},}
as discussed in Sec.~\ref{sec:results}.
\\

\paragraph*{$({\rm 7})$ Dependence of $R$ on $\mathcal{V}_3$ ---}
See Fig.~\ref{fig:score}(k):
At $g=1$, $s_{(1)}[R,|\epsilon^{\rm X}_{p_x}|] = 1$ and $|\epsilon^{\rm X}_{p_x}|$ is not in \jbmd{close} competition with other variables.
At $g=2$, $s_{(2)}[R,|\epsilon^{}_{o}|] = 1$ and $|\epsilon^{}_{o}|$ is not in competition with other variables.
At $g=3$ \jbmb{[Eq.~\eqref{eq:R-3var}]}, $s_{(3)}[R,\epsilon^{}_{e}] = 1$ and $\epsilon^{}_{e}$ is not in competition with other variables.
This corresponds to the scenario (S1).
Thus, $|\epsilon^{\rm X}_{p_x}|$, $|\epsilon_o|$ and $\epsilon_e$ correspond respectively to $x_{i^{\rm opt}_{1}}$, $x_{i^{\rm opt}_{2}}$ and $x_{i^{\rm opt}_{3}}$ unambiguously.
\\

\paragraph*{$({\rm 8})$ Dependence of $|\epsilon^{\rm X}_{p_x}|$ on $\mathcal{V}_1$ ---}
See Fig.~\ref{fig:score}(l):
At $g=1$, $s_{(1)}[|\epsilon^{\rm X}_{p_x}|,R_{\rm A'}] = 1$ and $R_{\rm A'}$ is not in \jbmd{close} competition with other variables.
At $g=2$ \jbmb{[Eq.~\eqref{eq:R-g1}]}, $s_{(2)}[|\epsilon^{\rm X}_{p_x}|,R_{\rm X}] = 1$ and $R_{\rm X}$ is not in competition with other variables.
This corresponds to the scenario (S1).
Thus, $R_{\rm A'}$ and $R_{\rm X}$ correspond respectively to $x_{i^{\rm opt}_{1}}$ and $x_{i^{\rm opt}_{2}}$ unambiguously.
\\

\paragraph*{$({\rm 9})$ Dependence of $|\epsilon_{o}|$ on $\mathcal{V}_1$ ---}
See Fig.~\ref{fig:score}(m):
At $g=1$, $s_{(1)}[|\epsilon_{o}|,n_{\rm AB}] = 1$ and $n_{\rm AB}$ is not in competition with other variables.
At $g=2$ \jbmb{[Eq.~\eqref{eq:R-g2}]}, $s_{(2)}[|\epsilon_{o}|,R_{\rm A}] = 1$ and $R_{\rm A}$ is not in competition with other variables.
This corresponds to the scenario (S1).
Thus, $n_{\rm AB}$ and $R_{\rm A}$ correspond respectively to $x_{i^{\rm opt}_{1}}$ and $x_{i^{\rm opt}_{2}}$ unambiguously.
\\

\paragraph*{$({\rm 10})$ Dependence of $\epsilon_{e}$ on $\mathcal{V}_1$ ---}
See Fig.~\ref{fig:score}(n):
At $g=1$, $s_{(1)}[\epsilon_{e},|Z_{\rm X}|] = 1$ and $|Z_{\rm X}|$ is not in competition with other variables.
At $g=2$ \jbmb{[Eq.~\eqref{eq:R-g3}]}, $s_{(2)}[\epsilon_{e},n_{\rm AB}] = 1$ and $n_{\rm AB}$ is not in competition with other variables.
This corresponds to the scenario (S1).
Thus, $|Z_{\rm X}|$ and $n_{\rm AB}$ correspond respectively to $x_{i^{\rm opt}_{1}}$ and $x_{i^{\rm opt}_{2}}$ unambiguously.

\section{\jbmc{Hole doping dependence of screening}}
\label{app:xfcl}

\renewcommand{\theequation}{F\arabic{equation}}

\begin{figure}%
\includegraphics[scale=0.17]{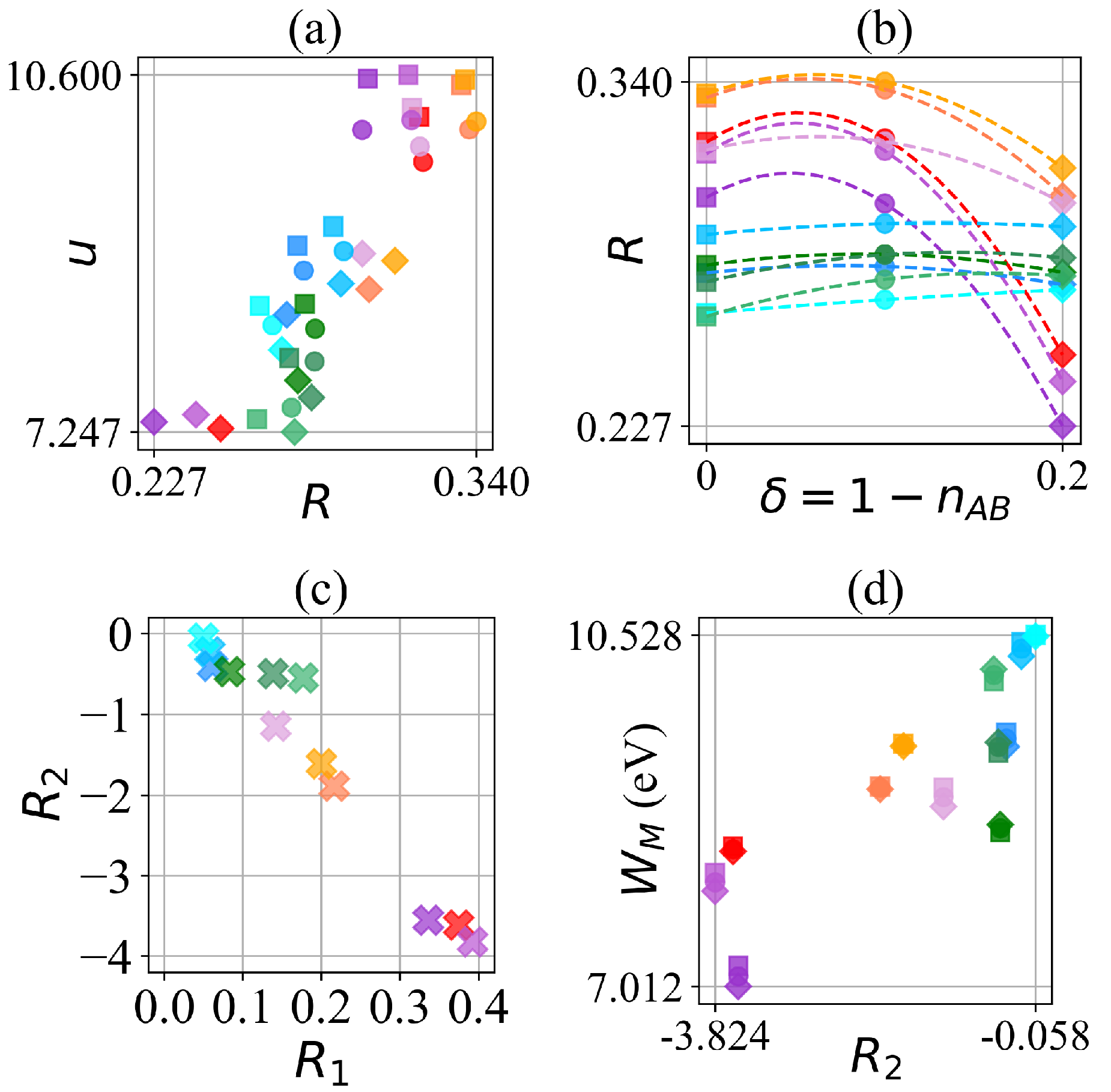}
\caption{
Panel (a): Values of $u$ and $R$ taken from Fig.~\ref{fig:chart-t1u}.
Panel (b): Values of $R$ as a function of hole doping $\delta = 1 - n_{\rm AB}$. The dashed curves show the quadratic interpolation of the $\delta$ dependence of $R$ by using Eq.~\eqref{eq:quad}.
Panel (c): Values of the coefficients $R_1$ and $R_2$ in Eq.~\eqref{eq:quad}.
Panel (d): Values of $W_{\rm M}$ as a function of $R_2$.
For each color point, the corresponding CF is shown in Fig.~\ref{fig:chart-comp}.
In the panel (c), the colors of the crosses correspond to the values of A, X and A' in Fig.~\ref{fig:chart-comp}.
}
\label{fig:xfcl}
\end{figure}

Here, as a complement to \jbmd{Sec.~\ref{sec:results-R} and} the MOD2 of $u$ and $R$ in Eqs.~\eqref{eq:intro-u} and~\eqref{eq:intro-R},
we discuss the $n_{\rm AB}$ dependence of $u$ and $R$ \jbmd{when the CF at $\delta=0$ is fixed}.
The values of $u$ and $R$ \jbmd{for all compounds in the training set} are shown in Fig.~\ref{fig:xfcl}(a).
For X = F, Cl and especially (X, A) = (F, Ba), (Cl, Ba) and (Cl, Sr), 
we observe a sharp decrease in $u$ between 10\% and 20\% hole doping (i.e. between $n_{\rm AB}=0.9$ and $n_{\rm AB}=0.8$).
This decrease in $u$ is caused by the sharp decrease in $R$ as seen in Fig.~\ref{fig:xfcl}(a,b).
Although the decrease in $R$ with increasing $\delta$ (or decreasing $n_{\rm AB}$) is consistent with the MOD2 in Eq.~\eqref{eq:intro-R},
the sharper decrease in $R$ for X = F, Cl compared to X = O is not captured by Eq.~\eqref{eq:intro-R}.
For completeness, we discuss the origin of \jbmd{the sharper decrease in $R$ for X = F, Cl} in detail here.

The decrease in the $\delta$ dependence of $R$ may be quantified by considering the quadratic interpolation
\begin{equation}
R(\delta) = R_0 + R_1 \delta + R_2 \delta^2,
\label{eq:quad}
\end{equation}
where $R_0$, $R_1$ and $R_2$ are determined entirely by the \textit{ab initio} values of $R$ at $\delta=0.0,$ $0.1$ and $0.2$:
$R_0$ is $R$ at $\delta=0$ in Fig.~\ref{fig:xfcl}(b), and the values of $R_1$ and $R_2$ are given in Fig.~\ref{fig:xfcl}(c).
The decrease in $R$ is encoded in $R_2 < 0$ \jbmd{for all compounds},
and the sharp decrease in $R$ for (X, A) = (F, Ba), (Cl, Ba) and (Cl, Sr)
is reflected in the high value of $|R_2|$ compared to the other compounds.
Other compounds with X = F, Cl also have higher values of $|R_2|$ compared to X = O.
Namely, we have $|R_2| \leq 0.55$ for X = O, 
$|R_2| \geq 1.14$ for X = F, Cl,
and $|R_2| \geq 3.55$ for (X, A) = (F, Ba), (Cl, Ba) and (Cl, Sr).

\jbmd{Possible causes of the higher value of $|R_2|$ are (i) the sharp decrease in $|\epsilon_o|$ from $\delta=0.1$ to $\delta=0.2$ for X = F, Cl (see Fig.~\ref{fig:chart-cryst}), and also (ii)} the lower value of the M space bandwidth $W_{\rm M}$ \jbmd{for X = F, Cl.}
\jbmd{On (ii), we see in Fig.~\ref{fig:xfcl}(d)} that the three compounds with (X, A) = (F, Ba), (Cl, Ba) and (Cl, Sr) have the lowest $W_{\rm M}$ at $\delta = 0.2$.
\jbmd{The above discussed dependence of $|R_2|$ on $W_{\rm M}$ (ii) is} interpreted as follows.
If $W_{\rm M}$ is lower, then the charge transfer energies between the occupied states in the M space
and the empty states in the M space (namely, the empty part of the AB band) are smaller.
Thus, the intra-M space cRPA screening will be stronger.
A rough scaling of the intra-M space screening is $1/W_{\rm M}$ [see Eq.~\eqref{eq:chiou}],
so that the smaller $W_{\rm M}$, the more $R$ will decrease when $W_{\rm M}$ further decreases.
[And, \jbmd{for X = F, Cl,} $W_{\rm M}$ decreases with increasing $\delta$ as seen in Fig.~\ref{fig:xfcl}(d).]
On the other hand, in X = F, Cl, the screening channel between the occupied states and empty states outside M
is relatively weak compared to X = O due to the larger $\epsilon_{e}$ (see Fig.~\ref{fig:chart-cryst}).
This suggests the intra-M space screening dominates over the other screening channels for X = F, Cl.
\jbmd{(Note that, besides the decrease in $W_{\rm M}$, the decrease in $|\epsilon_{o}|$ with increasing $\delta$ also participates in increasing the intra-M space screening.)}

The higher $|R_2|$ in X = F, Cl compared to X = O implies a nontrivial point on the origin of the SC \jbmd{in oxychlorides and oxyfluorides}.
[Confirmation of this point requires to improve the derivation of the AB LEH at the c$GW$-SIC+LRFB level, which is left for future studies.]
Even though $u$ and $R$ are higher in the undoped compound for X = F, Cl compared to X = O [according to Eq.~\eqref{eq:intro-u} and~\eqref{eq:intro-R}], $u$ may become lower in the hole doped compound at optimal hole doping \jbmd{for X = F, Cl} compared to X = O due to the higher $|R_2|$.
In particular, $u$ may fall into the weak-coupling regime ($u \simeq 6.5-8.0$) at $\delta \simeq 0.2$ for X = F, Cl\jbmd{, which may correspond to the optimal hole doping}.
For instance, in Table~\ref{tab:expsc}, the experimental $T_{c}$ in Ca$_{2-x}$Na$_x$CuO$_2$Cl$_2$ is $T_{\rm c}^{\rm exp} \simeq 27$ K \cite{Kim2006}, and this is realized at $\delta=0.18$, which is close to $\delta=0.2$ at which the sharp decrease in $R$ happens. If $u$ is in the weak-coupling regime at $\delta=0.18$, then the lower $T_c$ compared to other cuprates is caused by the decrease in $F_{\rm SC}$ with decreasing $u$ in the weak-coupling regime.

\section{\jbmd{Density of states near the Fermi level in hole-doped oxychlorides}}
\label{app:doseo}
\renewcommand{\theequation}{G\arabic{equation}}

In the item (9) (Dependence of $|\epsilon_{o}|$ on $\mathcal{V}_{1}$) in Sec.~\ref{sec:results-R},
we mention the increase in $|\epsilon_{o}|$ with decreasing $R_{\rm A}$ in the MOD2 of $|\epsilon_{o}|$ 
[Eq.~\eqref{eq:R-g2}].
Here, we discuss the interpretation of the increase in $|\epsilon_{o}|$ with decreasing $R_{\rm A}$.
To do so, we consider as an example
the three CFs in the training set
such that $\delta = 0.2$,
X = Cl and 
A = 
Ba$_{2-\delta}$Cs$_{\delta}$ (i), 
Sr$_{2-\delta}$Rb$_{\delta}$ (ii), and 
Ca$_{2-\delta}$K$_{\delta}$ (iii).
We choose the three above CFs (i-iii), 
because (i) has the lowest value of $|\epsilon_{o}|$ among the compounds in the training set
($|\epsilon_{o}|=0.134$ eV; see the purple diamond marker in Fig.~\ref{fig:chart-cryst}), 
and the value of $R_{\rm A}$ is progressively reduced from (i) to (iii).
[We have $R_{\rm A}=1.508$ \AA \ for (i), $R_{\rm A}=1.354$ \AA \ for (ii), and $R_{\rm A}=1.178$ \AA \ for (iii)].

\begin{figure}[!htb]
\includegraphics[scale=0.45]{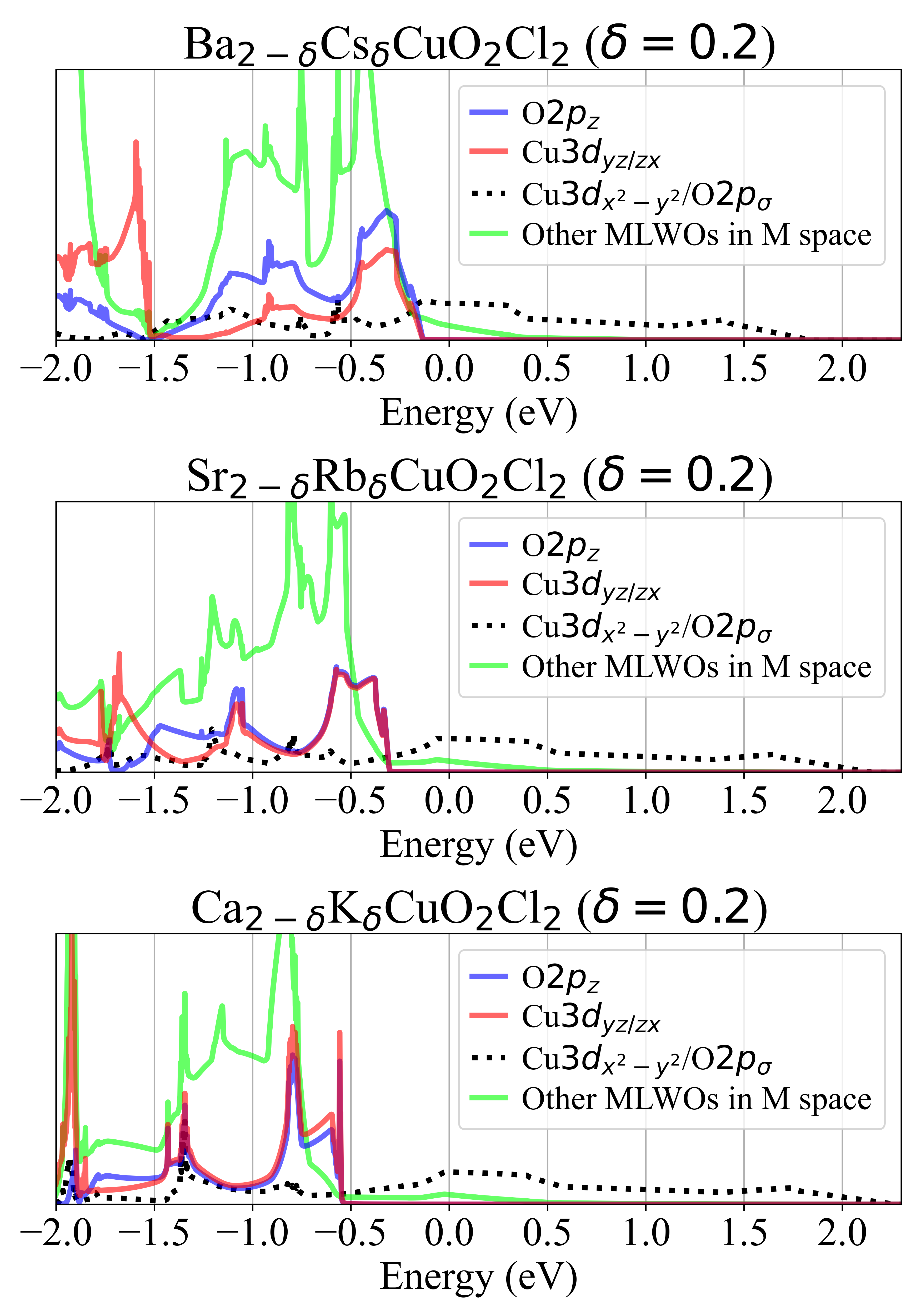}
\caption{
\jbmd{Partial densities of states within the M space and near the Fermi level for 
Ba$_{2-\delta}$Cs$_{\delta}$CuO$_2$Cl$_2$,
Sr$_{2-\delta}$Rb$_{\delta}$CuO$_2$Cl$_2$,
and Ca$_{2-\delta}$K$_{\delta}$CuO$_2$Cl$_2$,
at hole doping $\delta = 0.2$.
}
}
\label{fig:doseo}
\end{figure}

The interpretation of the increase in $|\epsilon_{o}|$ with decreasing $R_{\rm A}$
is based on two observations:
(a) The value of $|\epsilon_{o}|$ mainly depends on the O$2p_{z}$ and Cu$3d_{yz/zx}$ MLWOs,
and
(b) the O$2p_{z}$ and Cu$3d_{yz/zx}$ MLWOs are deeper in energy when $R_{\rm A}$ decreases.
Below, we discuss (a) and (b).

First, we discuss (a).
Because $\epsilon_{o}$ is the highest energy of the occupied bands outside the AB band,
the value of $|\epsilon_{o}|$ mainly depends on the orbitals that form the density of states %
near the Fermi level,
excluding the Cu$3d_{x^2-y^2}$ and O$2p_{\sigma}$ orbitals.
Thus, we examine the character of the occupied bands near the Fermi level, for the three above CFs.
To do so, we represent the partial density of states (pDOS) for several types of MLWOs in Fig.~\ref{fig:doseo}.
Near the Fermi level, the Cu$3d_{x^2-y^2}$/O$2p_{\sigma}$ pDOS corresponds to the AB subspace,
and the character of the highest occupied bands outside the AB band is given by the pDOS other than Cu$3d_{x^2-y^2}$/O$2p_{\sigma}$.
Outside the AB band, the density of states near the Fermi level is dominated by the O$2p_{z}$ pDOS and Cu$3d_{yz/zx}$ pDOS (see Fig.~\ref{fig:doseo}).
Thus, the value of $|\epsilon_{o}|$ mainly depends on the O$2p_{z}$ and Cu$3d_{yz/zx}$ MLWOs.
In particular, $|\epsilon_{o}|$ increases if the O$2p_{z}$ and Cu$3d_{yz/zx}$ orbitals are deeper in energy.

Now, we discuss (b), i.e. why the O$2p_{z}$ and Cu$3d_{yz/zx}$ orbitals are deeper in energy when $R_{\rm A}$ decreases.
If $R_{\rm A}$ is reduced, then $a$ is reduced [see Eq.~\eqref{eq:main_a_vs1}],
and the in-plane O is closer to the A cation.
Thus, the MP$^{+}$ from the A cation that is felt by the in-plane O is stronger.
[See Fig.~\ref{fig:cation_madelung}(a) for an illustration.]
Thus, the O$2p$ MLWOs in the M space are stabilized,
i.e. their onsite energy is reduced.
This shifts the O$2p$ pDOS downwards, 
i.e. farther from the Fermi level.
With respect to the O atom in the unit cell,
the positions of the four nearest A cations (in Cartesian coordinates)
are $\pm a/2 {\bf y} \pm d^{z}_{\rm A} {\bf z}$ (${\bf y}$ and ${\bf z}$ are unitary vectors along the $y$ and $z$ directions that are considered in Table~\ref{tab:atompos}):
The O$2p_{\sigma}$ orbital extends along $x$ direction and thus avoids the A cations,
whereas the O$2p_{z}$ orbital extends along $z$ direction,
so that the O$2p_{z}$ electrons are closer to the A cation compared to the O$2p_{\sigma}$ electrons.
This explains why the O$2p_{z}$ electrons are prominently affected by the MP$^{+}$ from the A cation.
Similarly, the Cu$3d_{yz/zx}$ orbitals extend along the $z$ direction, and may be prominently affected by the MP$^{+}$ from the A cation.
The above discussion is supported by the values of the onsite energies of the MLWOs:
From (i) to (iii), 
$\epsilon^{\rm O}_{p}$ decreases by $0.97$ eV, whereas
$\epsilon^{\rm O}_{p_z}$ decreases by $1.16$ eV,
so that the O$2p_z$ orbital is indeed more stabilized than the O$2p_{\sigma}$ orbital.
Also,
$\epsilon^{\rm Cu}_{x}$ decreases by $0.26$ eV, whereas
$\epsilon^{\rm Cu}_{yz/zx}$ decreases by $0.48$ eV,
so that the Cu$3d_{yz/zx}$ orbital is indeed more stabilized than the Cu$3d_{x^2-y^2}$ orbital.

\end{appendices}


\begin{thebibliography}{51}
\bibitem [{Mor()}]{Moree2024Supplemental_PRX}%
  \BibitemOpen
  \href@noop {} {}\bibinfo {note} {The {S}upplemental {M}aterial may be obtained from the authors upon request.}\BibitemShut {Stop}%
\end{thebibliography}
\end{document}